\documentclass[journal=jctcce,manuscript=article]{achemso}

\usepackage[version=3]{mhchem} 
\usepackage{lipsum}
\usepackage{accents}
\usepackage{mathrsfs}
\usepackage{physics}
\usepackage{braket}
\usepackage{mathtools}
\usepackage{amsfonts}
\usepackage{amsmath}
\usepackage{caption}
\usepackage{subcaption}
\usepackage{float}
\usepackage{geometry}
\usepackage{natbib}
\usepackage{setspace}
\usepackage{xkeyval}
\usepackage{multirow}
\usepackage{makecell}
\usepackage{stackengine}
\usepackage{nicefrac}
\usepackage{titlesec}
\titleformat{\section}{\large\bfseries}{\thesection}{1em}{}
\usepackage[acronym]{glossaries}
\usepackage[usenames,dvipsnames]{color}
\DeclareMathAlphabet{\mathpzc}{OT1}{pzc}{m}{it}
\DeclareFontFamily{U}{mathx}{\hyphenchar\font45}
\DeclareFontShape{U}{mathx}{m}{n}{
      <5> <6> <7> <8> <9> <10>
      <10.95> <12> <14.4> <17.28> <20.74> <24.88>
      mathx10
      }{}
\DeclareSymbolFont{mathx}{U}{mathx}{m}{n}
\DeclareFontSubstitution{U}{mathx}{m}{n}
\DeclareMathAccent{\widecheck}{0}{mathx}{"71}
\DeclareMathAccent{\wideparen}{0}{mathx}{"75}
\DeclareMathAccent{\widebar}{0}{mathx}{"73}

\newcommand{\angstrom}{\text{\normalfont\AA}}

\newcommand{\bPsi}{\boldsymbol{\Psi}}

\newcommand{\cb}{\color{black}}
\newcommand{\cn}{\color{black}}

\newcommand{\bk}{\boldsymbol{k}}

\newcommand{\bx}{\boldsymbol{\textbf{x}}}

\newcommand{\bC}{\boldsymbol{\textbf{C}}}
\newcommand{\bCbar}{\boldsymbol{\widebar{\textbf{C}}}}
\newcommand{\bChat}{\boldsymbol{\widehat{\textbf{C}}}}
\newcommand{\bD}{\boldsymbol{\textbf{D}}}

\newcommand{\bE}{\boldsymbol{\textbf{E}}}

\newcommand{\bH}{\boldsymbol{\textbf{H}}}

\newcommand{\bM}{\boldsymbol{\textbf{M}}}

\newcommand{\bO}{\boldsymbol{\textbf{O}}}

\newcommand{\bQ}{\boldsymbol{\textbf{Q}}}
\newcommand{\bR}{\boldsymbol{\textbf{R}}}
\newcommand{\bS}{\boldsymbol{\textbf{S}}}

\newcommand{\bPhi}{\boldsymbol{\Phi}}

\newcommand{\bhPhi}{\widehat{\bPhi}}
\newcommand{\barbelow}[1]{\stackunder[1.4pt]{$#1$}{\rule{1.5ex}{.075ex}}}
\newcommand{\bPhicheck}{\boldsymbol{\barbelow \Phi}}
\newcommand{\bPhihatcheck}{\boldsymbol{\barbelow \bhPhi}}
\newcommand{\bPsicheck}{\boldsymbol{\barbelow \Psi}}

\newcommand{\dx}{\,d\bx}

\newcommand{\Hphiop}{\mathpzc{H}^{\phi}}

\newcommand{\ordercomplexity}{\mathcal{O}}

\newcommand{\QE}{\texttt{QE}}
\newcommand{\Lob}{\texttt{LOBSTER}}
\newcommand{\DFTFE}{\texttt{DFT-FE}}
\newcommand{\oop}{\acrshort{pfoop}}
\newcommand{\ohp}{\acrshort{pfohp}}
\newcommand{\hop}{\acrshort{pfhop}}
\newcommand{\hhp}{\acrshort{pfhhp}}
\newcommand{\oder}{\acrshort{pfode}}
\newcommand{\hder}{\acrshort{pfhde}}
\newcommand{\pfop}{\acrshort{pfop}}
\newcommand{\pfhp}{\acrshort{pfhp}}
\newcommand{\ipohp}{\acrshort{ipohp}}
\newcommand{\pa}{\acrshort{pa}}
\newcommand{\cop}{\acrshort{cop}}
\newcommand{\chp}{\acrshort{chp}}

\makeglossaries
\newacronym{pfop}{\textcolor{black}{\texttt{pOA}}}{Projected  orbital  analysis }
\newacronym{pfhp}{\textcolor{black}{\texttt{pHA}}}{Projected  Hamiltonian  analysis} 
\newacronym{pfoop}{\textcolor{black}{\texttt{pOOP}}}{projected  orbital overlap population}
\newacronym{pfohp}{\textcolor{black}{\texttt{pOHP}}}{projected orbital Hamilton population}
\newacronym{ipohp}{\texttt{IPOHP}}{Integrated projected orbital Hamilton population} 
\newacronym{pfhop}{\textcolor{black}{\texttt{pHOP}}}{projected  Hamiltonian overlap population}
\newacronym{pfhhp}{\textcolor{black}{\texttt{pHHP}}}{projected Hamiltonian Hamilton population}
\newacronym{pfode}{\texttt{pODE}}{projected orbital density error}
\newacronym{pfhde}{\texttt{pHDE}}{projected Hamiltonian density error}
\newacronym{pa}{\texttt{PA}}{Pseudo-atomic}
\newacronym{cop}{\texttt{pCOOP}}{projected crystal orbital-overlap population}
\newacronym{chp}{\texttt{pCOHP}}{projected crystal orbital-Hamilton population}

\newcommand{\comment}[1]{}
\author{Kartick Ramakrishnan}
\author{Sai Krishna Kishore Nori}
\affiliation[Indian Institute of Science, Bangalore]
{Department of Computational and Data Sciences, Indian Institute of Science, Bangalore}

\author{Seung-Cheol Lee}
\affiliation[]
{Indo Korea Science and Technology Center, Bangalore, India}
\author{Gour P Das}
\affiliation[TCG-Crest]
{Research Institute for Sustainable Energy (RISE),  
TCG Center for Research and Education in Science and Technology,   
Salt Lake, Kolkata, India}
\author{Satadeep Bhattacharjee}
\affiliation[]
{Indo Korea Science and Technology Center, Bangalore, India}
\author{Phani Motamarri}
\affiliation[Indian Institute of Science, Bangalore]
{Department of Computational and Data Sciences, Indian Institute of Science, Bangalore}
\email{phanim@iisc.ac.in}

\title
 {Chemical bonding in large systems using projected population analysis from real-space density functional theory calculations}

\abbreviations{pFOOP, pFOHP, pFHOP, pFHHP, pFOP, pFHP, IFOHP, pFODE, pFHDE, DFTFE, QE }

\begin{document}








\begin{abstract}
We present an efficient and scalable computational approach for conducting projected population analysis from real-space finite-element (FE) based Kohn-Sham density functional theory calculations (\DFTFE). This work provides an important direction towards extracting chemical bonding information from large-scale DFT calculations on materials systems involving thousands of atoms while accommodating periodic, semi-periodic or fully non-periodic boundary conditions. Towards this, we derive the relevant mathematical expressions and develop efficient numerical implementation procedures that are scalable on multi-node CPU architectures to compute the projected overlap and Hamilton populations. The population analysis is accomplished by projecting either the self-consistently converged FE discretized Kohn-Sham orbitals, or the FE discretized Hamiltonian onto a subspace spanned by localized atom-centered basis set. The proposed methods are implemented in a unified framework within \DFTFE~code where the ground-state DFT calculations and the population analysis are performed on the same FE grid.  We further benchmark the accuracy and performance of this approach on representative material systems involving periodic and non-periodic DFT calculations with \Lob, a widely used projected population analysis code. Finally, we discuss a case study demonstrating the advantages of our scalable approach to extract the quantitative chemical bonding information of hydrogen chemisorbed in large silicon nanoparticles alloyed with carbon, a candidate material for hydrogen storage.
\end{abstract}
\printglossary[type=\acronymtype]

\section{Introduction}

 Approaches based on overlap~\cite{Mulli1,Mulli2,Hoff2} and Hamilton population analysis~\cite{Hoff2,Hoff1} are widely used to extract chemical bonding information in covalent material systems. The overlap population analysis is based on partitioning the number of electrons among distinct atoms and the orbitals around them, whereas the total electronic energy of a molecule or a crystal is partitioned in Hamilton population analysis. In the case of solid-state systems, these approaches are referred to as Crystal Orbital Overlap Population (COOP) originally discussed by Hughbanks and Hoffmann~\cite{Hoff1} and Crystal Orbital Hamilton Population (COHP) originally suggested by Dronskowski and Bl\"{o}chl~\cite{1993,steinberg2018crystal,RDronskowski2005}. Traditionally, these methods~\cite{cc1,cc2} were used within the framework of tight-binding linear combination of atomic orbitals (LCAO) or linearized muffin tin orbital approaches (LMTO)~\cite{lmto} which have minimal and a well localized atom-centered basis set. Building on these techniques are approaches like ``Balanced crystal orbital overlap population" (BCOOP)~\cite{BCOOP} and ``Crystal orbital bond index" (COBI)~\cite{COBI} for robust extraction of chemical bonding behaviour in solid-state materials. BCOOP was proposed in the context of less localized basis sets which are close to linear dependency, while COBI approach was proposed for studying multi-center interactions -via- a multi-center bond index. Another established way for analyzing chemical bonding in both molecular and solid-state systems is using the localized orbitals constructed as unitary transformations of extended single-particle eigenstates. For instance, maximally localized Wannier functions~\cite{MLWF} (the solid-state equivalent of Foster-Boys orbitals in quantum chemistry) and the recent Pipek-Mezey Wannier functions~\cite{pmwf} have been used for studying bonding characterization of crystalline and disordered materials.

Over the last few decades, plane-waves have become the popular choice of basis sets for electronic structure calculations due to the the systematic convergent nature of the basis set, offering spectral convergence rates to compute the ground-state properties of interest. Recent focus for extracting chemical bonding behaviour has been on projected population analysis~\cite{Deringer,maintz,LOBSTER,WOOP} as these methods combine the advantages of plane-wave-based DFT methods for accurately computing the electronic structure and minimal localized atom-centered basis for understanding chemical bonding properties as a post-processing step. In these methods, Kohn-Sham DFT eigenfunctions obtained from a plane-wave calculation are projected onto a subspace spanned by the localized atomic-orbital basis to compute energy-resolved quantities like overlap and Hamilton populations. 
The most popular and widely used code based on such a projected population analysis approach is \Lob\cite{LOBSTER}. Here, the Kohn-Sham eigenfunctions obtained from a Projector Augmented Wave (PAW) based DFT calculations using popular plane-wave based codes (e.g. VASP~\cite{vasp}, Quantum Espresso~\cite{qe}) are projected onto a subspace spanned by a localized atom-centered basis. While such a strategy has been largely successful, this approach has certain limitations to extract chemical bonding information. Firstly, plane-wave-based techniques often restrict the simulation domains to be periodic, which is incompatible with many application problems (e.g.: defects, nano-particles, charged systems). Furthermore, plane-wave basis provides uniform spatial resolution and is computationally inefficient in the study of defects, and isolated systems (e.g. molecules, clusters etc.) where a higher resolution is necessary to describe particular regions of interest and a coarse resolution suffices elsewhere. Moreover, plane-wave basis are extended in real-space and involve non-local communication between processors, affecting the scalability of computations on massively parallel computing architectures thereby restricting the material system sizes that can be simulated to a few hundreds of atoms. \Lob, which uses plane-wave discretized Kohn-Sham wavefunctions as an input to conduct projected population analysis suffers from the above limitations and is restricted only to bulk material systems (periodic systems) up to a maximum of hundred atoms and cannot be executed on more than 1 CPU node. 
Further, the use of multiple codes, such as the ground-state DFT calculation employing a plane-wave based code like VASP or quantum espresso, and the subsequent population analysis using \Lob~code, makes the process cumbersome and time-consuming.  Kundu \textit{et al.}~\cite{WOOP} recently proposed a population analysis where the Kohn-Sham occupied eigenspace obtained from a plane-wave DFT calculations are projected onto a localized Wannier orbital basis~\cite{MLWF}, thereby minimizing the projection error from plane-wave to localized atom-centered basis (spill factor~\cite{sanchez-portal}) due to the completeness of the Wannier functions. However, such an approach still suffers from the plane-wave basis limitations and further adds to the complexity of population analysis by requiring the use of three codes to complete three tasks (ground-state DFT calculation by a plane-wave code, Wannierization (using \texttt{wannier90}), and finally the population analysis code).  Currently, there are no computational methods available that can perform chemical bonding analysis from large-scale density functional theory (DFT) calculations using a systematically convergent basis set, while also having the ability to handle complex material systems with fully periodic, non-periodic, or semi-periodic boundary conditions. The aim of the current work is to address this gap and provide a solution to this problem. 
\par


Addressing the aforementioned limitations, we introduce here a real-space finite-element (FE) based density functional theory (\DFTFE) approach\cite{motamarri2020,motamarri2022} to conduct projected population analysis. FE basis set is a systematically convergent basis set comprising of a piece-wise polynomial of order $p$ and a strictly local basis set on which various electronic fields are represented. In contrast to widely used plane-wave based DFT calculations, the use of FE basis for DFT enables large-scale calculations (up to tens of thousands of electrons) and accommodates periodic, semi-periodic, and non-periodic boundary conditions. Additionally, the local character of the FE basis provides an inherent benefit in terms of parallel scalability of DFT-FE calculations in comparison to widely used DFT codes and has been tested up to ~100,000 cores on many-core CPUs~\cite{motamarri2020} and ~24,000 GPUs on hybrid CPU-GPU architectures~\cite{GB19,das2019,motamarri2022}. The proposed population analysis methodology developed within the framework of \DFTFE~inherits these advantages and enables scalable chemical bonding analysis in complex material systems. Furthermore, this methodology is developed as a unified approach that enables both Kohn-Sham DFT ground-state calculations and population analysis to be carried out within the same computational framework using the FE basis. \cb The framework opens up the possibility of extracting chemical bonding information for the first time in sizeable complex material systems critical in many technologically relevant applications, enabling efficient investigation of chemical bonding interactions in various scenarios, such as large-scale nanoparticles, layered materials with adsorbate-adsorbent interactions, complex defect-impurities interactions, bonding interactions between the migrating ion and the underlying solid electrolyte lattice in the presence of an electric field, and many more.\cn



We propose two methodologies for computing overlap and Hamilton populations -via- projected population analysis -- (a) projected orbital population analysis (\pfop), relying on orthogonally projecting the self-consistently converged FE discretized Kohn-Sham DFT eigenfunctions onto a subspace spanned by a minimal atomic-orbital basis set and is similar in spirit to \Lob~\cite{LOBSTER}, and (b) projected Hamiltonian population analysis (\pfhp), relying on orthogonally projecting the self-consistent FE discretized Hamiltonian onto the atomic-orbital subspace, a method motivated from the fact that many of the reduced scaling electronic structure codes targeted towards large-scale DFT calculations tend to avoid explicit computation of DFT eigenvectors with no explicit access to these eigenvectors for projection. 

The computational framework developed to implement the above methods hinges on the following key steps: (i) perform Kohn-Sham DFT ground-state calculation in \DFTFE~to compute the finite-element discretized eigenfunctions spanning the Kohn-Sham occupied eigenspace, (ii) construct the subspace spanned by the localized atom-centered orbitals $\mathbb{V}_{\phi}^{N_{orb}}$ (available as numerical data or analytical expressions) -via- interpolating these orbitals on the underlying finite-element grid, (iii) orthogonally project the occupied Kohn-Sham eigenfunctions onto $\mathbb{V}_{\phi}^{N_{orb}}$ in the case of \pfop, while orthogonally projecting the self-consistent Kohn-Sham FE discretized Hamiltonian onto $\mathbb{V}_{\phi}^{N_{orb}}$ in the case of \pfhp, (iv) compute the atom-centered orbital overlap matrix using the Gauss-Lobatto-Legendre quadrature rule, (v) compute the coefficient matrices corresponding to the representation of projected Kohn-Sham wavefunctions in the atom-centered orbital basis $\mathbb{V}_{\phi}^{N_{orb}}$ in the case of \pfop, while diagonalizing the projected Hamiltonian to obtain the eigenvector matrix in the subspace $\mathbb{V}_{\phi}^{N_{orb}}$ in the case of \pfhp,~(vi) using these coefficient matrices, evaluate the projected orbital overlap and Hamilton population in the case of \pfop, and evaluate the projected Hamiltonian overlap and Hamilton population in the case of \pfhp.

We evaluate the accuracy and performance of the proposed methods (\pfop~and \pfhp) on representative benchmark examples involving isolated molecules (CO, H$_2$O, O\textsubscript{2}, Si-H nanoparticles) and a periodic system involving carbon diamond supercell. We first benchmark the results from \pfop~method with that obtained from \Lob~code, and we find an excellent agreement with \Lob~for the material system sizes feasible to run on \Lob. We also demonstrate the significant advantage of the \pfop~approach in terms of computational time compared to \Lob~even on 1 CPU-node on these material systems. Furthermore, we take advantage of our parallel implementation of \pfop~using \texttt{MPI} and illustrate the reduction in wall-time of the population analysis by $\sim$ 70\% when scaled up to 1120 CPU cores from 280 CPU cores on a Si nanoparticle system containing 1090 atoms. We remark that these large-scale calculations are not currently feasible using \Lob.  Subsequently, we compare the accuracy and performance of \pfhp~approach with that of \pfop. The results obtained by the \pfhp~approach agree very well with that obtained by the \pfop~approach. We further show the advantage of using \pfhp~in  computational wall time compared to \pfop~on large-scale systems ($\approx$ 1100-2100 atoms) by employing 280 - 4500 CPU cores. Finally, we discuss a case study demonstrating the usefulness of the proposed computational framework in conducting large-scale bonding analysis. To this end, we consider the case of the chemisorption of hydrogen in silicon nanoparticles alloyed with carbon, a candidate material for hydrogen storage~\cite{Galli}. Towards this, we conduct projected population analysis and estimate the Si-Si and Si-H bond strength in increasing system sizes of Si nanoparticles with and without alloying ranging from 65 atoms to around 1000 atoms, and argue the ease of Si-Si dimerization with the increase in size of alloyed Si nanoparticles favouring the release of H$_2$.
\par
The remainder of our manuscript is structured as follows: Section 2 discusses the mathematical background and relevant finite-element(FE) discretization aspects required for describing the projected population analysis within the FE formalism in the subsequent sections. Projected  orbital population analysis (\pfop) is discussed in Section 3, highlighting the aspects of mathematical formulation, accuracy validation and performance comparison results with \Lob. Section 4 discusses the details of projected Hamiltonian population analysis (\pfhp) and highlights the advantages of \pfhp~over \pfop. We subsequently discuss a case study illustrating the usefulness of large-scale chemical bonding analysis in Section 5, concluding with a short discussion and outlook in Section 6.

\section{Mathematical background}
In this section, we introduce the notations, discuss the key mathematical preliminaries and the relevant finite-element (FE) discretization aspects required for subsequently describing the projected population analysis within the FE formalism in sections 3 and 4.


Let $\mathbb{H}$ denote an infinite-dimensional Hilbert space, where we assume the Kohn-Sham eigenfunctions of the continuous problem exist. $\mathbb{H}$ is equipped with inner product $\braket{\cdot|\cdot}$ over the field of complex numbers $\mathbb{C}$, and consequently, a norm $\Vert \cdot \Vert$ induced from the inner product is defined. Let $\mathpzc{H}$ be the Hermitian operator representing the  Kohn-Sham Hamiltonian of interest defined on the $M$-dimensional subspace $\mathbb{V}^M \subset \mathbb{H}$. In other words,  $\mathpzc{H} \in \mathbb{C}^{M \times M}$ represents the discretized Kohn-Sham Hamiltonian operator in $\mathbb{V}^M$  spanned by a suitably chosen systematically converging basis set --- plane waves~\cite{qe,vasp}, finite element basis~\cite{Pask_2005,Tuschida,motamarri2013}, finite difference approach~\cite{PARSEC,SPARC}, wavelets~\cite{bigdft} etc., all which can be employed to numerically solve the partial differential equation representing the Kohn-Sham DFT eigenvalue problem.  Consequently, the discretized spin unpolarized DFT eigenvalue problem to be solved for $N$-smallest eigenvalue-eigenvector pairs is given by
\begin{equation}\label{eqn:devp}
    \mathpzc{H}\ket{\psi_i} = \epsilon_i\ket{\psi_i} \;\;\;\; \text{for}\;\;\;i= 1,2,...N\;\;\;\text{with}\;\;\; N \geq \frac{N_e}{2}
\end{equation}
where $\ket{\psi_i} \in \mathbb{V}^M$ denotes the eigenfunction of $\mathpzc{H}$ and $N_e$ is the number of electrons in the given material system.

Extracting chemical bonding behaviour using projected population approaches requires us to define a $N_{orb}$-dimensional subspace $\mathbb{V}^{N_{orb}}_{\phi}$ ($N_{orb} < M$), spanned by the localized non-orthogonal atom-centered auxiliary basis set $\{\ket{\phi_{\mu}}\}$. These basis are constructed for the given configuration of atoms in the material system and are chosen to be minimal such that the occupied Kohn-Sham wavefunctions $\ket{\psi_i}$ are well-represented in $\mathbb{V}^{N_{orb}}_{\phi}$ while providing accurate insights into the chemical bonding behavior. Various types of atom-centered localized basis functions have been used in the past, such as  Slater-type orbital expansions of Hartree-Fock wavefunctions by Bunge et.al~\cite{Bunge:1993} (henceforth referred to as STO basis by Bunge), 
 functions fitted to PAW wavefunctions~\cite{LOBSTER} have all been used in the past as a choice for these minimal atomic-orbital basis sets. Pseudo-atomic (\pa) orbitals constructed from norm-conserving pseudopotentials~\cite{ONCV} also constitute a convenient choice of atom-centered basis sets for chemical bonding analysis, as demonstrated in the current work.

 
 
 In the current work, the discretized Kohn-Sham eigenvalue problem in eq \eqref{eqn:devp} is represented in finite-element (FE) basis~\cite{brenner2002}, a strictly local and a piece-wise continuous Lagrange polynomial basis interpolated over Gauss-Lobatto-Legendre nodal points. We refer to our prior work~\cite{motamarri2022,motamarri2020,motamarri2013} for more details on the spectral FE discretization of Kohn-Sham DFT eigenvalue problem. To this end, the representation of various fields employed in computing projected population analysis subsequently --- the  Kohn-Sham  wavefunctions ($\braket{\bx|\psi_{i}} = \psi_i(\bx)$) and the localized atom-centered functions ($\braket{\bx|\phi_{\mu}} = \phi_\mu(\bx)$) in the FE basis is given by
 

\begin{equation}\label{eqn:fem}
\psi_{i}(\bx) = \sum_{j=1}^{M} N^{h}_j(\bx) \psi^{j}_{i}\,\,,\;\;\;\;\phi_\mu(\bx) = \sum_{j=1}^{M} N^{h}_j(\bx) \phi_{\mu}^{j} \,,
\end{equation}     
where $N^{h}_{j}:1 \leq j \leq M$ denote the $M$ finite-element (FE) basis functions spanning the $M$-dimensional space $\mathbb{V}^{M}$. These are strictly local Lagrange polynomials of degree $p$ generated using the nodes of the FE triangulation $\mathcal{T}^h$, with the characteristic mesh size denoted by $h$. Further in eq \eqref{eqn:fem}, $\psi^{j}_{i}$ and  $\phi_{\mu}^{j}$ denote the coefficients in the expansion of the $i^{th}$ discretized Kohn-Sham wavefunction ($\psi_{i}(\bx)$) and the $\mu^{th}$ atom-centered localized basis function ($\phi_\mu(\bx)$). These coefficients constitute the nodal values of the discretized fields represented using the FE triangulation $\mathcal{T}^h$ since the FE basis functions  $N^{h}_{j}(\bx)$ satisfy the Kronecker-delta property i.e. $N^{h}_{j}(\bx_k) = \delta_{jk}$ where $\bx_k$ denotes the $k^{th}$ nodal point of $\mathcal{T}^h$. The nodal values $\psi^{j}_{i}$ are computed by solving the FE discretized Kohn-Sham DFT eigenvalue problem given in eq~\eqref{eqn:devp}. Computationally efficient and scalable methodologies to solve this problem on massively parallel many-core architectures have been discussed in Motamarri et.al~\cite{motamarri2020} and on hybrid CPU-GPU architectures in  Das et.al~\cite{motamarri2022}. Furthermore, in the current work, the nodal values $\phi_{\mu}^{j}$ are computed from the atom-centered orbital data, which is usually available as analytical expressions or in the form of numerical data. 

Finally, we introduce the atom-centered orbital overlap matrix $\bS$ with matrix entries $S_{\alpha\beta}=\braket{\phi_{\alpha} | \phi_{\beta}}$, a key quantity in evaluating projected populations as discussed in the subsequent sections. Using the FE representation of $\phi_\mu(\bx)$ in eq~\eqref{eqn:fem}, the matrix entries of $\bS$  evaluated in a FE discretized setting is given by
\begin{align}
    \bS = \bPhicheck^{\dagger}\bPhicheck \;\;\;\text{with}\;\bPhicheck=\bM^{1/2}\bPhi 
\label{eq:S}
\end{align}
where $\bPhi$ denotes a $M\times N_{orb}$ matrix whose columns are the components of $\phi_{\mu}(\bx)$ in FE basis (see eq~\eqref{eqn:fem}) and the $M \times M$ matrix $\bM$ denotes the FE basis overlap matrix with entries given by $M_{pq}=\int_{\Omega}{N_p(\bx)N_q(\bx)d\bx}$. Efficient computation of $\bM^{1/2}$ and, subsequently the $\bS$ matrix is crucial for evaluating projected populations in the FE setting. We refer the reader to the supporting information (see section S1.1) for more details about the computation of $\bM^{1/2}$ and $\bS$ matrices in a parallel computing environment.



\label{section: mathformulation}

\small{\section{Projected orbital population analysis (\pfop)}}\label{section:pfop}
\textit{Projected orbital population analysis}, henceforth referred to as \pfop~relies on the orthogonal projection of numerically computed Kohn-Sham DFT eigenfunctions onto a subspace spanned by localized atomic-orbitals to extract the chemical bonding behavior, and is in the spirit of Sanchez-Portal \textit{et al.} \cite{sanchez-portal} and Deringer \textit{et al.}\cite{Deringer}. In this section, we begin by discussing the mathematical formulation and, subsequently, the related expressions in a FE setting required for implementing \pfop. We then assess the accuracy and performance of the  proposed implementation with \Lob, a widely used package for conducting projected orbital population analysis. Further, for clarity and simplicity, we assume that the Kohn-Sham DFT eigenproblem in eq~\eqref{eqn:devp} is solved in a simulation domain with fully non-periodic boundary conditions or a supercell employing periodic/semi-periodic boundary conditions with Gamma point to sample the Brillouin zone. The extension to periodic unit-cell involving Brillouin zone integration -via- multiple k-point sampling is not explicitly considered in this section. However, the expressions and the benchmark results for $\bk$-dependent projected population analysis within the framework of \pfop~are discussed in supporting information (see sections S1.1 and S2.1)


\subsection{Mathematical formulation}\label{subsection:pfopMathformulation}
To begin, we introduce the orthogonal projection operator $\mathpzc{P}^{\phi} : \mathbb{V}^M \rightarrow \mathbb{V}^{N_{orb}}_{\phi} $  which can be written as  $\mathpzc{P}^{\phi} = \sum_{\alpha, \beta = 1}^{N_{orb}}{\ket{\phi_{\alpha}} \left({S}^{-1}\right)_{\alpha \beta} \bra{\phi_{\beta}} }$, with atomic orbital overlap matrix $S_{\alpha \beta} = \braket{\phi_{\alpha} | \phi_{\beta}}$ as introduced before. Denoting the orthogonal projection of Kohn-Sham eigenfunction $\ket{\psi_i} \in \mathbb{V}^M$ onto the subspace $\mathbb{V}^{N_{orb}}_{\phi}$ to be $\ket{\psi_i^{\phi}}$, we have $\ket{\psi_i^{\phi}} = \mathpzc{P}^{\phi} \ket{\psi_i}$ for $i = 1,2 \cdots N$. We note that the projected Kohn-Sham wavefunctions $\{\ket{\psi_i^{\phi}}\}$ need not form an orthonormal set and hence  L\"{o}wdin symmetric orthogonalization\cite{LSO} is employed to orthonormalize the projected Kohn-Sham wavefunctions. To this end, we denote the orthonormalized projected wavefunction as $\ket{\tilde{\psi}^{\phi}_i}$ where, $\ket{\tilde{\psi}^{\phi}_i}= \sum_{j}^{N_{orb}}{O^{-1/2}_{ij}\ket{\psi_j^{\phi}}}$, with $O_{ij} = \braket{\psi^{\phi}_i | \psi^{\phi}_j}$, denoting the matrix elements of the overlap matrix $\textbf{O}$ corresponding to $\{\ket{\psi_i^{\phi}}\}$.

\paragraph{Projected orbital overlap population (\oop):}\label{para:oop}
Recalling that $\braket{\tilde{\psi}_j^{\phi}|\tilde{\psi}_j^{\phi}}$ equals 1 and the fact that the number of electrons $N_e$ in a given material system is related to the density of states $\delta(\epsilon - \epsilon_j)$, we can write $N_e=\sum^{N}_{j=1}{\int_{-\infty}^{\infty}{\braket{\tilde{\psi}_j^{\phi}|\tilde{\psi}_j^{\phi}}f(\epsilon,\epsilon_F)\delta(\epsilon - \epsilon_j)d\epsilon}}$, where $f$ denotes orbital occupancy function usually given by the Heaviside function with a value $1$ if $\epsilon < \epsilon_F$ (Fermi-energy) and $0$ otherwise. Using the relations $\ket{\tilde{\psi}_j^{\phi}}= \sum_{j}^{N_{orb}}{O^{-1/2}_{ij}\ket{\psi_j^{\phi}}}$ and  $\ket{\psi_j^{\phi}} =  \mathpzc{P}^{\phi} \ket{\psi_j}$, the above expression relating $N_e$  and $\delta(\epsilon - \epsilon_j)$ can be expressed in terms of Kohn-Sham wavefunctions $\ket{\psi_j} \in \mathbb{V}^M$ and the localized atom-centered basis  $\ket{\phi_{\mu}} \in \mathbb{V}^{N_{orb}}$ in the following way:
\begin{align}
    N_e &= \sum_{\nu,\nu'}^{N_{orb}}{\sum_{\mu,\mu'}^{N_{orb}}{\sum_{k,q}^{N}{{\sum^{N}_{j}{O^{-1/2}_{jk}O^{-1/2}_{qj}\int_{-\infty}^{\infty}{f(\epsilon,\epsilon_F)\braket{\psi_q |\phi_\mu}S^{-1}_{\mu\mu'}\braket{\phi_{\mu'}|\phi_\nu}S^{-1}_{\nu\nu'}\braket{\phi_{\nu'}|\psi_k}\delta(\epsilon - \epsilon_j)d\epsilon}}}}}} \nonumber\\
    & = \sum_{I\alpha}{\sum_{J\beta}{\sum_{k,q}{{\sum_{j}{O^{-1/2}_{jk}O^{-1/2}_{qj}\int_{-\infty}^{\infty}{f(\epsilon,\epsilon_F)\braket{\psi_q | \phi^{I\alpha}}S_{I\alpha J\beta}\braket{\phi^{J\beta} | \psi_k}\delta(\epsilon - \epsilon_j)d\epsilon}}}}}} 
    \label{eq:doscoop2}
\end{align}
In the above eq~\eqref{eq:doscoop2}, $\ket{\phi^{\mu}}$ denotes the dual of the basis function $\ket{\phi_{\mu}}$, satisfying the property $\braket{\phi^{\mu}|\phi_{\nu}} =  \braket{\phi^{\nu}|\phi_{\mu}} = \delta_{\mu\nu}$ where, $\ket{\phi^{\mu}}$ is given by $\ket{\phi^{\mu}}= \sum_{\nu}{S^{-1}_{\nu\mu}\ket{\phi_\nu}}$. Furthermore, a multi-index $\mu = \{I\alpha\}$ is introduced above to denote the localized atom-centered basis function $\ket{\phi^{\mu}}$ as $\ket{\phi^{I\alpha}}$ where $\alpha$ denotes the index of the atomic orbital centered at a nuclear position $\textbf{R}_I$. The orbital overlap population deals with the distribution of the total number of electrons $N_e$ among the atoms in a given material system and can be motivated from the above equation. To this end, projected-orbital overlap population~\cite{Hoff1,Mulli1}  $\oop_{IJ}(\epsilon)$ associated with a source atom $I$ and a target atom $J \neq I$ is extracted from  eq~\eqref{eq:doscoop2} to be defined as:
\begin{equation}\label{eq:pOOP}
 \oop_{IJ}(\epsilon) = \sum_{j}\sum_{\alpha \beta}{{\mathfrak{Re}\left\{\left(\sum_q{O^{-1/2}_{qj}\braket{ \psi_q| \phi^{I\alpha}}}\right) S_{I\alpha J\beta}\left(\sum_k{O^{-1/2}_{jk}\braket{\phi^{J\beta} | \psi_k}}\right)\right\} \delta(\epsilon - \epsilon_j)}}   
\end{equation}
where, $\mathfrak{Re}(z)$ refers to the real-part of a complex number $z$. Introducing the finite-element (FE) discretization of various fields in eq~\eqref{eq:pOOP}, we deduce the relevant matrix expressions required to evaluate the overlap population in the FE setting. To begin, we define a matrix $\bCbar$ of size $N_{orb} \times N$ whose entries $\bar{C}^{j}_{I\alpha} =  \sum_{q}{O^{-1/2}_{jq}\braket{\phi^{I\alpha} | \psi_q}}$ are the coefficients of $\ket{\tilde{\psi}_j^{\phi}}$ expressed in the localized atom centered basis set $\{\ket{\phi_{\mu}}\}$. The FE representation of various fields in the expression for $\bar{C}^{j}_{I\alpha}$ allows $\bCbar$ to be recast in the matrix form as follows:
\begin{equation}
    \bCbar = \bS^{-1}\bPhicheck^{\dagger}\bPsicheck \bO^{-1/2}\;\;\text{where} \;\; \bPsicheck = \bM^{1/2}\bPsi  
    \label{eq:Cbar}
\end{equation}
where $\bPsi$ denotes a $M\times N$ matrix whose column vectors are the components of $\ket{\psi_j}$ in FE basis, while the $M\times N_{orb}$ atomic-orbital matrix $\bPhi$ and the $M\times M$ FE basis overlap matrix $\bM$ are  defined in the previous section. We note that the rows of the matrix $\bCbar$ are stored in the order of atoms and their corresponding atom-centered orbitals for a given atom in succession. To elaborate, $\mu^{th}$ row of $\bCbar$ corresponds to $I^{th}$ atom and an atom-centered index $\alpha$ associated with this atom $I$, while the $j^{th}$ column of this matrix corresponds to the index of projected Kohn-Sham wavefunction ($j = 1 \cdots N$). More details related to the derivation and efficient computation of the matrix expressions for $\bCbar$, $\bO$ in the FE setting can be found in the supporting information (section S1.1). Finally \textit{projected orbital overlap population} ($\oop$) associated with a source atom $I$ and a target atom $J$ is evaluated by extracting the appropriate entries of the matrices $\bCbar$, $\bS$ and the expression is given by
\begin{equation}\label{eq:pfoop}
    \oop_{IJ} (\epsilon) = \sum_{j}{\sum_{\alpha \beta}{\mathfrak{Re}\left\{\widebar{C}_{I\alpha}^{j*}\widebar{C}_{J\beta}^{j} S_{I\alpha J \beta}\right\} \delta(\epsilon - \epsilon_j) }}
\end{equation}


\paragraph{Projected  orbital Hamilton Population (\ohp):}\label{para:ohp}
Recalling that the electronic band energy  ($E_{\text{band}}$) in DFT is related to the expectation value of the Kohn-Sham Hamiltonian $\mathpzc{H} \in \mathbb{C}^{M \times M}$ with respect to its occupied eigenstates $\ket{\psi_j} \in \mathbb{V}^M$, we have
$E_{\text{band}} = \sum_{j=1}^{N}{\int_{-\infty}^{\infty}{f(\epsilon,\epsilon_F)\bra{\psi_j}\mathpzc{H}\ket{\psi_j}}\delta(\epsilon-\epsilon_j)d\epsilon}$.
 Following Maintz et.al,\cite{maintz} we now define the projected Hamiltonian operator $\mathpzc{H}^p$ on the subspace $\mathbb{V}^{N_{orb}}_{\phi}$ in terms of $\ket{{\tilde{\psi}}_j^{\phi}}$ as
$\mathpzc{H}^p = \sum_{j=1}^{N}{\ket{{\tilde{\psi}}_j^{\phi}}\epsilon_j\bra{{\tilde{\psi}}_j^{\phi}}}$ where $\epsilon_j$ denotes the DFT eigenvalues (see eq~\eqref{eqn:devp}). Using this definition of $\mathpzc{H}^p$, we can see that $\bra{\psi_j}\mathpzc{H}\ket{\psi_j} = \bra{{\tilde{\psi}}_j^{\phi}}\mathpzc{H}^p\ket{{\tilde{\psi}}_j^{\phi}}$ and hence the band energy $E_{\text{band}}$ can be written as: $E_{\text{band}} = \sum_{j=1}^{N}{\int_{-\infty}^{\infty}{f(\epsilon,\epsilon_F)\bra{{\tilde{\psi}}_j^{\phi}}\mathpzc{H}^p\ket{{\tilde{\psi}}_j^{\phi}}\delta(\epsilon-\epsilon_j)d\epsilon}}$.

Along the lines of Maintz et.al\cite{maintz}, we consider the orthogonalized basis $\{\ket{\widehat{\phi}_{\nu}}\}$ obtained -via-  L\"{o}wdin symmetric orthonormalization\cite{LSO} of $\{\ket{{\phi}_{\nu}}\}$ where the two basis are related by the expression $\ket{\widehat{\phi}_{\mu}}=\sum_{\nu} S^{-1/2}_{\mu \nu}\ket{\phi_{\nu}}$. Subsequently, the projection operator $\mathpzc{P}^{\phi}$ expressed in terms of $\{\ket{\widehat{\phi}_{\nu}}\}$ i.e $\mathpzc{P}^{\phi} = \sum_{i=1}^{N_{orb}}{\ket{\widehat{\phi}_{\mu}}\bra{\widehat{\phi}_{\mu}}}$ can be used to recast the above equation corresponding to $E_{\text{band}}$ as:
\begin{align}\label{eqn:bandenergy}
 E_{\text{band}} &= {\sum_{\mu,\nu=1}^{N_{orb}}{{\sum_{k,q,j=1}^{N}{{{O^{-1/2}_{jk}O^{-1/2}_{qj}\int_{-\infty}^{\infty}{f(\epsilon,\epsilon_F) 
 \braket{\psi_q | \widehat{\phi}_\mu}\bra{\widehat{\phi}_\mu}\mathpzc{H}^p\ket{\widehat{\phi}_\nu}\braket{\widehat{\phi}_\nu | \psi_k}\delta(\epsilon - \epsilon_j)d\epsilon}}}}}}} \nonumber \\
 &= {\sum_{I\alpha,J\beta}^{N_{orb}}{{\sum_{k,q,j}^{N}{{{O^{-1/2}_{jk}O^{-1/2}_{qj}\int_{-\infty}^{\infty}{f(\epsilon,\epsilon_F) 
 \braket{\psi_q | \widehat{\phi}_{I\alpha}}H^{p}_{I\alpha,J\beta}\braket{\widehat{\phi}_{J\beta} | \psi_k}\delta(\epsilon - \epsilon_j)d\epsilon}}}}}}}
 \end{align}
 where, the composite index notation $\mu = \{I\alpha\}$ $[\nu = \{J\beta\}]$ has been used for the basis functions $\{\ket{\widehat{\phi}_{\mu}}\}$  to denote $\alpha^{th}$ $[\beta^{th}]$ basis function centered at the atomic position $\textbf{R}_I$  $[\textbf{R}_J]$ and $H^{p}_{I \alpha, J \beta}$ denotes the matrix element of $\mathpzc{H}^p$ represented in the $\{\ket{\widehat{\phi}_{\mu}}\}$ basis. The orbital Hamilton population analysis deals with the partitioning of the band energy $E_{\text{band}}$ among the constituent atoms in a given material system, and projected-orbital Hamilton population~\cite{1993} $\mathtt{pOHP}$ can be defined taking recourse to eq~\ref{eqn:bandenergy}. To this end, $\mathtt{pOHP}_{IJ}(\epsilon)$ associated with a source atom $I$ and a target atom $J \neq I$ is extracted from eq~\ref{eqn:bandenergy} to be defined as
\begin{equation}\label{eqn:pOHP}
 \ohp_{IJ}(\epsilon) = \sum_{j}{\sum_{\alpha \beta}{\mathfrak{Re}\left\{ \left(\sum_q{O^{-1/2}_{qj}\braket{ \psi_q| \widehat{\phi}_{I\alpha}}}\right) H^{p}_{I\alpha,J\beta}\left(\sum_k{O^{-1/2}_{jk}\braket{\widehat{\phi}_{J\beta} | \psi_k}}\right)\right\} \delta(\epsilon - \epsilon_j)}}   
\end{equation}

where, $\mathfrak{Re}(z)$ refers to the real-part of a complex number $z$. Introducing the finite-element (FE) discretization of various fields in eq~\eqref{eqn:pOHP}, we deduce the relevant matrix expressions required to evaluate the Hamilton population in the FE setting. Consequently, we  define the $N_{orb} \times N$ matrix $\bChat$~whose entries $\widehat{C}_{I\alpha}^{j} = \sum_{k}{O^{-1/2}_{jk}\braket{\widehat{\phi}_{I\alpha} |\psi_k}}$ are the coefficients of $\ket{\tilde{\psi}_j^{\phi}}$ expressed in the orthonormalized atom-centered basis set $\{\ket{\widehat{\phi}_{\mu}}\}$. The FE representation of various fields in $\widehat{C}_{I\alpha}^{j}$ allows one to recast $\bChat$ in terms of the matrices $\bCbar$ (refer eq.~\ref{eq:Cbar}) and $\bS$ (refer eq.~\ref{eq:S}) i.e. $\bChat = \bS^{1/2}\bCbar$. In all our subsequent discussions and computations in this work, we choose $N=N_{orb}$ and we compute the $N_{orb}\times N_{orb}$ projected Hamiltonian matrix $\bH^{p}$ using the coefficient matrix $\widehat{\bC}$ as $\bH^{p} = \widehat{\bC} \bD \widehat{\bC}^{T}$ where the matrix $\bD$ is diagonal and comprises of the Kohn-Sham eigenvalues $\epsilon_i$ obtained from Kohn-Sham DFT problem solved in the FE basis. We refer the reader to supporting information (section S1.1) for more details on the derivation and efficient computation of the matrix expressions for $\bH^p$, $\bO$,  $\bChat$ within the FE framework. Finally, the \textit{projected orbital Hamilton population} (\ohp) associated with the partitioning of energy between a source atom $I$ and a target atom $J \neq I$ is evaluated by extracting the appropriate entries of the matrices $\widehat{\bC}$, $\bH^{p}$ and the expression is given by

\begin{equation} \label{eq:pfohp}
\ohp_{IJ}(\epsilon) = \sum_j{\sum_{\alpha,\beta}{\mathfrak{Re}\left\{\widehat{C}^{j*}_{{I\alpha}}H^{p}_{I\alpha,J\beta}\widehat{C}^{j}_{{J\beta}}\right\}\delta(\epsilon-\epsilon_j)}} \end{equation}


\paragraph{Total computational complexity estimate of \pfop:}\label{para:hp complexity} The current implementation of the projected orbital population analysis (\pfop) assumes $N=N_{orb}$ and thereby the total computational complexity can be estimated to be $\sim 4 M_{loc} N^2 + 26 N^3$ (refer to supporting information S1.1 ). Here, $M_{loc}$ is the ratio of the total number of FE degrees of freedom ($M$) to the number of MPI tasks ($P$) on a parallel computing system. $M_{loc}$ can be reduced by increasing the value of $P$. Consequently, the second term in the computational complexity becomes dominant when the number of MPI tasks $P$ exceeds $4M/26N$.

\paragraph{Spilling factor:}\label{para:spillfactors} The total spilling factor or charge spilling factor, as introduced by Sanchez-Portal et al.~\cite{sanchez-portal} describes the ability of the atom-centered localized basis spanning $\mathbb{V}^{N_{orb}}_{\phi}$ to represent the FE discretized Kohn-Sham eigenfunctions $\ket{\psi_j}$, the self-consistent solution of the FE discretized Kohn-Sham eigenvalue problem. The charge spilling factor is given by the average of the $L_2$ projection errors of the Kohn-Sham occupied eigenstates while the total spilling factor is computed as the average of $L_2$ projection errors of the Kohn-Sham eigenstates considered for the projection. To this end, we compute the absolute total spilling factor $\mathcal{S}$ and absolute charge spilling factor $\mathcal{S}_c$ in the spirit of Stefan Maintz et al.~\cite{maintz2016lobster} and are expressed in terms of the diagonal entries of the matrix $\bO$ (see eq~6 in SI) as
\begin{equation}
  \mathcal{S} = \frac{1}{N} \sum_{i=1}^{N} | 1 - \braket{\psi^{\phi}_i | \psi^{\phi}_i} | = \frac{1}{N} \sum_{i=1}^{N} | 1 - O_{ii} |, \;\;\;\;\;\;\;\; \mathcal{S}_c = \frac{1}{N_{\text{occ}}} \sum_{i=1}^{N_{\text{occ}}} | 1 - O_{ii} |
\end{equation}
\paragraph{Projected  orbital density error (\oder):}\label{para:oder} We compute the $L_2$ norm of the difference between the ground-state electron density ($\rho(\bx)$) computed from \DFTFE~using the FE discretized occupied Kohn-Sham eigenfunctions \{$\ket{\psi_i}$\} and the electron-density ($\rho^{o}(\bx)$) computed from the  L\"owdin symmetric orthonormalized\cite{LSO} projected Kohn-Sham wavefunctions  $\{\ket{\tilde{\psi}^{\phi}_i}\} \in \mathbb{V}^{N_{orb}}_{\phi}$. To this end, \oder~ is evaluated as
\begin{equation}\label{eq:oder}
  \oder = \frac{||\rho(\bx) - \rho^{o}(\bx)||_2}{||\rho(\bx)||_2}\;\; \text{where}\;\;\rho(\bx) = \sum_{i=1}^{N_{occ}}
  \braket{\bx|\psi_i}\braket{\psi_i|\bx},\;
  \;\;\rho^{o}(\bx) = \sum_{i=1}^{N_{occ}}
  \braket{\bx|\tilde{\psi}^{\phi}_i}\braket{\tilde{\psi}^{\phi}_i|\bx}
 \end{equation}
\vspace{-0.25in}
\begin{figure}[H]
\includegraphics[scale=0.4]{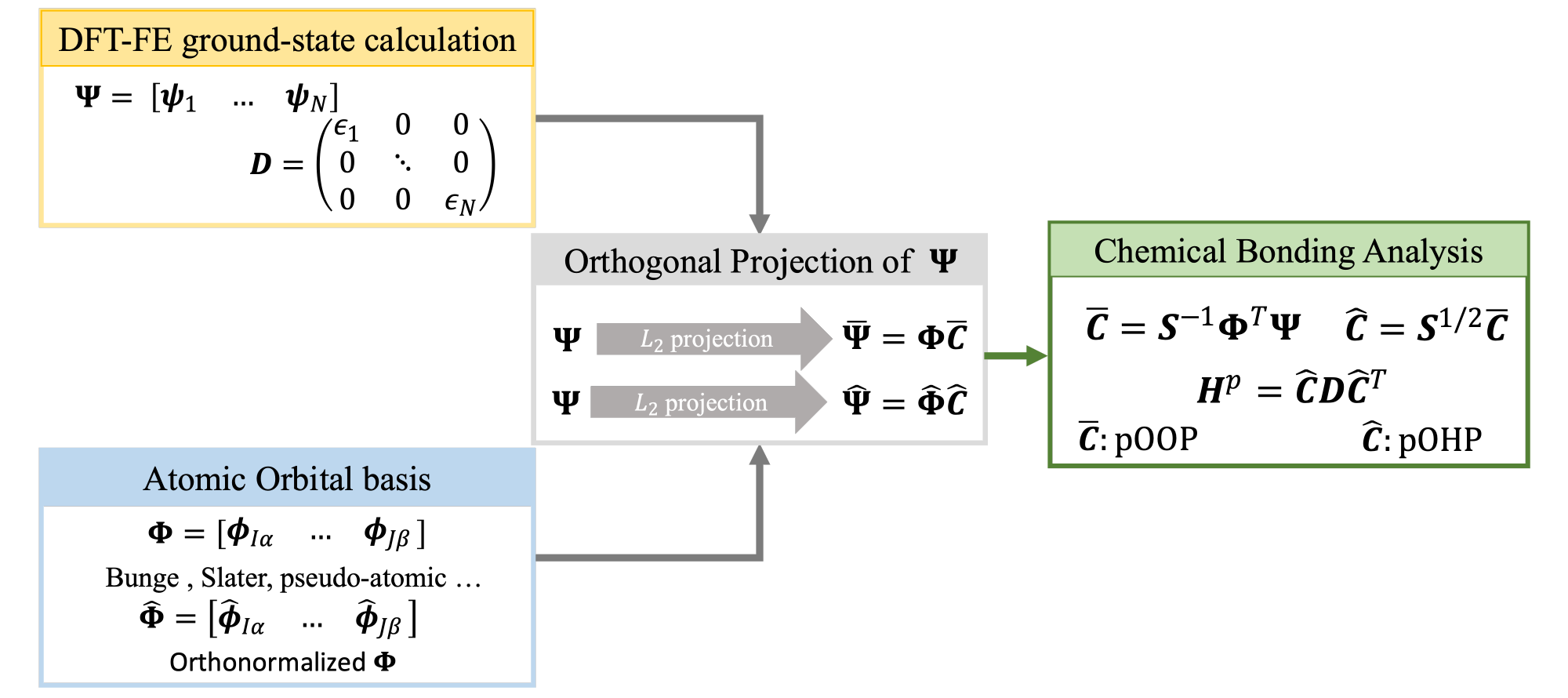}
\vspace{-0.1in}
\caption[short]{\small{Overview of the implementation strategy for projected orbital population analysis (\pfop) within the finite-element (FE) framework. The strategy involves projecting the self-consistently converged Kohn-Sham eigenfunctions obtained from \DFTFE~onto a localized atom-centered basis represented on the FE grid. Subsequently, the overlap matrix($\bS$), projected Hamiltonian ($\bH^{p}$) and the coefficient matrices $\bCbar$, $\widehat{\bC}$ are evaluated which in-turn are used to compute projected orbital overlap population $\mathtt{pOOP}$ and projected orbital Hamilton  population $\mathtt{pOHP}$ respectively.} }\label{Flowchart pOPA}
\end{figure}


\subsection{Results: Accuracy and performance benchmarking}
We now assess the accuracy and performance of the proposed \pfop~implementation within the framework of \DFTFE~\cite{motamarri2020,motamarri2022} by comparing with \Lob. To this end, we employ Quantum Espresso\cite{qe} (\QE) with PAW \cite{PAW} pseudopotentials to perform DFT calculations for all the benchmark systems, and the resulting ground-state DFT wavefunctions and eigenvalues are used as an input for conducting the population analysis using \Lob. The calculations using \QE~are performed by employing the internal implementation of GGA \cite{GGA} exchange-correlation of the PBE \cite{PBE} form. While in \DFTFE~calculations, we employ optimized norm-conserving Vanderbilt (ONCV) \cite{ONCV} pseudopotentials from pseudoDojo database\cite{pseudoDojo} to conduct pseudopotential DFT calculations using GGA \cite{GGA} exchange-correlation of the PBE \cite{PBE}  form employing $\mathtt{Libxc}$ package\cite{Libxc}~. We employ non-periodic boundary conditions for isolated systems and periodic boundary conditions for crystalline systems in \DFTFE.  The population analysis methodology discussed in this section is implemented in the \DFTFE~code, and we note the numerical implementation of population analysis in \DFTFE~takes advantage of parallel computing architectures via \texttt{MPI}(Message Passing Interface), enabling chemical bonding analysis of large-scale systems in a unified computational framework. 

For all the benchmark calculations reported here, the cutoff energies in \QE~and mesh sizes in \DFTFE~are chosen such that a discretization error of $O(10^{-5}) \frac{E_h}{\text{atom}}$~  in ground state energy and a force discretization error of $O(10^{-4})\frac{E_h}{\text{bohr}}$~in ionic forces is achieved. The simulations and computational times reported in this work are performed on the CPU nodes of the supercomputer PARAM Pravega\footnote{PARAM Pravega is one of India's fastest supercomputers stationed at Indian Institute of Science comprising of 584 Intel Xeon Cascade-Lake based CPU nodes (28,032 Cores)}.

In this subsection, the computations of $\mathtt{pOOP}$ and $\mathtt{pOHP}$ as described in eq~\ref{eq:pfoop} and eq~\ref{eq:pfohp} are validated using \cop~and \chp~obtained from \Lob~in terms of both accuracy and performance. To this end, we project the self-consistently converged Kohn-Sham (KS) wavefunctions obtained from \DFTFE~onto a subspace spanned by (i) STO basis by Bunge \cite{Bunge:1993} and (ii) pseudo-atomic (\pa) orbitals constructed from the ONCV\cite{ONCV} pseudopotentials. On the other hand, the converged ground-state DFT wavefunctions obtained using PAW formalism in \QE~are used as input to \Lob, which are inturn projected onto a subspace spanned by localized atom-centered basis functions known as \texttt{pbevaspfit2015}\cite{maintz2016lobster}. In our benchmark studies, we consider (i) isolated systems comprising of CO, spin-polarized O\textsubscript{2}, H\textsubscript{2}O molecules, Si-H nanoparticle with 65 atoms (Si\textsubscript{29}H\textsubscript{36}), and (ii) $2\cross 2 \cross 2$ periodic carbon diamond supercell. To simulate isolated systems in \DFTFE, we consider the simulation domain large enough to allow the wavefunctions to decay to zero on the boundary by employing non-periodic boundary conditions.
In contrast, in \QE, which always employs periodic boundary conditions, we consider a simulation domain with sufficient vacuum to minimize the image-image interactions. In the benchmark problem involving a bulk material system, we consider a $2\cross 2 \cross 2$ diamond supercell employing periodic boundary conditions  using a $\Gamma$-point to sample the Brillouin zone in both \DFTFE~and \QE. We now describe the comparative study of the projection spill factors, population analysis energy diagrams and the computational costs between the proposed implementation in \DFTFE~and \Lob. 
 
\paragraph{Accuracy validation of \pfop:}\label{para:op accuracy} The projected-orbital population analysis (\pfop) implemented in DFT-FE is compared with \Lob~for the case of Si\textsubscript{29}H\textsubscript{36} and periodic $2\cross2\cross2$ supercell of carbon. Table~\ref{tab:SpillFactorsResults} shows the comparison of absolute spilling factor $\mathcal{S}$ and the absolute charge spilling factor $\mathcal{S}_c$ for these systems, and we note that our spill factors are in close agreement with that obtained from \Lob. Furthermore, we observe that the spill factors $\mathcal{S}$ and $\mathcal{S}_c$ obtained in our \pfop~approach employing pseudo-atomic (\pa) orbitals are smaller in comparison to the spill factors obtained using   STO basis by Bunge. This can be attributed to the fact that the subspace spanned by \pa~orbitals is a better representation of the FE discretized ground-state Kohn-Sham eigenfunctions obtained from \DFTFE~using ONCV pseudopotential calculations. We use these \pa~orbitals for our subsequent comparisons with \Lob~in this section, and comparisons with  STO basis by Bunge  are discussed in the  Supporting Information (see section S2.1) .

\begin{figure}[H]
\centering
\includegraphics[scale=0.5]{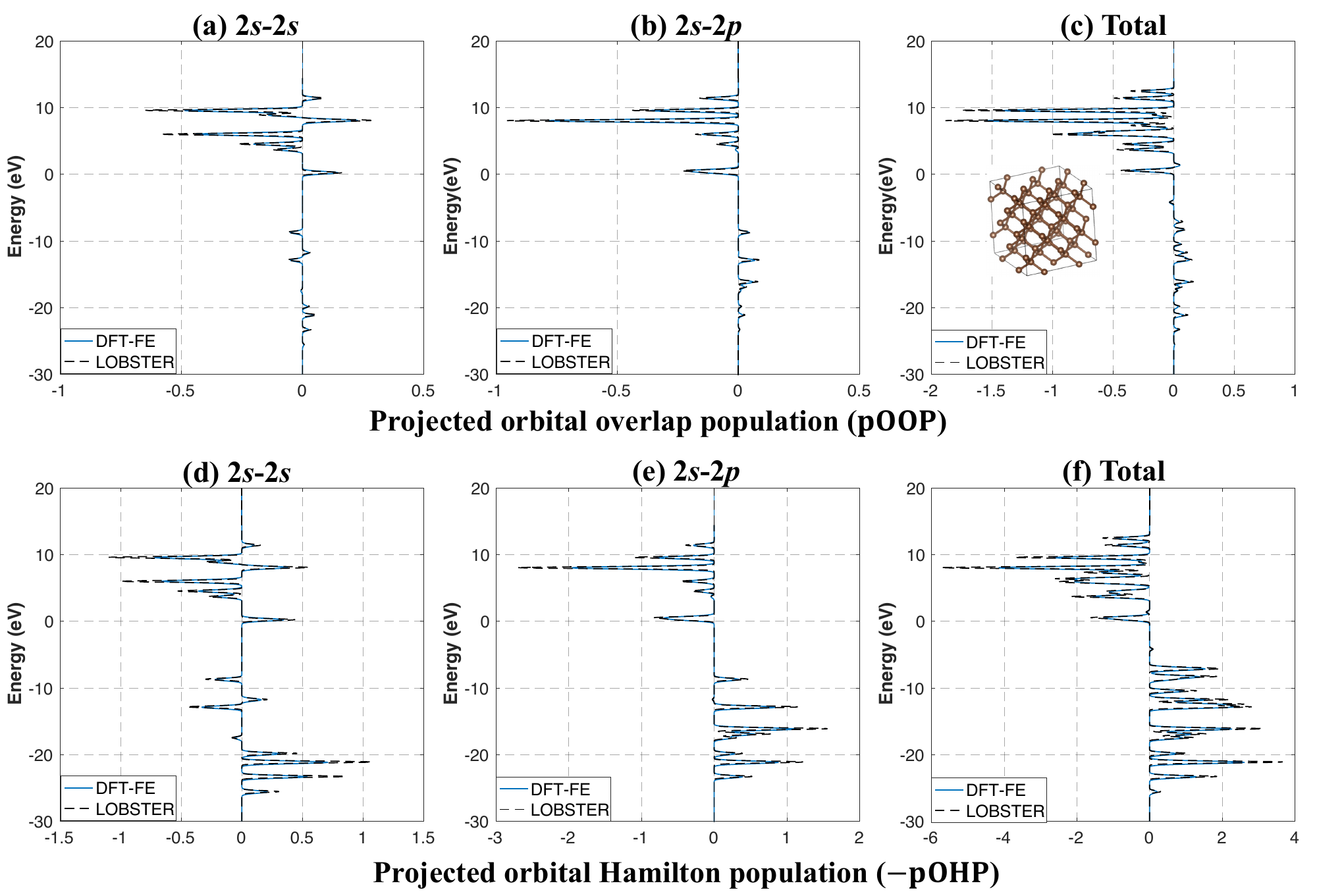}
\centering
\caption[short]{\footnotesize{\cb Comparison between \pfop~implemented in \DFTFE~using \pa~orbitals, and \Lob~for nearest neighbour C atoms in carbon diamond supercell. The top row shows the projected orbital overlap population, with the sub-figures (a) and (b) in this row showing the contributions of C$_{2s}$-C$_{2s}$ and C$_{2s}$-C$_{2p}$  to the total orbital overlap population that is plotted in sub-figure (c). The bottom row shows the negative of the projected orbital Hamilton population. The sub-figures in the bottom row (d) and (e) show the contributions of C$_{2s}$-C$_{2s}$ and C$_{2s}$-C$_{2p}$ to the total orbital Hamilton population that is plotted in sub-figure (f). Energy-scale is shifted such that Fermi level ($\epsilon_F$) is zero. \cn \textbf{Case study:} Carbon diamond $2\cross2\cross2$ supercell with periodic boundary conditions using a $\Gamma$ point. }}\label{fig:C periodic PA comparison}
\end{figure}

\begin{figure}[H]
\includegraphics[scale=0.5]{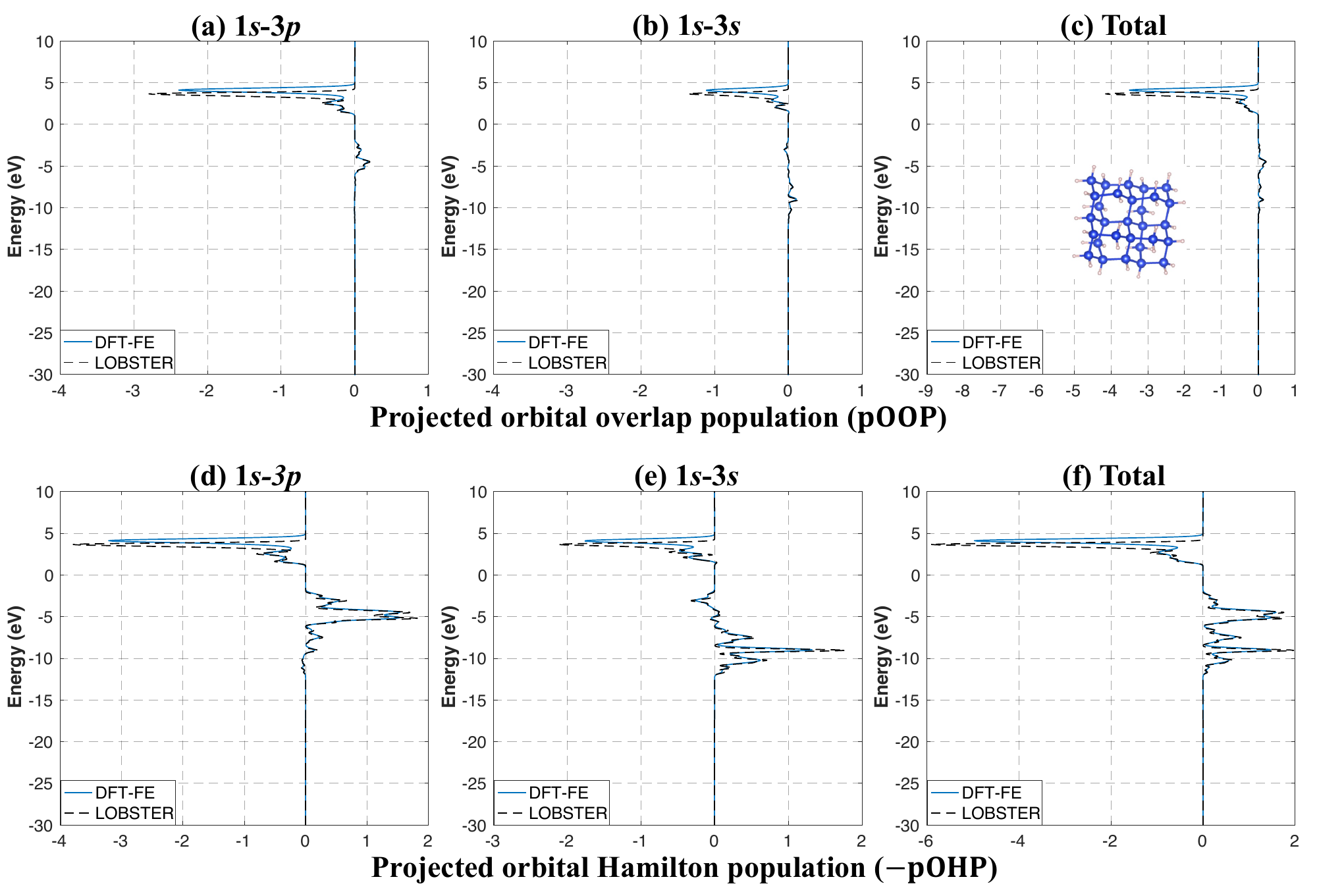}
\caption[short]{\footnotesize{\cb Comparison between \pfop~implemented in \DFTFE~using \pa~orbitals, and \Lob~for nearest neighbor Si-H atoms in Si nanoparticle. The top row shows the projected orbital overlap population, with the  sub-figures (a) and (b) in this row showing the contributions of H$_{1s}$-Si$_{3p}$ and H$_{1s}$-Si$_{3s}$ to the total orbital overlap population that is plotted in sub-figure (c). The bottom row shows negative of the projected orbital Hamilton population. The sub-figures in the bottom row (d) and (e) show the contributions of H$_{1s}$-Si$_{3p}$ and H$_{1s}$-Si$_{3s}$ to the total orbital Hamilton population that is plotted in sub-figure (f). Energy-scale is shifted such that Fermi level ($\epsilon_F$) is zero. \cn  \textbf{Case study:} Single-fold Si\textsubscript{29}H\textsubscript{36} nanoparticle}}\label{fig:Si29H36 PA comparison}
\end{figure}

Next, we illustrate the comparisons of population energy diagrams in Figures \ref{fig:C periodic PA comparison} and \ref{fig:Si29H36 PA comparison}, both for the case of  periodic $2\cross2\cross2$ carbon diamond supercell and Si\textsubscript{29}H\textsubscript{36} nanoparticle. In the case of carbon diamond supercell, a pair of nearest neighbour carbon atoms are picked as the source and target atom, and the corresponding contributions of $s-s$ and $s-p$ orbitals are plotted in Figure~\ref{fig:C periodic PA comparison} both for overlap population and Hamilton population. A comparison of total populations is also illustrated in this figure. These results indicate that the energy diagrams obtained with \Lob~match very well with our current approach. The location of the bonding and anti-bonding peaks are identical to that obtained from \Lob, with a slight difference in the amplitude of the peaks that can be attributed to the use of different pseudopotentials in \Lob~and \DFTFE. Similar agreements are observed in the case of Si\textsubscript{29}H\textsubscript{36} nanoparticle in which the Si atom and the nearest H atom are picked as the source and target atom, and the corresponding contributions of H$_{1s}$-Si$_{3p}$ and H$_{1s}$-Si$_{3s}$ orbitals are plotted in Figure~\ref{fig:Si29H36 PA comparison} (see inset in the figure) both for overlap population and Hamilton population. Comparisons of spill factors and the population energy diagrams for benchmark systems involving molecules -- CO, spin-polarized O$_2$, H$_2$O are discussed in section S2.1, and we observe a close agreement of the \oop~and \ohp~with that obtained using \Lob.  Comparisons with \Lob~involving $\bk$-dependent population analysis are discussed in supporting information in the case of $1\cross1\cross1$ orthogonal unit-cell of carbon diamond structure (see S2.1). 

\begin{table}[]
    \centering
    \begin{tabular}{|c|c|c|c|c|c|c|}
    \hline
     \multirow{2}{*}{System} & \multicolumn{2}{c|}{\Lob}  & \multicolumn{2}{c|}{\DFTFE~Bunge}  & \multicolumn{2}{c|}{\DFTFE~\pa}  \\
     \cline{2-7}
     &$\mathcal{S}_c$&$\mathcal{S}$&$\mathcal{S}_c$&$\mathcal{S}$&$\mathcal{S}_c$&$\mathcal{S}$ \\
    \hline\hline
    \makecell{Carbon diamond \\ $2\cross2\cross2$ periodic supercell} & 0.010 & 0.092 & 0.011 & 0.101 & 0.003 & 0.068 \\
    \hline
    Si\textsubscript{29}H\textsubscript{36} nanoparticle & 0.012 & 0.308 & 0.039 & 0.314 & 0.017 & 0.271 \\
    \hline\hline
    \end{tabular}
    \caption{\footnotesize{Comparison of absolute charge spill factor($\mathcal{S}_c$) and absolute spill factor($\mathcal{S}$) obtained using projections carried out in \DFTFE~and \Lob. \DFTFE~Bunge indicates the projection of finite-element discretized Kohn-Sham eigenfunctions to  STO basis by Bunge  and \DFTFE~\pa~basis indicates projection onto pseudo-atomic orbitals.  Projection in \Lob~uses \texttt{pbeVaspfit2015} as auxiliary atom-centered basis.}}
    \label{tab:SpillFactorsResults}
\end{table}




\paragraph{Performance comparison:}\label{para:op perf}  The computational CPU times in terms of node-sec (wall time taken on 1 compute node), measuring the computation of overlap population and Hamilton population, are tabulated in Table~\ref{tab: Timing Comparison} for proposed \pfop~approach and~\Lob. The benchmark systems involving $2 \times 2 \times 2$ carbon diamond supercell, Si-H nanoparticles --- Si$_{29}$H$_{36}$ and Si$_{145}$H$_{150}$ are chosen for comparison. The computational times indicate significant advantage for the proposed implementation compared to \Lob. For instance, computational gains up to $2\times$ are observed for the case of carbon diamond $2\times2\times2$ periodic supercell, while in the case of  Si$_{29}$H$_{36}$ nanoparticle, a speed up of $14 \times$ is observed. This higher speedup in the case of Si nanoparticle using the proposed \pfop~approach can be attributed to the use of non-periodic boundary conditions in our implementation while \Lob~is restrictive in terms of the boundary conditions one can employ and uses periodic boundary conditions even for Si nanoparticle, an isolated system. Furthermore, it was computationally prohibitive to conduct population analysis using \Lob~for the Si$_{145}$H$_{150}$ nanoparticle. We further demonstrate the advantage of our parallel implementation by measuring the wall times of the population analysis conducted on a large Si$_{580}$H$_{510}$ nanoparticle containing 1090 atoms. Table~\ref{tab:Scalingstudy comparison of pFOP} shows a reduction in wall time of about $1.7\times$ with an increase in the number of CPU cores from 280 to 1120.

\begin{table}[]
    \centering
    \begin{tabular}{|c|c|c|}
    \hline \hline
    Material system & \makecell {\pfop (sec)} & \Lob (sec) \\
    \hline \hline
    \makecell{Carbon diamond \\ $2\cross2\cross2$ supercell} & 8.26 & 17.5 \\
   \hline
    Si$_{29}$H$_{36}$ nanoparticle & 1.72 & 24.2 \\
    \hline
    Si$_{145}$H$_{150}$ nanoparticle & 21.46 & - \\
    \hline
    \end{tabular}
    \caption{\footnotesize{Comparison of computation CPU time measured in node-secs, between projected orbital population analysis (\pfop) implementation in \DFTFE~and \Lob. The timing includes the reading and construction of the atomic orbital basis($\bPhi$) and the calculation of the appropriate matrices for conducting the population analysis ($\bCbar$,$\widehat{\bC},\bH^{p}$)}}
    \label{tab: Timing Comparison}
\end{table}
\begin{table}[H]
    \centering
    \begin{tabular}{|c|c|c|}
    \hline \hline
    Number of cores  & Wall-time (seconds) \\
    \hline \hline
    280  & 33.85 \\
    \hline
    560  & 25.04 \\
    \hline
    1120  & 20.48 \\
    \hline
    \end{tabular}
    \caption{\footnotesize{Wall times of \pfop~for various number of CPU cores. \textbf{Case study:} Si\textsubscript{580}H\textsubscript{510} nanoparticle (1090 atoms with 2830 valence electrons). Total degrees of freedom (DoFs) per atom is around 8400.}}
    \label{tab:Scalingstudy comparison of pFOP}
\end{table}

\small{\section{Projected Hamiltonian population analysis (\pfhp)}}\label{section:pfhp}
We propose here \textit{Projected Hamiltonian population analysis}, henceforth referred to as \pfhp~as an alternate approach different from \pfop~described previously, to conduct both overlap and Hamilton population analysis. The necessity of this alternate approach is motivated by the fact that many of the reduced scaling electronic structure codes targeted towards large-scale DFT calculations tend to avoid explicit computation of DFT eigenvectors having no access to these eigenvectors for projection. To this end, in this approach, we orthogonally project the self-consistently converged discretized DFT Hamiltonian onto the subspace spanned by the minimal atomic-orbital basis set to extract the chemical bonding information from DFT calculations. This is in contrast to the previous approach where the self-consistently converged Kohn-Sham eigenfunctions are projected. As will be demonstrated subsequently, population analysis -via- \pfhp~shows computational advantage over \pfop~in wall times with increase in the number of MPI tasks for large material system sizes. In this section, we discuss the mathematical formulation and derive the relevant expressions in a finite-element setting required for implementing \pfhp.  We subsequently compare the accuracy and performance of the proposed implementation with that of \pfop, which was discussed earlier. For clarity, the extension to periodic unit-cell involving Brillouin zone integration -via- multiple k-point sampling is not explicitly considered here.

\subsection{Mathematical formulation}\label{subsection:pfhpMathformulation}
Projected-population (both overlap and Hamilton populations) in this approach is computed by orthogonally projecting the Kohn-Sham discretized  Hamiltonian operator $\mathpzc{H}$ onto the atomic-orbital subspace $\mathbb{V}^{N_{orb}}_{\phi}$. The computed  projected Hamiltonian $\mathpzc{H}^{\phi} \equiv \mathpzc{P}^{\phi}\mathpzc{H}\mathpzc{P}^{\phi}: \mathbb{V}^{N_{orb}}_{\phi}  \rightarrow \mathbb{V}^{N_{orb}}_{\phi}$ is subsequently diagonalized to compute the orthonormal eigenvectors ($\ket{\tilde{\psi}^{E}_{i}}$) in the subspace $\mathbb{V}^{N_{orb}}_{\phi}$. These eigenvectors of the projected Hamiltonian,  $\ket{\tilde{\psi}^{E}_{i}} \in \mathbb{V}^{N_{orb}}_{\phi}$ are thereby employed to compute both overlap and Hamilton populations. The proposed approach \pfhp~is in contrast with \pfop~discussed above, (see section~\ref{section:pfop}) wherein the discretized Kohn-Sham wavefunction $\ket{\psi_i} \in \mathbb{V}^M$ is orthogonally projected (L$_2$ projection) onto the subspace $\mathbb{V}^{N_{orb}}_{\phi}$ to compute the overlap and Hamilton populations. \pfhp~is similar in spirit to the iterative orthogonal projection techniques\cite{iterorth} employed in the solution of large-scale matrix eigenvalue problems of the form $\mathpzc{A} \ket{x} = \lambda \ket{x}$, wherein, one seeks an approximate eigenvector, eigenvalue pair ($\ket{\tilde{x}},\tilde{\lambda}$) of $\mathpzc{A}$ in a carefully constructed lower-dimensional subspace such that the residual vector $\ket{r} = \mathpzc{A} \ket{\tilde{x}} - \tilde{\lambda} \ket{\tilde{x}}$ is orthogonal to this subspace. This orthogonality condition, also known as the Galerkin condition, is equivalent to diagonalizing the lower dimensional matrix (obtained by orthogonally projecting $\mathpzc{A}$ into the subspace), which approximates the eigenvalues and eigenvectors of the original matrix $\mathpzc{A}$ better than any vectors lying in the subspace.

\paragraph{Projected  Hamiltonian Hamilton population (\hhp):}\label{para:hhp} We note that $\ket{\tilde{\psi}^{{E}}_{i}}$, the eigenvectors of $\mathpzc{H^{\phi}}$ lie in the atomic-orbital subspace $\mathbb{V}^{N_{orb}}_{\phi}$ and, hence we express
$\ket{\tilde{\psi}^{{E}}_i}$ as a linear combination of the orthonormalised atomic orbital basis $\{\ket{\widehat{\phi}_{\mu}}\}$ i.e. $\ket{\tilde{\psi}^{E}_i} = \sum_{\mu}{\widehat{E}^{i}_{\mu}\ket{\widehat{\phi}_\mu}}$. Subsequently, following the similar arguments used in deriving eq~\eqref{eqn:bandenergy}, we can define the Hamilton population $\mathtt{pHHP}_{IJ}(\epsilon)$ associated with a  source atom $I$ and a target atom $J \neq I$ as
\begin{equation}
\hhp_{IJ}(\epsilon) = \sum_{j}{\sum_{\alpha \beta}{\mathfrak{Re}\left\{\left(\braket{ \tilde{\psi}^E_j| \widehat{\phi}_{I\alpha}}\right) H^{\phi}_{I\alpha,J\beta}\left(\braket{\widehat{\phi}_{J\beta} |\tilde{\psi}^E_j}\right)\right\} \delta(\epsilon - \epsilon_j)}} 
\label{eq:phhptot}
\end{equation}
where $H^{\phi}_{I\alpha,J\beta}$ is the matrix element of $\mathpzc{H^{\phi}}$ and  $\mathfrak{Re}(z)$ refers to the real-part of a complex number $z$.
Introducing the finite-element discretization of the various fields in eq \ref{eq:phhptot}, we now deduce the relevant matrix expressions required to evaluate the Hamilton population in FE setting. To begin, we use the definition of projected Hamiltonian $\Hphiop$, whose  matrix element $H^{\phi}_{I\alpha,J\beta}$ can be evaluated as $H^{\phi}_{I\alpha,J\beta}=\bra{\widehat{\phi}_{I\alpha}}\mathpzc{H}^{\phi}\ket{\widehat{\phi}_{J\beta}}$ and further can be expressed in the matrix form as $\bH^{\phi} = \bS^{-1/2}\bPhicheck^{\dagger}\bH \bPhicheck \bS^{-1/2}$ where $\bH$ denotes the matrix corresponding to the finite-element discretized Kohn-Sham Hamiltonian operator $\mathpzc{H}$ introduced in Section ~\ref{section: mathformulation}. Upon diagonalization of $\bH^{\phi}$ we have $\bH^{\phi} = \widehat{\bE} \widebar{\bD} \widehat{\bE}^{\dagger}$, where $\widehat{\bE}$ denotes the $N_{orb} \times N_{orb}$ eigenvector matrix. We note that the $j^{th}$ column of $\widehat{\bE}$ represents the coefficients of $\ket{\tilde{\psi}^{{E}}_i}$ with respect to $\{\ket{\widehat{\phi}_{\mu}}\}$ basis. 
Similar to section~\ref{section:pfop}, we introduce a composite-index $\mu = \{I\alpha\}$  to denote the orthonormalized atomic-orbital basis function $\ket{\widehat{\phi}_{\mu}}$ as $\ket{\widehat{\phi}_{I\alpha}}$ where $\alpha$ denotes the index of the atomic orbital centered at a nuclear position $\textbf{R}_I$. Further, $\mu^{th}$ row of $\widehat{\bE}$ corresponds to $I^{th}$ atom and an atom-centered index $\alpha$ associated with this atom $I$. We refer the reader to supporting information (see S1.2) for details on the derivation and efficient computation of the matrix expressions related to $\bH^{\phi}$ and $\widehat{\bE}$.
Finally, the \textit{projected  Hamiltonian Hamilton population} (\hhp) associated with the partitioning of band energy between a source atom $I$ and a target atom $J\neq I$ is evaluated by extracting the appropriate entries of the matrices $\widehat{\bE}$, $\bH^{\phi}$ and the expression is given by
\begin{equation} \label{eqn:pFHHPTotal}
\hhp_{IJ}(\epsilon) = \sum_j{\sum_{\alpha\beta}{ \mathfrak{Re}\left \{  \widehat{E}^{j*}_{{I\alpha}}H^{\phi}_{I\alpha,J\beta}\widehat{E}^{j}_{{J\beta}} \right \} } \delta(\epsilon-\epsilon_j)}
\end{equation}

\paragraph{Projected Hamiltonian overlap population (\hop):}\label{para:hop} Overlap population in the current approach is computed using the non-orthogonal localized atom-centered orbitals $\{\ket{\phi_\nu}\}$. To this end, we first compute the $N_{orb}$ linear combination coefficients of the expansion of $\ket{\tilde{\psi}_i^{E}}$ in terms of these basis $\{\ket{\phi_\nu}\}$ i.e., $\ket{\tilde{\psi}^E_j} = \sum_{\mu}\bar{E}^{j}_{{\mu}}\ket{\phi_{\mu}}$ for $j = 1 \cdots N$. Following the similar arguments used in arriving at eq~\eqref{eq:doscoop2},
we can define the overlap population  $\mathtt{pHOP}_{IJ}(\epsilon)$ between the source atom $I$ and $J$ as:
\begin{equation}
\hop_{IJ}(\epsilon) = \sum_{j}{\sum_{\alpha \beta}{ \mathfrak{Re}\left\{\left(\braket{ \tilde{\psi}^E_j| \phi_{I\alpha}}\right) S_{I\alpha,J\beta}\left(\braket{\phi_{J\beta} |\tilde{\psi}^E_j}\right)\right\} \delta(\epsilon - \epsilon_j)}}    
\end{equation}
where, $\mathfrak{Re}(z)$ refers to the real-part of a complex number $z$. Introducing finite-element discretization of the various fields in the above equation, we define the $N_{orb} \times N_{orb}$ matrix $\widebar{\bE}$ which can be computed as $\widebar{\bE} = \bS^{-1/2}\widehat{\bE}$ where the matrix $\widehat{\bE}$ is the eigenvector matrix of $H^{\phi}$ introduced previously. For the derivation and the computational cost associated with the computation of $\widebar{\bE}$, we refer to supporting information S1.2. Finally, the \textit{projected Hamiltonian overlap population} (\hop) associated with a source atom $I$ and a target atom $J$ is evaluated by extracting the appropriate entries of the matrices $\bS$, $\widebar{\bE}$ and the expression is given by: 
\begin{equation} \label{eq:pHOPTot}
\hop_{IJ}(\epsilon) = \sum_j{{\sum_{\alpha\beta}{\mathfrak{Re}\left\{\bar{E}_{I\alpha}^{j*}\bar{E}_{J\beta}^{j}S_{I\alpha J\beta}\right \} \delta(\epsilon - \epsilon_j)}}}
\end{equation}

\paragraph{Total computational complexity estimate of \pfhp:}\label{para:php complexity} The current implementation of the projected Hamiltonian population analysis (\pfhp) assumes $N=N_{orb}$ and thereby the total computational complexity is estimated to be $\sim 4 M_{loc} N^2 + 16 N^3$. Since $M_{loc} = M/P$, the second term in the computational complexity becomes dominant when the number of MPI tasks $P$ is greater than $4M/16N$ and starts to become computationally efficient than the \pfop~approach described in the previous subsection.
\paragraph{Projected Hamiltonian density error (\hder):}\label{para:hder} To understand the loss of information due to the projection of the finite-element discretized Hamiltonian onto $\mathbb{V}^{N_{orb}}_{\phi}$, we introduce the projected density error (\hder). Here, we compute the L$_2$ norm error between the ground-state electron density ($\rho(\bx)$) computed from FE discretized occupied Kohn-Sham eigenfunctions \{$\ket{\psi_i}$\} solved using \DFTFE~ and the electron-density ($\rho^{H}(\bx)$) computed from the occupied eigenfunctions $\{\ket{\tilde{\psi}_j^{E}}\} \in \mathbb{V}^{N_{orb}}_{\phi}$ associated with the projected Hamiltonian $\bH^{\phi}$. To this end, \hder~ is evaluated as
\begin{equation}\label{eq:hder}
  \hder = \frac{||\rho(\bx) - \rho^{H}(\bx)||_2}{||\rho(\bx)||_2}\;\; \text{where}\;\;\rho(\bx) = \sum_{i=1}^{N_{occ}}
  \braket{\bx|\psi_i}\braket{\psi_i|\bx},\;
  \;\;\rho^{H}(\bx) = \sum_{i=1}^{N_{occ}}
  \braket{\bx|\tilde{\psi}_i^{E}}\braket{\tilde{\psi}_i^{E}|\bx}
\end{equation}
\vspace{-0.25in}
\begin{figure}[H]
\includegraphics[scale=0.4]
{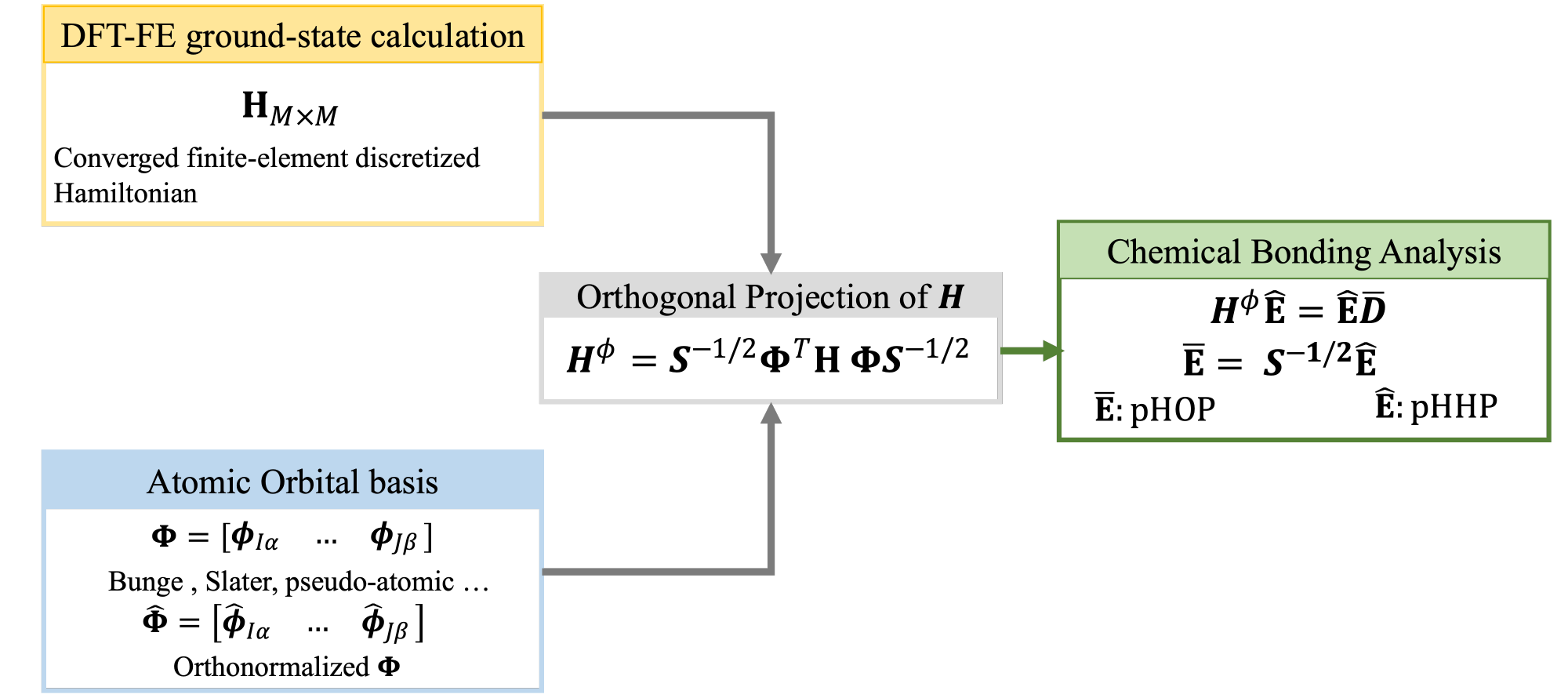}
\vspace{-0.1in}
\caption[short]{Overview of the implementation strategy for projected Hamiltonian population analysis (\pfhp) within the finite-element (FE) framework. The employed strategy involves projecting the self-consistently converged FE discretized Hamiltonian obtained from \DFTFE~onto the atomic-orbital basis and, subsequently diagonalizing this projected Hamiltonian $\bH^{\phi}$ to compute the eigenvector matrix $\widehat{\bE}$.  Overlap matrix ($\bS$) and the coefficient matrices $\widebar{\bE}$, are finally evaluated, which in-turn are used to compute projected Hamiltonian overlap population $\mathtt{pHOP}$ and projected Hamiltonian Hamilton population $\mathtt{pHHP}$}\label{chart pHPA}
\end{figure}

\subsection{Results: Accuracy and performance benchmarking}
We assess here the performance and accuracy of the proposed \pfhp~procedure implemented within the \DFTFE~framework. To this end, we project the self-consistently converged Kohn-Sham finite-element (FE) discretized Hamiltonian obtained from \DFTFE~into a subspace spanned by pseudo-atomic (\pa) orbitals  and conduct the population analysis as discussed in Section~\ref{subsection:pfhpMathformulation}. We discuss here a comparative study of the population analysis conducted using this approach and \pfop~reported in the subsection ~\ref{subsection:pfopMathformulation}. 

\begin{figure}[ht]
\includegraphics[scale=0.5]{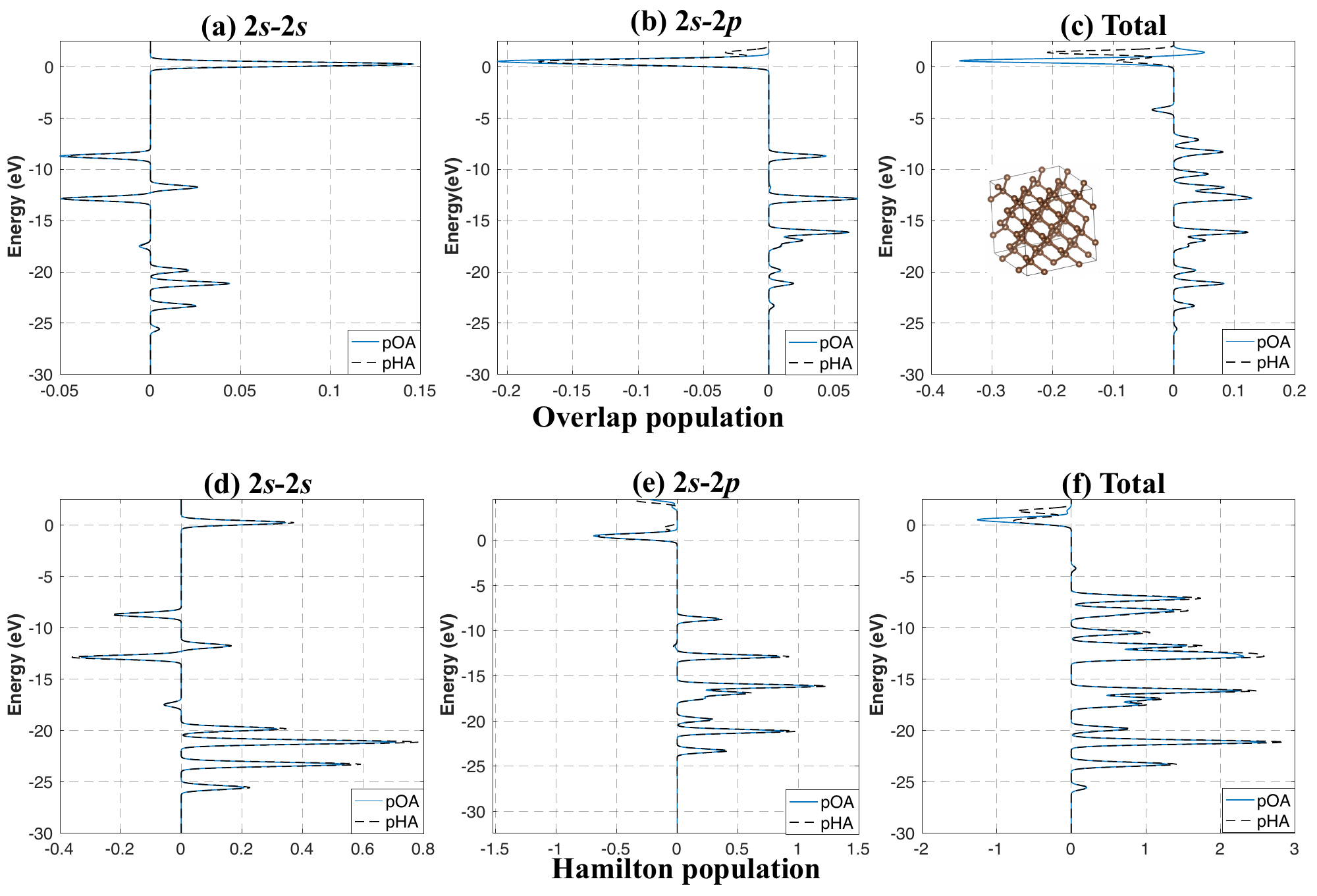}
\caption[short]{\footnotesize{\cb Comparison of overlap and Hamilton populations between the two proposed methods of projected population analysis (\pfop~and \pfhp) for nearest C atoms in carbon diamond supercell. The top row shows the overlap population obtained using both these methods. The sub-figures (a) and (b) in this row show the contributions of C$_{2s}$-C$_{2s}$ and C$_{2s}$-C$_{2p}$ to the total overlap population that is plotted in sub-figure (c). The bottom row shows the negative of the Hamilton population for both methods. The sub-figures in this bottom row (d) and (e) show the contributions of C$_{2s}$-C$_{2s}$ and C$_{2s}$-C$_{2p}$ to the total Hamilton population that is plotted in sub-figure (f). Energy-scale is shifted such that Fermi level ($\epsilon_F$) is zero. \cn
\textbf{Case study:} $2\cross2\cross2$ carbon diamond supercell with periodic boundary conditions at $\Gamma$-point for Brilloun zone sampling. }}\label{fig:C diamond new comparison}
\end{figure}

\begin{figure}[ht]
\includegraphics[scale=0.5]{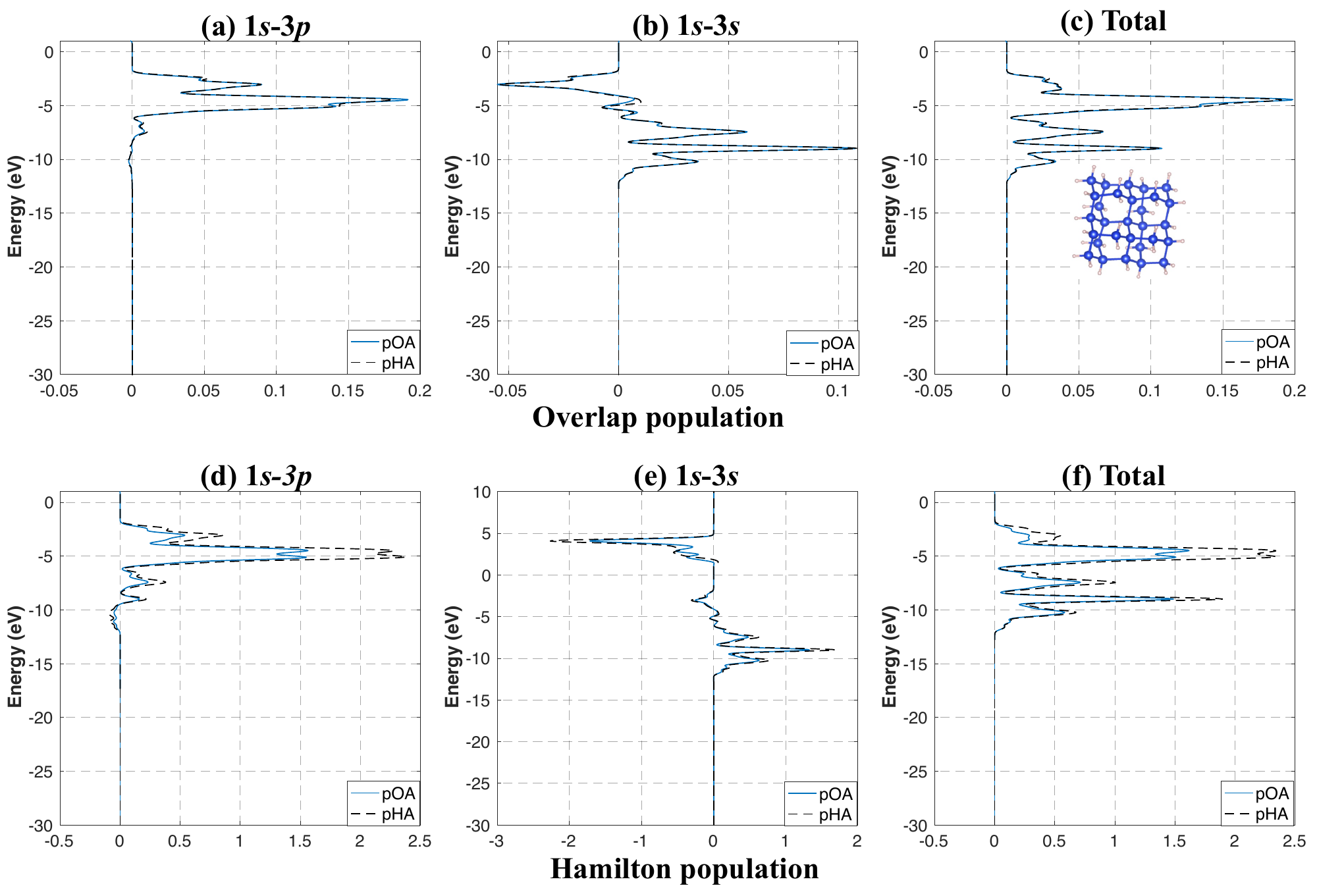}
\caption[short]{\footnotesize{\cb Comparison of overlap and Hamilton populations between the two proposed methods of projected population analysis (\pfop~and \pfhp) for nearest Si-H atoms in Si nanoparticle. The top row shows the overlap population obtained using both these methods. The sub-figures (a) and (b) in this row show the contributions of  H$_{1s}$-Si$_{3p}$ and H$_{1s}$-Si$_{3s}$ to the total overlap population that is plotted in sub-figure (c). The bottom row shows the negative of the Hamilton population for both methods. The sub-figures in the bottom row (d) and (e) show the contributions of H$_{1s}$-Si$_{3p}$ and H$_{1s}$-Si$_{3s}$ to the total Hamilton population that is plotted in sub-figure (f). Energy-scale is shifted such that Fermi level ($\epsilon_F$) is zero. \cn
\textbf{Case study:} single-fold Si\textsubscript{29}H\textsubscript{36} nanoparticle with non-periodic boundary conditions.}}\label{fig:1Fold Si29H36 new comparison}
\end{figure}

\paragraph{Accuracy validation:}\label{para:hp accuracy} To begin with, we plot the population energy diagrams corresponding to \hhp~and \hop~derived in eqns.~\ref{eq:phhptot} and~\ref{eq:pHOPTot} compare with \ohp~and \oop~on the same benchmark systems comprising of isolated systems and a periodic system as discussed in the previous subsection. As mentioned previously, we will refer to the approach of projected orbital population (\oop~and \ohp) as \pfop~and the approach of projected Hamiltonian population (\hop~and \hhp) as \pfhp. Figure~\ref{fig:C diamond new comparison} illustrates the comparison in the case of periodic $2 \times 2 \times 2$ carbon diamond supercell for two nearest carbon atoms picked as source and target atom (see inset in Figure~\ref{fig:C diamond new comparison}). Figure ~\ref{fig:1Fold Si29H36 new comparison} demonstrates the comparisons in the case of Si$_{29}$H$_{36}$ nanoparticle, an isolated system in which Si atom and the nearest H atom are picked as source and target atom (see inset in Figure~\ref{fig:1Fold Si29H36 new comparison}). As the results demonstrate, we see a very good match of the corresponding contributions of $s-s$ and $s-p$ orbitals for both overlap and Hamilton population conducted using both \pfop~ and \pfhp. Comparisons between both methods for benchmark systems involving molecules -- CO, spin polarized O$_2$, H$_2$O are discussed in supporting information section S2.1, and we observe a very close agreement.

We now compare the density error metrics (\oder~and \hder), a measure of loss of information during projections, as introduced in Section~\ref{para:oder} and Section~\ref{para:hder}. Recall from eq~\ref{eq:oder} and eq~\ref{eq:hder}, these metrics measure the error between the self-consistently converged ground-state electron-density computed from \DFTFE~and the electron-density computed using the projected wavefunctions in the subspace $\mathbb{V}_{\phi}^{N_{orb}}$. As shown in Table \ref{tab: Density Error}, we see a close agreement between \hder~ and \oder~ for a variety of benchmark material systems which include isolated systems (CO, H$_2$O, spin-polarized O$_2$, Si$_{29}$H$_{36}$, Si$_{58}$H$_{66}$, and Si$_{145}$H$_{150}$) and a periodic system (carbon $2\times 2 \times 2$ supercell). 

The comparative study discussed so far demonstrates the excellent match of numerical results obtained between the approaches \pfop~and \pfhp. As remarked before, \pfop~is similar in spirit to the projected population analysis approach implemented in \Lob~but the \pfhp~proposed in this work is different in spirit than the \pfop~approach and relies on the projection of finite-element discretized Kohn-Sham Hamiltonian to conduct population analysis. 

\begin{figure}[htp]
     \centering
     \begin{subfigure}[b]{0.49\textwidth}
         \centering
         \includegraphics[width=0.99\textwidth]{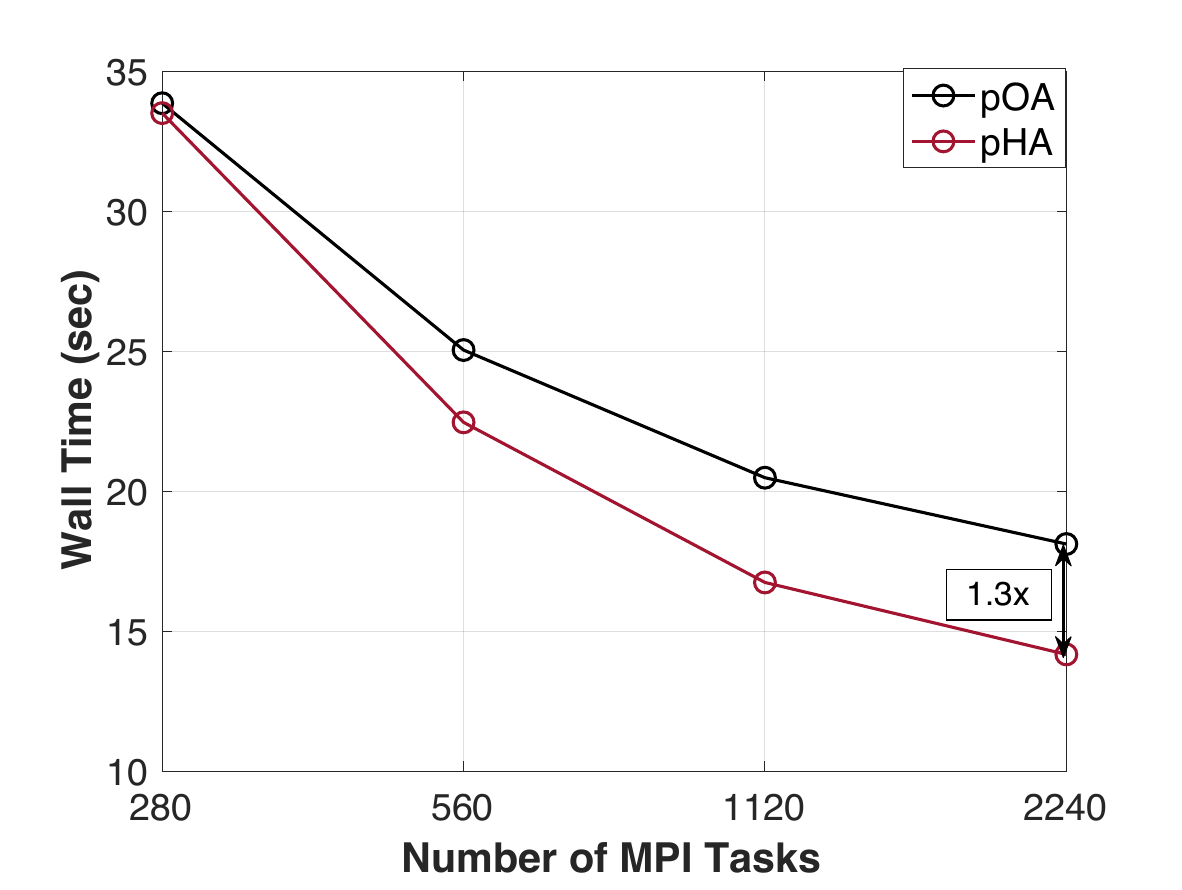}
         \caption[short]{\footnotesize{\textbf{Case study:} Si\textsubscript{580}H\textsubscript{510} (1090 atoms, 2830 wavefunctions)}}
         \label{fig:Scalingstudy comparison of pFOP with pFHP 20fold}
     \end{subfigure}
     \hfill
     \begin{subfigure}[b]{0.49\textwidth}
         \centering
         \includegraphics[width=0.99\textwidth]{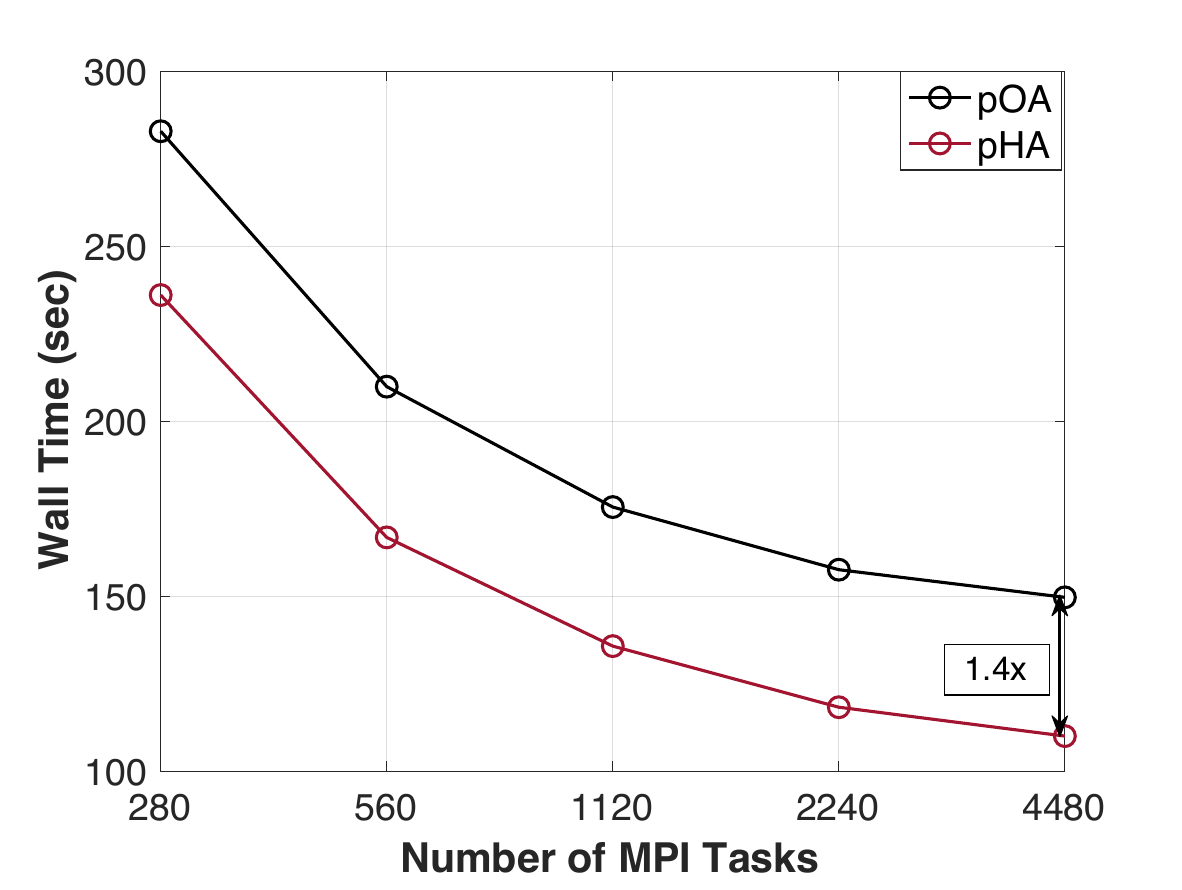}
         \caption[short]{\footnotesize{\textbf{Case study:} Si\textsubscript{1160}H\textsubscript{990} (2150 atoms, 5630 wavefunctions)}}
         \label{fig:Scalingstudy comparison of pFOP with pFHP 40fold}
     \end{subfigure}
        \caption[short]{\footnotesize{Wall-time comparison of \pfop~with \pfhp~on PARAM Pravega supercomputer (1 MPI task per core). Case study considered are 20-fold and 40-fold Si nanoparticles. Total DoFs ($M$) are 9130679 and 17254979 for Si\textsubscript{580}H\textsubscript{510} (20-fold) and Si\textsubscript{1160}H\textsubscript{990}(40-fold) respectively.}}
        \label{fig:three graphs}
\end{figure}

\paragraph{Performance comparisons:}\label{para:hp perf} We now demonstrate the computational advantage of conducting population analysis using \pfhp~for large-scale systems. Figures \ref{fig:Scalingstudy comparison of pFOP with pFHP 20fold} and \ref{fig:Scalingstudy comparison of pFOP with pFHP 40fold} show the computational wall times measuring the projected population analysis with the increasing number of MPI tasks comparing the two methods \pfop~and \pfhp. For this study, we consider two nanoparticles Si\textsubscript{580}H\textsubscript{510} and Si\textsubscript{1160}H\textsubscript{990} comprising of 1090 and 2150 atoms respectively. The results indicate a speed-up of $1.3\times$ - $1.4\times$ for \pfhp~with the increase in CPU cores beyond a certain number. These speed-ups are consistent with the computational complexity estimates derived in previous sections, i.e. $\ordercomplexity{(N^3)}$ cost becoming dominant beyond a certain number of cores with a lower prefactor for \pfhp.

\begin{table}[H]
    \centering
    \begin{tabular}{|c|c|c|}
    \hline \hline
    Material System & \oder & \hder \\
    \hline \hline
    \makecell{Carbon diamond \\ $2\cross2\cross2$ supercell} & 0.061 & 0.062 \\
    \hline
    CO & 0.118&0.119\\
    \hline
    H\textsubscript{2}O&0.136&0.121\\
    \hline
    O\textsubscript{2} $\uparrow$ (Spin-up density) & 0.097& 0.098 \\
    \hline
    O\textsubscript{2} $\downarrow$ (Spin-down density) & 0.062 & 0.064 \\
    \hline    
    Si$_{29}$H$_{36}$ nanoparticle & 0.179 &0.175 \\
    \hline
    Si$_{58}$H$_{66}$ nanoparticle & 0.172 & 0.168\\
    \hline
    Si$_{145}$H$_{150}$ nanoparticle & 0.165 & 0.162\\
    \hline
    \end{tabular}
    \caption{\footnotesize{Comparison of density errors (\oder~and \hder) computed from  projected orbital population (\pfop) and projected Hamiltonian population approach} (\pfhp) respectively}
    \label{tab: Density Error}
\end{table}

\section{Bonding insights in large systems:  Chemisorption in Si nanoparticles}
In this section, we demonstrate the advantage of the proposed computational framework for conducting population analysis to extract chemical bonding in large-scale material systems. We motivate the need for large-scale bonding analysis by considering the case-study of chemisorption of hydrogen in silicon nanoparticles, a candidate material for hydrogen storage \cite{Galli}, where the storage (release) of hydrogen is a result of the formation (breaking) of Si-H bond. As discussed in Williamson et.al\cite{Galli}, the release of hydrogen in these Si nanoparticles occurs due to the dimerization of dihedral SiH$_2$ co-facial pairs in a Si\textsubscript{29}H\textsubscript{36} unit to reduce to Si\textsubscript{29}H\textsubscript{24} unit by formation of an additional Si-Si bond. Furthermore, the authors also argued from a thermodynamic viewpoint that alloying these Si nanoparticles with C reduces the critical temperature of hydrogen absorption/desorption to an operating temperature compatible with fuel cell applications. In this case study,  we attempt to provide a chemical bonding viewpoint by conducting population analysis to extract bonding information in these Si nanoparticles. We compute a quantity known as integrated projected orbital Hamilton population $\mathtt{IPOHP} = \int_{\epsilon \leq \epsilon_F}{-\ohp(\epsilon)d\epsilon}$. A higher value of \ipohp~correlates with a stronger covalent bonding interaction between the source-target atoms and this quantity is similar to integrated \chp (\texttt{ICOHP}) computed in \Lob.  To this end, we argue the ease of dimerization of dihedral SiH$_2$ co-facial pairs in the Si nanoparticle by computing \ipohp~between Si-Si atoms in the adjacent dihedral SiH$_2$ pairs and the associated Si-H atoms in a given SiH$_2$ dihedral unit.  We examine \ipohp~values as a function of increasing Si nanoparticle size ranging from 65 atoms to 1090 atoms and further, with and without carbon alloying. In particular, we consider 1-fold, 2-fold, 5-fold and 20-fold structure of dihedral Si\textsubscript{29} and Si\textsubscript{24}C\textsubscript{5} units. Towards this, we build the 2-fold and further the 5-fold structure by connecting the Si\textsubscript{29}H\textsubscript{36} (Si\textsubscript{24}C\textsubscript{5}H\textsubscript{36}) units by their $(111)$ facets, as discussed in Williamson et.al\cite{Galli}.

\begin{figure}[H]
\centering
    \includegraphics[width=0.95\textwidth]{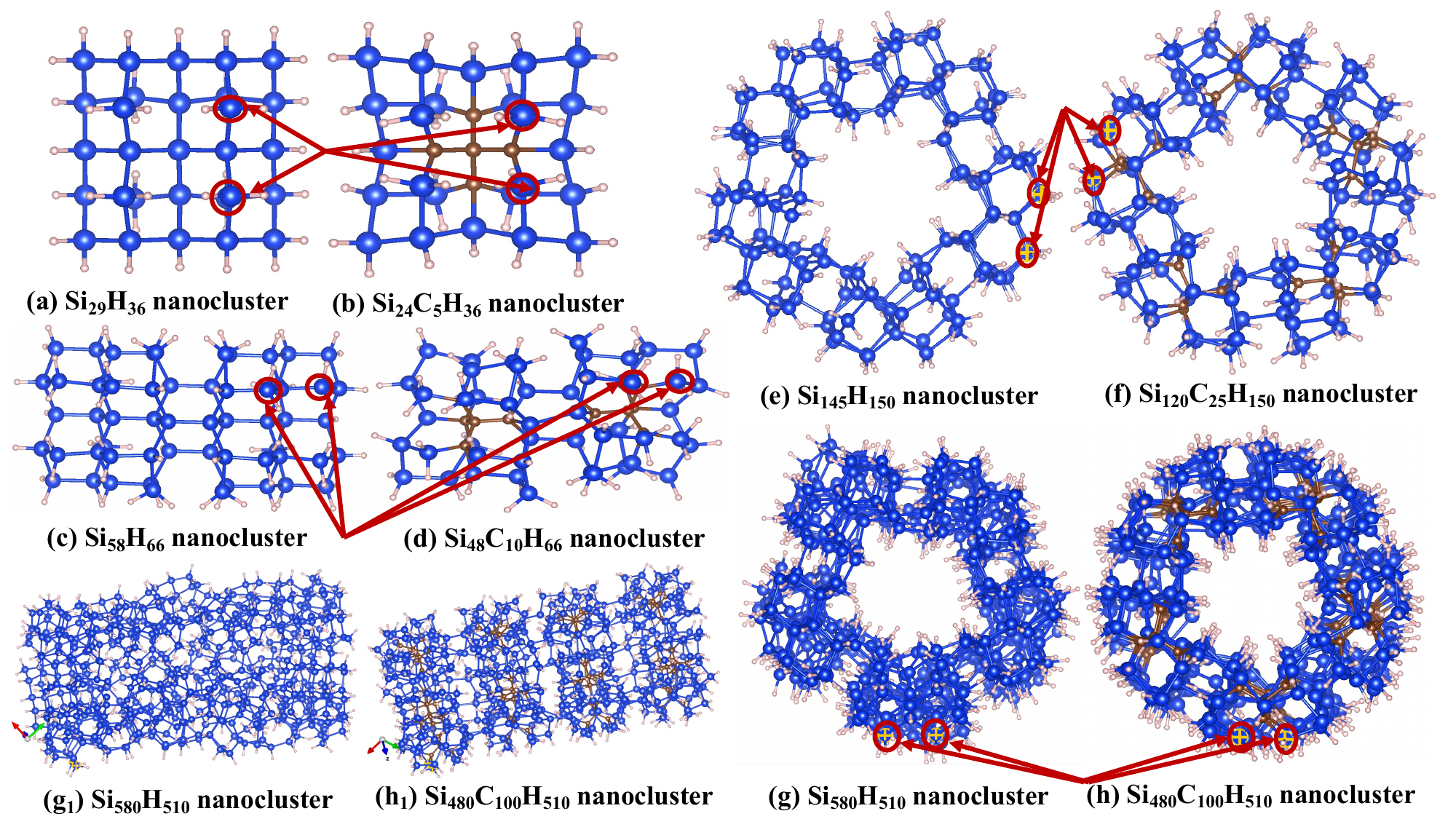}
\caption[short]{\footnotesize{Relaxed atomic configurations of (a,b) 1-Fold, (c,d) 2-fold, (e,f) 5-fold and (g,h) 20-fold nanoparticle structures. Mean of \ipohp~values computed in Table \ref{tab: ICOHP}~corresponds to  Si-Si interaction between nearest SiH$_2$ dihedral pairs sharing a core silicon/carbon atom. The silicon atoms highlighted in red illustrates one such pair considered. The structures were relaxed in DFT until the maximum force component on any atom reached a tolerance level of $5\cross 10^{-4}\frac{E_h}{\text{bohr}}$.}} \label{fig:Structures}
\end{figure}

\begin{figure}[H]
    \centering
    \includegraphics[width=0.99\textwidth]{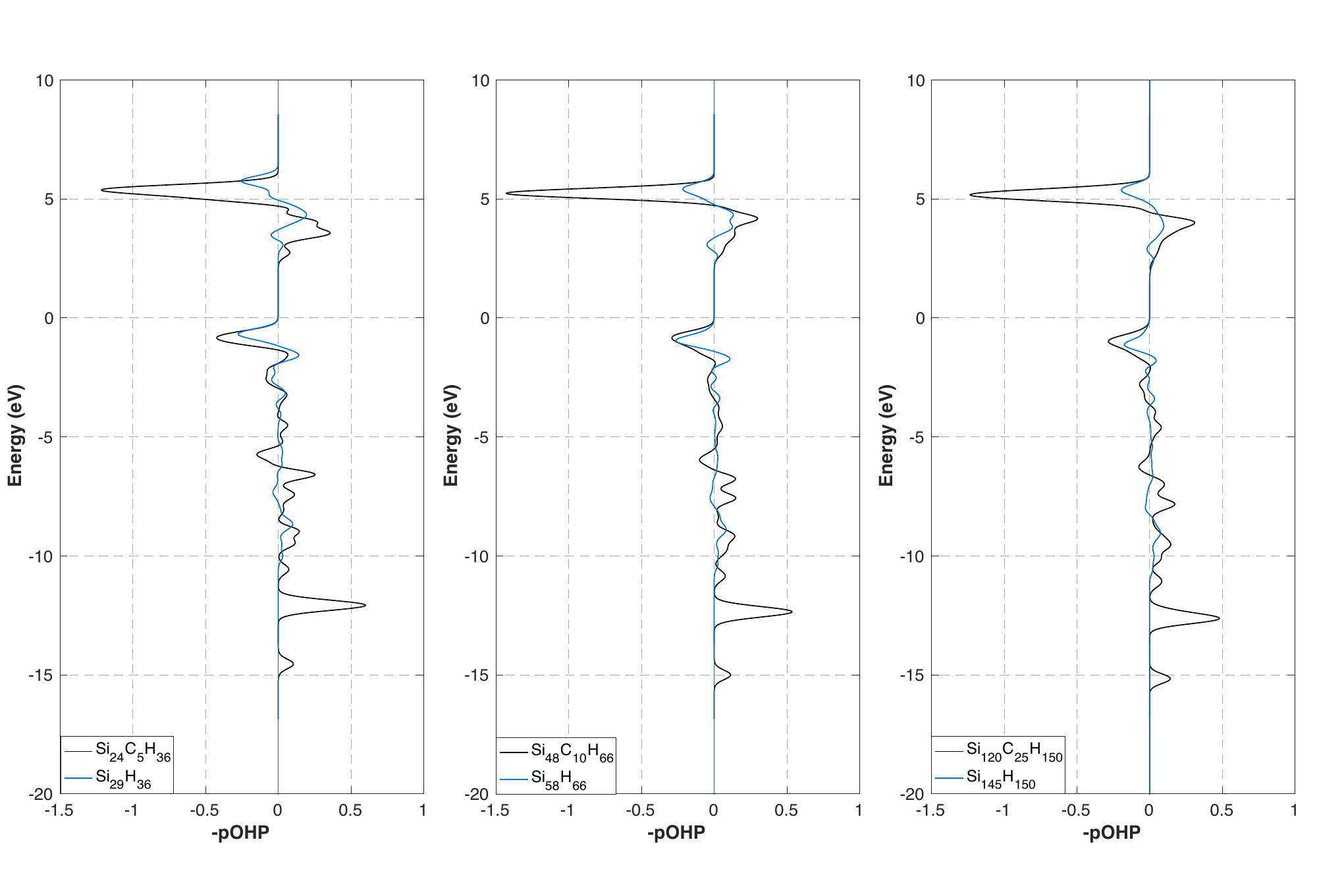}
    \caption{\footnotesize{Comparison of $-$\ohp~between  nearest non-bonded neighbours of Si-Si  in 1-Fold, 2-Fold and 5-Fold nanoparticles}}\label{fig:Si-Si}
\end{figure}

The atomic configurations corresponding to various sizes of Si\textsubscript{29} and Si\textsubscript{24}C\textsubscript{5} nanoparticles are obtained by performing geometry optimization in \DFTFE~code till the maximum atomic force in each direction reaches a tolerance of approximately $ 5 \times 10^{-4} \frac{E_h}{\text{bohr}}$. Figure \ref{fig:Structures} shows the relaxed atomic configuration of the various nanoparticles considered in this work. To determine the strength of the Si-Si bonding interaction between the nearest SiH$_2$ dihedral structures, we compute \ipohp~by conducting 
 \ohp~analysis. Figure~\ref{fig:Si-Si} shows the total Hamilton population energy diagrams for various sizes of Si nanoparticles, capturing this Si-Si interaction for one of the SiH$_2$ dihedral pairs (see atoms marked in Figure~\ref{fig:Structures}). We observe that the bonding and anti-bonding peaks are of equal magnitude for different sizes of Si nanoparticles without carbon, while for the nanoparticles alloyed with carbon, the bonding peaks are observed to be higher. Table \ref{tab: ICOHP} reports the mean of \ipohp~values corresponding to the Si-Si interaction between nearest SiH$_2$ dihedral pairs sharing a core silicon/carbon atom. The mean of the \ipohp~values corresponding to the weakest of the Si-H interaction in each of the dihedral unit of these pairs has also been tabulated in this table (Si-H1 and Si-H2). 
Equivalent Si-Si and Si-H atom pair was also picked from the nanoparticles without alloying for comparison. Consistent with the total Hamilton population energy plots of \ohp, \ipohp~values reported in Table \ref{tab: ICOHP} show a higher value ($\approx 4\cross$ higher) for nanoparticles alloyed with carbon in comparison to no alloying. This can be attributed to the carbon core in the alloyed nanoparticles drawing the Si atoms in the nearest SiH$_2$ dihedral units towards each other, thereby leading to a stronger Si-Si interaction. It is interesting to note the increase in Si-Si \ipohp~values with an increase in the curvature of the alloyed nanoparticles, indicating the strengthening of Si-Si interaction. This can possibly explain the ease of dimerization of dihedral SiH$_2$ co-facial pairs as the size increases from 1-fold to 5-fold structures. Similarly, due to the reduced separation between the dihedral SiH$_2$ groups, the corresponding \ipohp~values suggest a weakening of Si-H bond as seen in Table \ref{tab: ICOHP} for alloyed nanoparticles, a favourable condition for the release of H$_2$.

Maximizing the hydrogen storage capacity when designing hydrogen storage devices is important; hence, large nanoparticles are desirable. As described in Williamson et.al\cite{Galli}, one such structure is obtained by stacking four 5-fold Si nanoparticles resulting in a 20-fold nanoparticle containing 1090 atoms. \ipohp~values for this large nanoparticle with and without C alloying have been computed by conducting population analysis on the relaxed geometries. We observe a mean \ipohp~value of Si-Si interaction between the nearest SiH$_2$ dihedral pairs to be around 0.510 and 0.005 for the 20-fold nanoparticle with and without C alloying respectively, which are close to the values observed in the case of 5-fold nanoparticle. This is consistent with the fact that stacking does not significantly impact the curvature of the large nanoparticle compared to the 5-fold nanoparticle.

\begin{table}[]
    \centering
    \begin{tabular}{|c|c|c|c|}
    \hline
     System & Si-Si interaction  & Si-H1 interaction  & Si-H2 interaction  \\       
        \hline
         Si\textsubscript{29}H\textsubscript{36} & 0.009  & 4.505  & 4.504   \\
         \hline
         Si\textsubscript{24}C\textsubscript{5}H\textsubscript{36}  & 0.402 & 4.358   &  4.362   \\
         \hline \hline        
         Si\textsubscript{58}H\textsubscript{66} & 0.014 & 4.484  &  4.486  \\
         \hline
         Si\textsubscript{48}C\textsubscript{10}H\textsubscript{66}   &0.406  & 4.319 &   4.333  \\
         \hline   \hline       
         Si\textsubscript{145}H\textsubscript{150}& 0.005  & 4.498  &  4.498   \\
         \hline
         Si\textsubscript{120}C\textsubscript{25}H\textsubscript{150} &0.476  &4.297
   &    4.361 \\
         
         \hline         
    \end{tabular}
    \caption{\footnotesize{Comparison of mean \ipohp~for various SiH\textsubscript{2} dihedral pairs in 1-Fold, 2-Fold and 5-Fold nanoparticle of Si\textsubscript{29} and Si\textsubscript{24}C\textsubscript{5} units. Si-Si interaction is computed between nearest SiH$_2$ dihedral pairs. Si-H1 and Si-H2 denotes the weakest of the Si-H interaction in each of the dihedral unit of these pairs.}}
    \label{tab: ICOHP}
\end{table}


\section{Conclusions}
In the present work, we formulate and implement two methods for conducting projected population analysis (both overlap and Hamilton populations) to extract chemical bonding information from finite-element (FE) based density functional theory (DFT) calculations. The first method (\pfop) relies on the orthogonal projection of FE discretized Kohn-Sham DFT eigenfunctions onto a subspace spanned by localized atom-centered basis functions. In contrast, the second method (\pfhp) relies on the orthogonal projection of FE discretized Kohn-Sham Hamiltonian onto this subspace. These methods are implemented within the \DFTFE~code~\cite{motamarri2020,motamarri2022} and take advantage of \DFTFE's capability to conduct fast, scalable and systematically convergent large-scale DFT calculations, enabling large-scale bonding analysis on complex material systems not accessible before, without any restriction on the boundary conditions employed.  

First, we present the mathematical formulation, 
 and efficient finite-element strategies adopted to compute both the overlap and Hamilton population within the \textit{projected  orbital population analysis} (\pfop) framework. Following which, we assess the accuracy of the proposed method on representative material systems comprising of isolated molecules, nanoparticles and a periodic system with a large supercell. In all the cases, the proposed method shows excellent agreement with the population energy diagrams obtained using \Lob, a widely used orbital-population analysis code. The computational advantage of \pfop~over \Lob~is also clearly illustrated on few of these benchmark examples. Subsequently, we discuss an alternate approach for projected population analysis that does not rely on the availability of converged Kohn-Sham DFT eigenfunctions. This approach is motivated by the fact that many of the reduced scaling electronic structure codes targeted towards large-scale DFT calculations tend to avoid explicit computation of DFT eigenvectors with no access to these vectors for projection. This alternate method is referred to as \textit{projected Hamiltonian population analysis} (\pfhp). Accuracy and performance benchmarks of \pfhp~with \pfop~show similar trends in bonding behaviour with improved scalability for large-scale systems. Finally, we leverage the proposed population analysis approach in a case study to extract bonding insights in increasing sizes of Si nanoparticles up to 1000 atoms, a candidate material for hydrogen storage. This analysis demonstrates a correlation of Si-Si and Si-H bonding interactions with the nanoparticle size and argues the ease of Si-Si dimerization with the increase in the size of Si-C alloy nanocluster favouring the release of H$_2$.

\cb In summary, the proposed projected population analysis methods within the framework of finite-element discretization of DFT open the possibility of extracting chemical bonding information in large material systems critical to many technologically relevant applications. Our work demonstrates one such case by conducting a large-scale chemical bonding analysis in Si nanoparticles, a candidate material for hydrogen storage. Further, such analysis can also reveal bonding interactions between complex defects (e.g., dislocations, grain boundaries etc.) and solute impurities, offering atomistic insights into the stability of these defects, which has implications for understanding the strength-cum-ductility of structural materials. Another area of application is solid-state battery materials design, where such population analysis can aid in understanding ionic conductivity by revealing bonding interactions between migrating ions and the underlying solid electrolyte lattice in an electric field. Furthermore, using a single computational framework for both ground-state DFT calculations and population analysis allows for on-the-fly bonding analysis in \emph{ab initio} molecular dynamics simulations, yielding bonding interaction insights as a function of time. These are just a few examples among numerous possibilities the proposed methods can provide access to, offering a robust means for extracting chemical bonding information in various complex scenarios.\cn

%




\begin{acknowledgement}
The authors would like to thank Prof. Richard Dronskowski for many helpful discussions and his valuable suggestions. The authors gratefully acknowledge the seed grant from Indian Institute of Science and SERB Startup Research Grant from the Department of Science and Technology India (Grant Number: SRG/2020/002194) for the purchase of a GPU cluster, which also provided computational resources for this work. This work was also  supported by an NRF grant funded by MSIP, 470 Korea (No. 2009-0082471 and No. 2014R1A2A2A04003865), 471
the Convergence Agenda Program (CAP) of the Korea 472 Research Council of Fundamental Science and Technology 473 (KRCF), and the GKP (Global Knowledge Platform) project 474 of the Ministry of Science, ICT and Future Planning. The research used the resources of PARAM Pravega at Indian Institute of Science, supported by National Supercomputing Mission (NSM) R\&D for exa-scale grant (DST/NSM/R\&D\_Exascale/2021/14.02). P.M. also thanks Prathu Tiwari at Nvidia, Bangalore for helping us run a few of the geometry optimizations involving large-scale material systems on GPU clusters.
\end{acknowledgement}


\begin{suppinfo}

\setcounter{section}{0}
\setcounter{figure}{0}
\setcounter{table}{0}
\setcounter{equation}{0}
\setcounter{page}{1}

\section{Efficient finite-element implementation strategies}\label{section:S1}
This section discusses various aspects of the numerical implementation procedure employed to conduct projected population analysis using the Kohn-Sham DFT wavefunctions obtained -via- the solution of the finite-element discretized DFT eigenvalue problem (\DFTFE). To this end, the computations of various matrices involved in projected population analysis discussed in the main manuscript are implemented in the \DFTFE~code~\cite{motamarri2020,motamarri2022}, a massively parallel real-space code for large-scale density functional theory calculations based on adaptive finite-element discretization. Furthermore, the numerical implementation of population analysis in \DFTFE~also takes advantage of parallel computing architectures via \texttt{MPI}(Message Passing Interface), enabling chemical bonding analysis of large-scale systems in a unified computational framework to conduct both DFT calculations and the chemical bond analysis.

We begin by discussing the computation of the finite-element overlap matrix ($\bM$) and the atomic-orbital overlap matrix computed in FE basis ($\bS$). We then delve into the implementation strategies employed for computing the  projected  orbital population (\pfop) and the projected Hamiltonian population (\pfhp) discussed in Sections 3 and 4 of the main manuscript respectively.

\paragraph{FE basis overlap matrix ($\bM$):} Finite-element (FE) basis functions are non-orthogonal, and the associated overlap matrix $\bM$ is computed by evaluating the following integral over the simulation domain volume denoted by $\Omega$
\begin{align}\label{eqn:FEOverlap}
    M_{ij} &= \int_{\Omega} N_i(\bx)N_j(\bx) \dx =  \sum_{\Omega_e} \int_{-1}^{1} \int_{-1}^{1} \int_{-1}^{1} N_i(\xi,\eta,\zeta)\,N_j(\xi,\eta,\zeta)\, det(J_e)\, d\xi \,d\eta\, d\zeta\\ \nonumber
    &= \sum_{\Omega_e}\sum_{p,q,r=0}^{n_q} w_{p,q,r}\, N_i(\xi_p,\eta_q,\zeta_r)\, N_j(\xi_p,\eta_q,\zeta_r)\, det(J_e)
\end{align}
$(\xi,\eta,\zeta)$ above denote the barycentric coordinates~\cite{brenner2002}, $J_e$ denotes the Jacobian matrix corresponding to a finite-element $\Omega_e$, and $n_q$ denotes the number of quadrature points in each dimension employed to evaluate the integral in eq~\eqref{eqn:FEOverlap}. Gauss-Lobatto-Legendre (GLL) quadrature rules ~\cite{spectralGLL} are employed to evaluate the integrals in eq~\eqref{eqn:FEOverlap}. These rules have quadrature points coincident with the FE nodal points in the spectral finite-element discretization employed in this work rendering the matrix $\bM$ diagonal since the above equation is non-zero only if $i=j$. This diagonal FE basis overlap matrix has been employed in the \DFTFE~ code ~\cite{motamarri2013,motamarri2020,motamarri2022} to transform the generalized Kohn-Sham eigenvalue problem into standard eigenvalue problem allowing the use of efficient Chebyshev-filtered subspace iteration procedures to compute the Kohn-Sham eigenspace. This diagonal matrix $\bM$ will also play a crucial role in the computationally efficient evaluation of projected  populations within the FE setting, as discussed subsequently.

\paragraph{Atom-centered orbital overlap matrix ($\bS$):} The localized non-orthogonal atom-centered basis orbitals $\{\ket{\phi_{\mu}}\}$ are represented in a finite-element basis (see eq (3) of Section 2 in main manuscript) and hence the associated overlap matrix element $S_{\alpha \beta} = \braket{\phi_{\alpha} | \phi_{\beta}}$ is computed as follows:
\begin{equation}
    S_{\mu \nu} = \int \phi_{\mu}(\bx) \phi_{\nu}(\bx) \dx  = \sum_{p,q=1}^{M} \phi_{\mu}^p \left(\int_{\Omega} N_p(\bx) N_q(\bx) \dx\right) \phi_{\nu}^q 
\end{equation}
 The above equation is recast in a matrix form using the FE basis overlap matrix $\bM$ as shown below:
 \begin{equation}\label{eqn:aooverlap}
    \bS = \bPhi^{\dagger} \bM \bPhi = \bPhi^{\dagger}\bM^{1/2}\bM^{1/2}\bPhi = \bPhicheck^{\dagger}\bPhicheck \;\; \text{where} \;\; \bPhicheck = \bM^{1/2}\bPhi.
 \end{equation}
In the above, $\bPhi$ denotes a $M \times N_{orb}$ matrix whose column vectors are the components of $\ket{\phi_{\alpha}}$ in FE basis. Recalling that $\bM$ is diagonal, computation of $\bM^{1/2}$ becomes trivial and thereby, the evaluation of $\bPhicheck$ involves a point-wise scaling operation of the columns in $\bPhi$ with the diagonal entries of $\bM^{1/2}$. Finally, the computation of $\bS$ is reduced to matrix-matrix multiplication involving $\bPhicheck$ as described in eq~\eqref{eqn:aooverlap}. In a parallel implementation of the computation of $\bS$ using \texttt{MPI} on multi-node CPU  architectures, the matrix $\bPhicheck$ is distributed equally among the available \texttt{MPI} tasks into equipartitioned matrix $\bPhicheck_P$ of size $M_{\text{loc}} \times N_{orb}$ with $M_{loc} \approx M/P$, and $P$ denoting the number of \texttt{MPI} tasks. The domain decomposition of the underlying FE mesh across these \texttt{MPI} tasks achieves this equal distribution. We now compute the $N_{orb} \times N_{orb}$ matrix $\bS_P = \bPhicheck_P^{T}\bPhicheck_P$ associated with each core locally using BLAS level 3 optimized math kernel libraries. We finally add these local matrices $\bS_P$ by employing \texttt{MPI} collectives to compute the matrix-matrix product $\bPhicheck^{T}\bPhicheck$ in a distributed setting. To that effect, the computational complexity of computing $\bS$ using the above algorithm when running on $P$ \texttt{MPI} tasks is $\ordercomplexity{(M_{loc} N_{orb}^2)} \sim 2 M_{loc} N_{orb}^2$. As will be discussed subsequently, the projected population analysis requires the computation of $\bS^{-\beta}$ where $\beta = 1$ or $-1/2$ or $1/2$. Therefore, in the current work, we diagonalize $\bS$ matrix and evaluate $\bS^{-\beta} = \bQ \bD^{-\beta} \bQ^{T}$ where $\bQ$ is an eigenvector matrix with columns as eigenvectors of $\bS$ and $\bD$ is a diagonal matrix comprising of eigenvalues of $\bS$ in its diagonal. We note that an efficient implementation of divide-and-conquer algorithm~\cite{demmel,cuppen} available in \texttt{LAPACK} library is employed for diagonalization and the computational complexity of this algorithm for diagonalization of $\bS$ is $\ordercomplexity({N_{orb}^3}) \sim 4 N_{orb}^3$. Furthermore, the computational complexity of the matrix-matrix multiplication for evaluating $\bS^{-\beta}$ (after diagonalization) is $\ordercomplexity({N_{orb}^3}) \sim 2 N_{orb}^3$.

\subsection{Projected orbital population analysis (\pfop)}\label{subsection: pfop}
In this subsection we discuss the numerical implementation strategies 
and computational complexity of evaluating various matrices involved in conducting \pfop~implemented within the framework of \DFTFE. 
\paragraph{Projected Kohn-Sham wavefunction overlap matrix ($\bO$):} We first discuss the computation of coefficients of projected Kohn-Sham wavefunctions in the basis of $\{\ket{\phi_{\mu}}\}$ and then derive an expression for computing the overlap matrix associated with projected Kohn-Sham wavefunctions using this coefficient matrix ($\bC$) within the finite-element setting. To this end, we note that the projected Kohn-Sham wavefunction $\ket{{\psi}_i^{\phi}} \in \mathbb{V}^{N_{orb}}_{\phi}$ can be expressed as a linear combination of the atomic-orbital basis $\{\ket{\phi_{\mu}}\}$ i.e. $\ket{{\psi}_i^{\phi}} = \sum_{\alpha} C^{i}_{\alpha} \ket{\phi_\alpha}$ where $C^{i}_{\alpha} = \sum_{\nu}S^{-1}_{\alpha \nu} \braket{\phi_{\nu}|\psi_{i}}$. Introducing the finite-element discretization for $\ket{{\psi}_i}$ and $\ket{\phi_{\alpha}}$ (see eq (2) in section 2 of main manuscript), we have
\begin{equation}
 C^{i}_{\alpha} = \sum_{\nu}\sum_{p}\sum_{q} S^{-1}_{\alpha \nu} \, \phi^{p}_{\nu}\left(\int_{\Omega} N_p(\bx)  N_q(\bx)  \dx \right) \psi_i^{q}   
\end{equation}
The above equation is recast in a matrix form in terms of the matrix $\bM$ as 
\begin{equation}\label{eqn:coeffmat}
    \bC = \bS^{-1}\bPhi^{\dagger}\bM\bPsi = \bS^{-1}\bPhi^{\dagger}\bM^{1/2} \bM^{1/2} \bPsi = \bS^{-1}\bPhicheck^{\dagger}\bPsicheck \;\;\text{where} \;\; \bPsicheck = \bM^{1/2}\bPsi
\end{equation}
In the above, $\bPsi$ denotes a $M \times N$ matrix whose column vectors are the components of $\ket{\psi_{i}}$ in FE basis while $\bPhi$ is a $M \times N_{orb}$ matrix comprising of the atomic-orbital data as defined above. Further, the matrix $\bC$ is of dimension $N_{orb} \times N$ and is evaluated efficiently on a parallel computing system by first evaluating the matrix-matrix product $\bPhicheck^{T}\bPsicheck$ locally on each \texttt{MPI} task and summing the contributions across various \texttt{MPI} tasks. When running on $P$ \texttt{MPI} tasks, the computational complexity of this step is $\ordercomplexity{(M_{loc} N_{orb} N)} \sim 2 M_{loc} N_{orb} N$.  Subsequently, the resulting $N_{orb} \times N$ matrix is pre-multiplied with the inverse of the atomic-orbital overlap matrix $\bS$ computed above to evaluate the $\bC$ matrix finally. We now consider the evaluation of the overlap matrix $\bO$ associated with the projected Kohn-Sham wavefunctions $\psi^{\phi}_i$ by first recalling that $O_{ij} = \braket{\psi^{\phi}_i | \psi^{\phi}_j} = \bra{\psi_i}\mathpzc{P}^{\phi}\mathpzc{P}^{\phi}\ket{\psi_j} = \bra{\psi_i}\mathpzc{P}^{\phi}\ket{\psi_j}$. Using the definition of $\mathpzc{P}^{\phi} = \sum_{\mu, \nu = 1}^{N_{orb}}{\ket{\phi_{\mu}} {S}^{-1}_{\mu \nu} \bra{\phi_{\nu}}}$, one can rewrite the matrix elements of  $\bO$ as $O_{ij} = \sum_{\mu,\nu}\braket{\psi_i|\phi_\mu}S^{-1}_{\mu \nu} \braket{\phi_\nu|\psi_j}$. To this end, in a finite-element discretized setting, the matrix $\bO$ can be computed as
\begin{align}\label{eq:projoverlap}
    &O_{ij} = \sum_{\mu \nu}  \sum_{p q} \sum_{r s} \left[\psi_{i}^{p} \left(\int_{\Omega} N_p(\bx) N_q(\bx) \dx\right)\phi_{\mu}^q\right] S^{-1}_{\mu \nu} \left[\phi_{\nu}^{r} \left(\int_{\Omega} N_r(\bx) N_s(\bx) \dx\right)\psi_{j}^s\right] \nonumber \\
    &\implies \bO = \bPsi^{\dagger}\bM\bPhi \bS^{-1} \bPhi^{\dagger}\bM \bPsi = \bC^{\dagger} \bS \bC
\end{align}
The matrix expression for the projected Kohn-Sham wavefunction overlap matrix $\bO$ in the above equation uses the expression for the coefficient matrix $\bC$ in eq~\eqref{eqn:coeffmat}.
After the computation of $\bS$ and $\bC$ are computed using the expressions in eqns.~\eqref{eqn:aooverlap} and ~\eqref{eqn:coeffmat}, the matrix $\bO$  can be computed by performing matrix-matrix multiplications and the computational complexity for evaluating $\bO$ is $\ordercomplexity{(N_{orb}^2 N)} + \ordercomplexity{(N_{orb} N^2)} \sim 2 N_{orb}^2 N + 2 N^2 N_{orb}$
\paragraph{Computation of coefficient matrix $\bCbar$:}
We evaluate the coefficient matrix ($\bCbar$) corresponding to the coefficients of $\{\ket{\tilde{\psi}_j^{\phi}}\}$ in the basis of $\{\ket{\phi_{\mu}}\}$. Recall from Section 3 that $\widebar{C}^{j}_{\mu} =  \sum_{q} O_{jq}^{-1/2}C^{q}_{\mu}$. Hence in the matrix form, the coefficient matrix $\bCbar$ can be written as $\bCbar = \bC \bO^{-1/2}$ and can be evaluated with a computational complexity of $\ordercomplexity{(N_{orb}N^2)} \sim 2N_{orb}N^2$. Furthermore, we note that $\bO^{-1/2}$ is evaluated by diagonalizing $\bO$ and associated the computational complexity is $\ordercomplexity{(N^3)} \sim 4N^3$.  The matrix $\bCbar$ is of size $N_{orb} \times N$, and the rows of this matrix are stored in the order of atoms and their corresponding atom-centered orbitals for a given atom in succession. To elaborate, $\mu^{th}$ row of $\bCbar$ corresponds to $I^{th}$ atom and an atom-centered index $\alpha$ associated with this atom $I$, while the $j^{th}$ column of this matrix corresponds to the index of projected Kohn-Sham wavefunction ($j = 1 \cdots N$). 
\paragraph{Computation of coefficient matrix $\bChat$:}
We recall the relation between  L\"{o}wdin symmetric orthonormalized  atom-centered basis $\{\ket{\widehat{\phi}_\nu}\}$ and the non-orthogonal atom-centered basis $\{\ket{{\phi}_\nu}\}$ to be $\ket{\widehat{\phi}_{\mu}}=\sum_{\nu} S^{-1/2}_{\mu \nu}\ket{\phi_{\nu}}$. Recasting this relation in matrix form we get $\widehat{\bPhi} = \bPhi \bS^{-1/2}$ where $\widehat{\bPhi}$ denotes $M \times N_{orb}$ matrix whose column vectors are components of $\{\ket{\widehat{\phi}_\nu}\}$ in FE basis. Now, we note that  $\ket{{\tilde{\psi}}_i^{\phi}} \in \mathbb{V}^{N_{orb}}$ can be expressed as a linear combination of the basis $\{\ket{\widehat{\phi}_\nu}\}$ i.e $\ket{{\tilde{\psi}}_i^{\phi}}=\sum_{\nu} \widehat{C}_{\nu}^{i}\ket{\widehat{\phi}_\nu}$ where $C_{\nu}^{i} = \sum_{q} O_{jq}^{-1/2} \braket{\widehat{\phi}_\nu|\psi_q}$. Introducing finite-element discretization for $\ket{\widehat{\phi}_\nu}$ and $\ket{\psi_q}$, we have
\begin{equation}
 \widehat{C}_{\nu}^{i} = \sum_{q}\sum_{r}\sum_{s} O^{-1/2}_{iq} \widehat{\phi}^{r}_{\nu}\left(\int_{\Omega} N_r(\bx) N_s(\bx) \dx \right) \psi_q^s   
\end{equation}
Recasting the above equation in matrix form, we have
\begin{equation}
  \widehat{\bC} =  \widehat{\bPhi}^{\dagger} \bM \bPsi \bO^{-1/2} = \bPhihatcheck \bPsicheck \bO^{-1/2} \;\;\;  = \bS^{1/2}\bS^{-1}\bPhihatcheck^\dagger \bPsicheck \bO^{-1/2} \;\;\; = \bS^{1/2}\widebar{\bC} 
\end{equation}
where $\bPhihatcheck = \bM^{1/2}\widehat{\bPhi}$,  $\bPsicheck$ and $\widebar{\bC}$ are defined previously. $\widehat{\bC}$ in the above equation is of size $N_{orb} \times N$ and is computed -via- matrix-matrix multiplication involving $\widebar{\bC}$ and $\bS^{1/2}$ with a computational complexity of $\ordercomplexity{(N_{orb}^2 N)} \sim 2 N_{orb}^2 N$. Similar to other coefficient matrices described previously, the rows of this matrix are stored in the order of atoms and their corresponding atom-centered orbitals for a given atom in succession.
\paragraph{Computation of projected Hamiltonian matrix $\bH^{p}$:}
We compute the $N_{orb}\times N_{orb}$ projected Hamiltonian matrix $\bH^{p}$ using the coefficient matrix $\widehat{\bC}$ as $\bH^{p} = \widehat{\bC} \bD \widehat{\bC}^{\dagger}$ where the matrix $\bD$ is diagonal and comprises of the Kohn-Sham eigenvalues $\epsilon_i$ obtained from Kohn-Sham DFT problem solved in the finite-element basis. The computation of $\bH^{p}$ involves matrix-matrix multiplication after scaling $\widehat{\bC}^{T}$ with diagonal matrix $\bD$ and has the computational complexity of $\ordercomplexity{(N_{orb}^2 N) \sim 2 N_{orb}^2 N}$.
 \paragraph{$\bk$-dependent projected orbital population analysis:} \label{subsection: kdependentFE}
We now discuss the expressions for projected population analysis for conducting a $\bk$-dependent calculation within the finite-element framework of \pfop~for periodic systems. This is very similar to the expressions derived in section 3. However, we project here the Bloch wavefunction ($\ket{\psi_{i,\bk}}$) computed from the self-consistently converged solution of the FE-discretized Kohn-Sham Hamiltonian in \DFTFE~onto a basis spanned a linear combination of atomic orbitals $\{ \ket{\phi_{\mu,\bk}} \}$  that satisfy the Bloch theorem, given by $ \phi_{\mu,\bk}(\bx) = \sum_{\bR}{e^{i\bk\cdot\bR}\phi_\mu(\bx - \bR)}$ where $\bR$ denotes the lattice translation vector. Similar in spirit to \oop~and \ohp~from section 3, we extract the appropriate entries from the matrices $\bH^p(\bk), \bS(\bk),\bCbar(\bk), \bChat(\bk)$ to compute the $\bk$-dependent overlap and Hamilton population as given below:
\begin{equation}\label{eq:pfoopKpoint}
    \oop_{IJ} (\epsilon,\bk) = \sum_{j}{\sum_{\alpha \beta}{\mathfrak{Re}\left({\widebar{C}_{I\alpha}^{j*}(\bk)}\widebar{C}_{J\beta}^{j}(\bk) S_{I\alpha J \beta}(\bk)\right) \delta(\epsilon - \epsilon_{j,\bk}) }}
\end{equation}
\begin{equation} \label{eq:pfohpKpoint}
\ohp_{IJ}(\epsilon,\bk) = \sum_j{\sum_{\alpha,\beta}{\mathfrak{Re}\left({\widehat{C}^{j*}_{{I\alpha}}(\bk)H^{p}_{I\alpha,J\beta}(\bk)}\widehat{C}^{j}_{{J\beta}}(\bk)\right)\delta(\epsilon - \epsilon_{j,\bk})}} \end{equation}
where, $\mathfrak{Re}(z)$ refers to the real-part of a complex number $z$. Further, we refer to section SI-\ref{subsection: kdependentOPA} in this supporting information for validation of our implementation.

\subsection{Projected Hamiltonian population analysis(\pfhp)}\label{subsection: pfhp}
In this subsection, we discuss the numerical implementation strategies 
and computational complexity of evaluating various matrices involved in the implementation of \pfhp~within the framework of \DFTFE. 
\paragraph{Computation of projected Hamiltonian matrix $\bH^{\phi}$:}
Let $\bH$ denote the matrix corresponding to the self-consistently converged finite-element discretized Kohn-Sham Hamiltonian operator $\mathpzc{H}$ introduced earlier. We first begin with the computation of the projection of $\bH$ into the space $\mathbb{V}^{N_{orb}}_{\phi}$ spanned by   L\"{o}wdin symmetric orthonormalized   atomic-orbital basis $\{\ket{\widehat{\phi}_\nu}\}$. From section 3, we recall the relation between L\"{o}wdin orthonormalized atom-centered basis $\{\ket{\widehat{\phi}_\nu}\}$ and the non-orthogonal atom-centered basis $\{\ket{{\phi}_\nu}\}$ to be $\ket{\widehat{\phi}_{\mu}}=\sum_{\nu} S^{-1/2}_{\mu \nu}\ket{\phi_{\nu}}$. Recasting this relation in matrix form we get $\widehat{\bPhi} = \bPhi \bS^{-1/2}$ where $\widehat{\bPhi}$ denotes $M \times N_{orb}$ matrix whose column vectors are components of $\{\ket{\widehat{\phi}_\nu}\}$ in FE basis. To this end, the matrix elements of the projected Hamiltonian expressed in $\{\ket{\widehat{\phi}_\nu}\}$ basis is given by $H^{\phi}_{ij}=\bra{\widehat{\phi}_{i}}\mathpzc{H}^{\phi}\ket{\widehat{\phi}_{j}}=\bra{\widehat{\phi}_{i}}\mathpzc{H}\ket{\widehat{\phi}_{j}}$, which can, in turn, be recast in the matrix form as $\bH^{\phi} = \bS^{-1/2}\bPhicheck^{T}\bH \bPhicheck \bS^{-1/2}$. The dominant computational complexity of computing $\bH^{\phi}$ when running on $P$ MPI tasks is $ \ordercomplexity{(M_{loc}N_{orb}^2)} + \ordercomplexity{(N_{orb}^3)} + \ordercomplexity{(N_{orb}^3)} \sim 2M_{loc}N_{orb}^2 + 4N_{orb}^3$.

\paragraph{Computation of coefficient matrices $\widehat{\bE}$ and $\widebar{\bE}$:} We note that the diagonalization of the projected Hamiltonian $\bH^{\phi}$ results in $\bH^{\phi} = \widehat{\bE} \widebar{\bD} \widehat{\bE}^{T}$ where $\widehat{\bE}$ denotes the $N_{orb} \times N_{orb}$ eigenvector matrix. We note that the $j^{th}$ column of $\widehat{\bE}$ represents the coefficients of the eigenvector $\ket{\tilde{\psi}_j^{E}}$ of $\mathpzc{H}^{\phi}$ with respect to $\{\ket{\widehat{\phi}_\nu}\}$ basis and is used in the computation of the Hamilton population (see eq (14) of section 4.1). Further, the computation of overlap population (see eq (16) in section 4.1) requires the evaluation of the $N_{orb} \times N_{orb}$ matrix $\widebar{\bE}$ with $j^{th}$ column of $\widebar{\bE}$ representing the coefficients of the eigenvector $\ket{\tilde{\psi}_j^{E}}$ of $\mathpzc{H}^{\phi}$ with respect to $\{\ket{{\phi}_\nu}\}$ basis. Hence the matrix $\widebar{\bE}$ can be easily obtained from $\widehat{\bE}$ by taking recourse to basis transformation operation i.e., $\widebar{\bE} = \bS^{-1/2}\widehat{\bE}$. The computational complexity of diagonalizing $\bH^{\phi}$ is $\ordercomplexity{(N_{orb}^3)} \sim 4N_{orb}^3$ and the computation of  $\widebar{\bE}$ is of $\ordercomplexity{(N_{orb}^3)} \sim 2N_{orb}^3$ complexity.

\section{Results: Additional benchmarking studies} \label{section: S2}
\subsection{Projected orbital population analysis (\pfop)} \label{subsection:opa}
In this section, we describe \pfop~benchmarking studies involving the computation of \oop~and \ohp~on few representative material systems not discussed in the main manuscript. $\mathtt{pCOOP}$ and $\mathtt{pCOHP}$ obtained by using \Lob~are used for benchmarking. To this end, we report the absolute spill factor($\mathcal{S}$), charge spill factor($\mathcal{S}_c$) and the population energy diagrams of CO molecule, H\textsubscript{2}O molecule and spin-polarized O\textsubscript{2} molecule. From Table \ref{tab: Spill Factors}, we observe that the spill factors obtained from \pfop~using \pa~atom-centered orbitals are similar to that obtained from \Lob. Further, we compare the population energy diagrams resulting from the projection of the Kohn-Sham(KS) eigenfunctions obtained from \DFTFE~onto a space spanned by both (i) STO basis by Bunge and  (ii)pseudo-atomic(\pa) orbitals (constructed from ONCV pseudopotentials). Finally, we discuss the population energy diagrams of $2\cross 2\cross 2$ carbon diamond and Si\textsubscript{29}H\textsubscript{36} employing STO basis by Bunge as atom-centered orbitals.
\begin{table}[]
    \centering
    \begin{tabular}{|c|c|c|c|c|c|c|}
    \hline
     \multirow{2}{*}{System} & \multicolumn{2}{c|}{\Lob~}  & \multicolumn{2}{c|}{\DFTFE~ Bunge}  & \multicolumn{2}{c|}{\DFTFE~ \pa}  \\
     \cline{2-7}
     &$\mathcal{S}_c$&$\mathcal{S}$&$\mathcal{S}_c$&$\mathcal{S}$&$\mathcal{S}_c$&$\mathcal{S}$ \\
    \hline
    CO & 0.018 & 0.143 & 0.034 & 0.154 & 0.017 & 0.140 \\
    \hline
    H\textsubscript{2}O & 0.016 & 0.263 & 0.033 & 0.277 & 0.016 & 0.247 \\
    \hline
    O\textsubscript{2}$\uparrow$ & 0.015 & 0.136 & 0.036 & 0.152 & 0.009 & 0.133 \\
    O\textsubscript{2}$\downarrow$ & 0.012 & 0.135 & 0.029 & 0.150 & 0.008 & 0.132 \\
    \hline
    \end{tabular}
    \caption{\footnotesize{Comparison of absolute charge spill factor($\mathcal{S}_c$) and absolute spill factor($\mathcal{S}$) obtained using projections carried out in \DFTFE~and \Lob. \DFTFE~Bunge indicates the projection of finite-element discretized Kohn-Sham eigenfunctions to  STO basis by Bunge~ and \DFTFE~\pa~basis indicates projection onto pseudo-atomic orbitals.  Projection in \Lob~uses \texttt{pbeVaspfit2015} as auxiliary atom-centered basis.} }
    \label{tab: Spill Factors}
\end{table}
\vspace{-0.2in}
\paragraph{CO molecule:} Figures \ref{fig:CO PA comparison} and \ref{fig:CO Bunge comparison} show the comparison between the population energy diagrams obtained using the proposed \pfop~approach and \Lob. The results indicate excellent agreement with that obtained from \Lob. The C-O bond length in the CO molecule considered in this study is 1.14\angstrom.
\paragraph{H\textsubscript{2}O molecule:}  Figures \ref{fig:H2O PA comparison} and \ref{fig:H2O Bunge comparison} show the comparison between the population energy diagrams obtained using the proposed \pfop~approach and \Lob. The results indicate excellent agreement with that obtained from \Lob. The H\textsubscript{2}O molecule considered in this study has an O-H bond length of 0.971\angstrom. 
\paragraph{Spin polarized calculation on O\textsubscript{2} molecule:} Figures~\ref{fig:O2 PA up comparison} to~\ref{fig:O2 Bunge down comparison} show the comparison between the population energy diagrams obtained using the proposed \pfop~approach and \Lob~for the case of O$_2$ molecule. From Figure~\ref{fig:O2 PA up comparison} and Figure~\ref{fig:O2 Bunge up comparison}, we observe that the location and the number of peaks corresponding to \oop~and \ohp~are identical to that of \Lob~for the up spin channel. We observe a similar trend in Figure~\ref{fig:O2 PA down comparison} and Figure~\ref{fig:O2 Bunge down comparison} corresponding to the down spin channel. O\textsubscript{2} molecule considered in this study has an O-O bond length of 1.227\angstrom. 
\paragraph{Periodic $2\cross2\cross2$ supercell of carbon:}  Figure \ref{fig:C periodic Bunge comparison} shows the comparison of \pfop~obtained from using STO basis by Bunge  with that obtained from \Lob.  We pick the nearest neighbour carbon atoms as source and target atoms, and the corresponding $2s$-$2s$ and $2s$-$2p$ orbital interactions are plotted in Figure \ref{fig:C periodic Bunge comparison}. The corresponding C-C bond length is around 1.55\angstrom.
\paragraph{Si\textsubscript{29}H\textsubscript{36} nanocluster:} Figure~\ref{fig:Si29H36 Bunge comparison} illustrates the comparison of \pfop~obtained from using STO basis by Bunge with that obtained from \Lob. The $1s$-$3s$ and $1s$-$3p$ interaction between Si-H in one of the SiH$_2$ dihedrals are plotted in this Figure~\ref{fig:Si29H36 Bunge comparison}. The corresponding Si-H bond length is 1.49 \angstrom, and we observe that the results are in very good agreement with \Lob.   
\begin{figure}[H]
\includegraphics[scale=0.325]{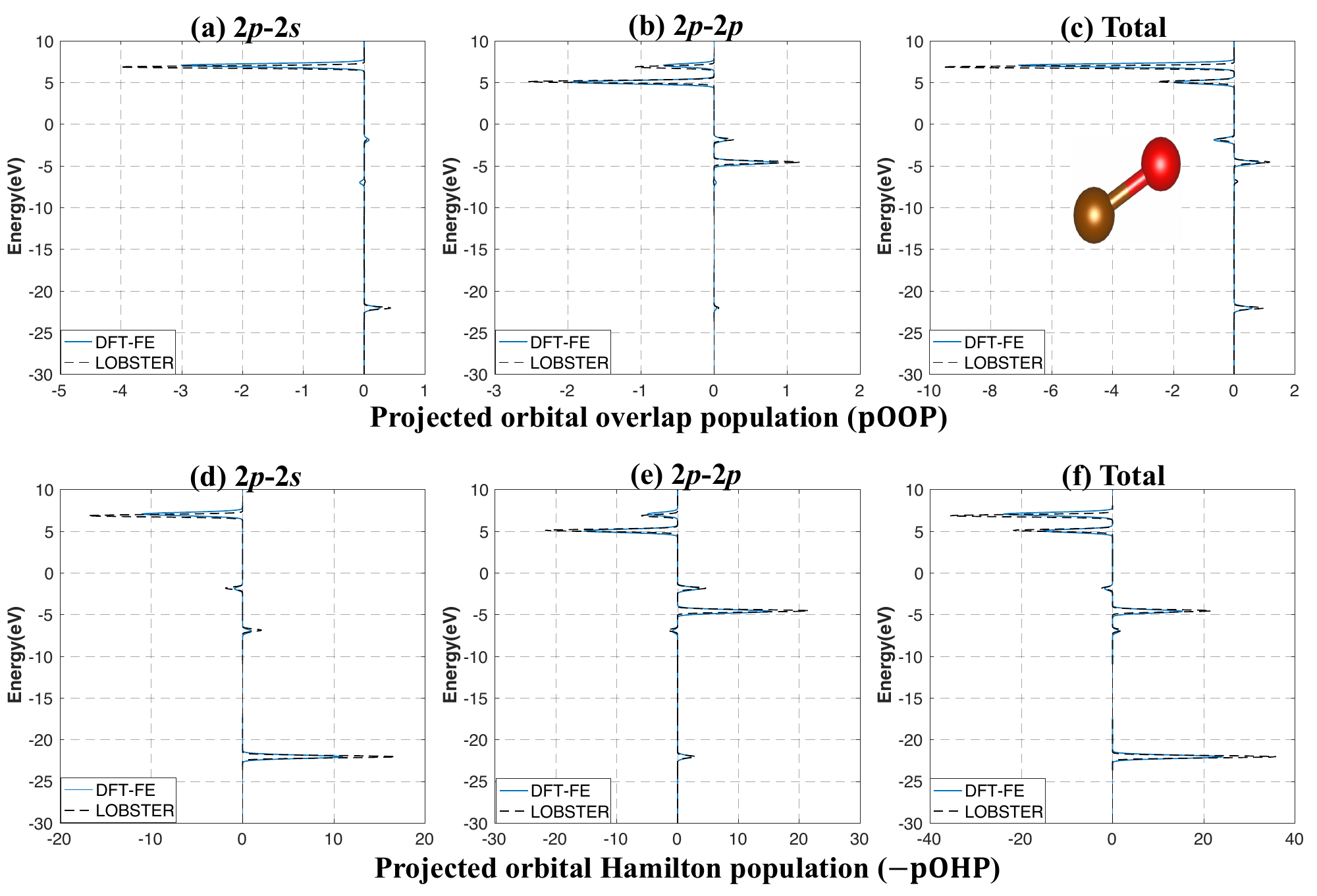}
\caption[short]{\footnotesize{Comparison between \pfop~implemented in \DFTFE~using \pa~orbitals, and \Lob~for C-O atom pair in CO molecule. The top row shows the projected orbital overlap population, with the  sub-figures (a) and (b) in this row showing the contributions of C$_{2p}$-O$_{2s}$ and C$_{2p}$-O$_{2p}$ to the total orbital overlap population that is plotted in sub-figure (c). The bottom row shows negative of the projected orbital Hamilton population. The sub-figures in the bottom row (d) and (e) show the contributions of C$_{2p}$-O$_{2s}$ and C$_{2p}$-O$_{2p}$ to the total orbital Hamilton population that is plotted in sub-figure (f). Energy-scale is shifted such that Fermi level ($\epsilon_F$) is zero. \textbf{Case study:} CO molecule}}\label{fig:CO PA comparison}
\end{figure}
\begin{figure}[H]
\includegraphics[scale=0.325]{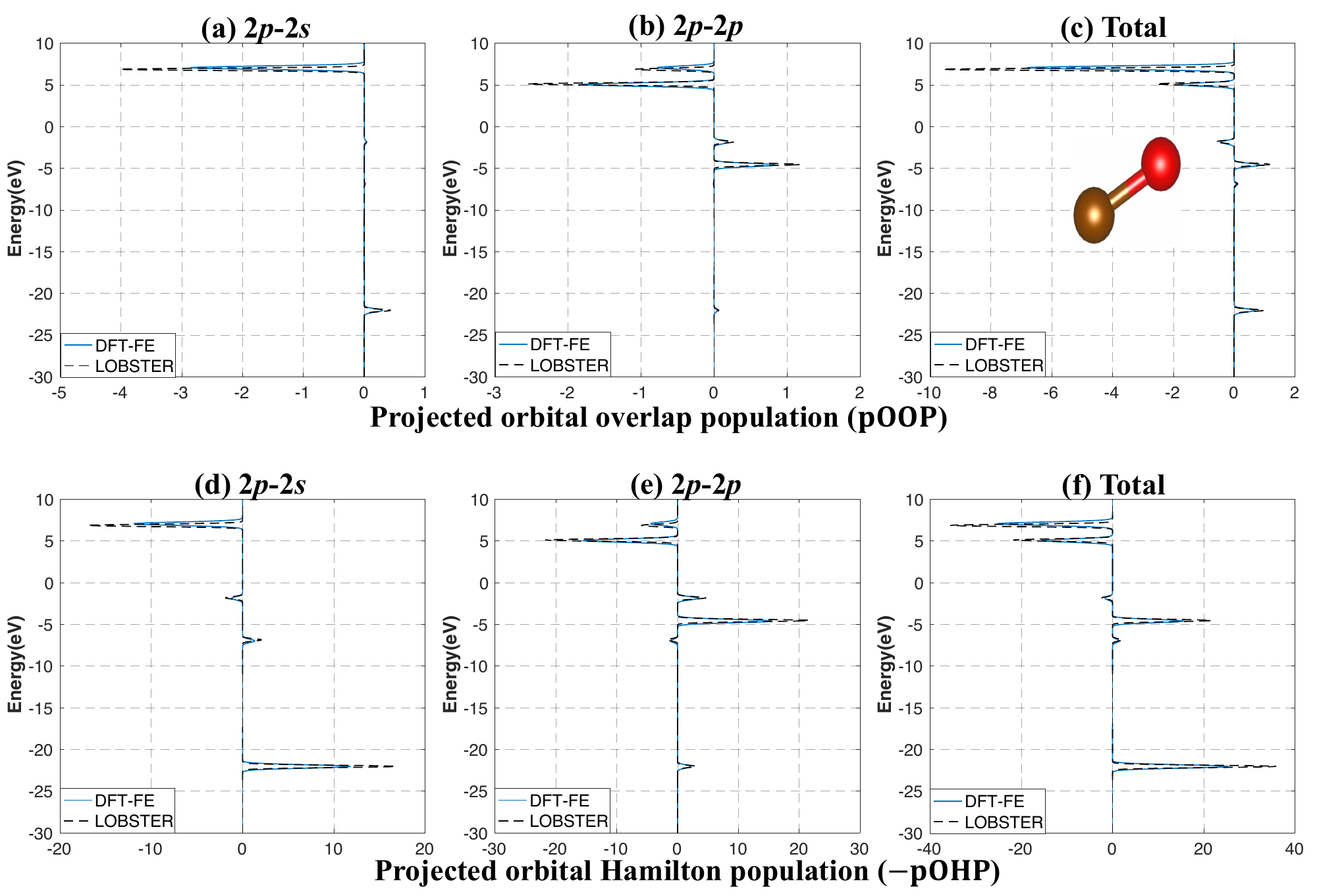}
\caption[short]{\footnotesize{Comparison between \pfop~implemented in \DFTFE~using STO basis by Bunge, and \Lob~for C-O atom pair in CO molecule. The top row shows the projected orbital overlap population, with the  sub-figures (a) and (b) in this row showing the contributions of C$_{2p}$-O$_{2s}$ and C$_{2p}$-O$_{2p}$ to the total orbital overlap population that is plotted in sub-figure (c). The bottom row shows negative of the projected orbital Hamilton population. The sub-figures in the bottom row (d) and (e) show the contributions of C$_{2p}$-O$_{2s}$ and C$_{2p}$-O$_{2p}$ to the total orbital Hamilton population that is plotted in sub-figure (f). Energy-scale is shifted such that Fermi level ($\epsilon_F$) is zero. \textbf{Case study:} CO molecule}}\label{fig:CO Bunge comparison}
\end{figure}

\begin{figure}[H]
\includegraphics[scale=0.325]{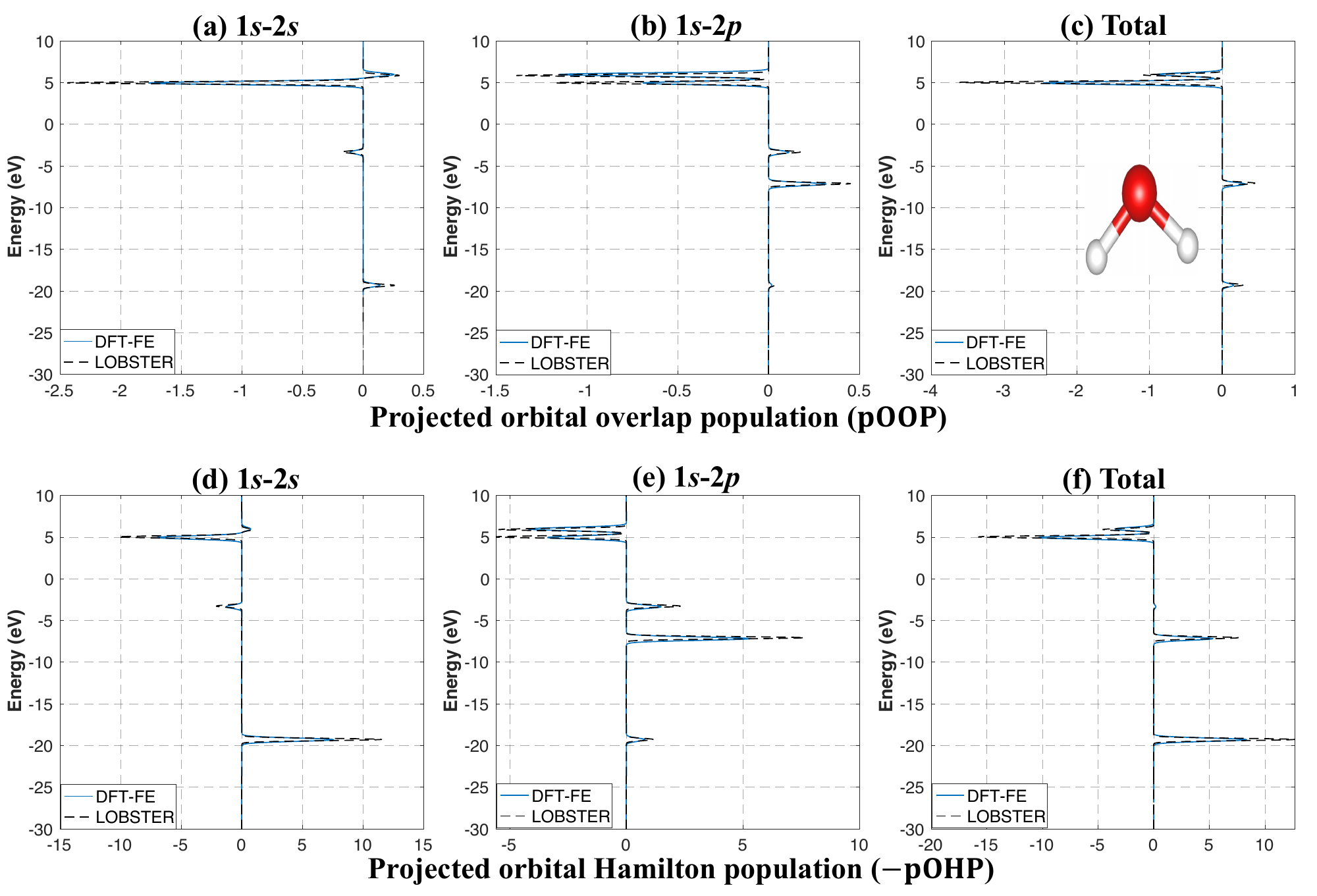}
\caption[short]{\footnotesize{Comparison between \pfop~implemented in \DFTFE~using \pa~orbitals, and \Lob~for H-O atom pair in H\textsubscript{2}O molecule. The top row shows the projected orbital overlap population, with the  sub-figures (a) and (b) in this row showing the contributions of H$_{1s}$-O$_{2s}$ and H$_{1s}$-O$_{2p}$ to the total orbital overlap population that is plotted in sub-figure (c). The bottom row shows negative of the projected orbital Hamilton population. The sub-figures in the bottom row (d) and (e) show the contributions of H$_{1s}$-O$_{2s}$ and H$_{1s}$-O$_{2p}$ to the total orbital Hamilton population that is plotted in sub-figure (f). Energy-scale is shifted such that Fermi level ($\epsilon_F$) is zero. \textbf{Case study:} H\textsubscript{2}O molecule}}\label{fig:H2O PA comparison}
\end{figure}
\begin{figure}[H]
\includegraphics[scale=0.325]{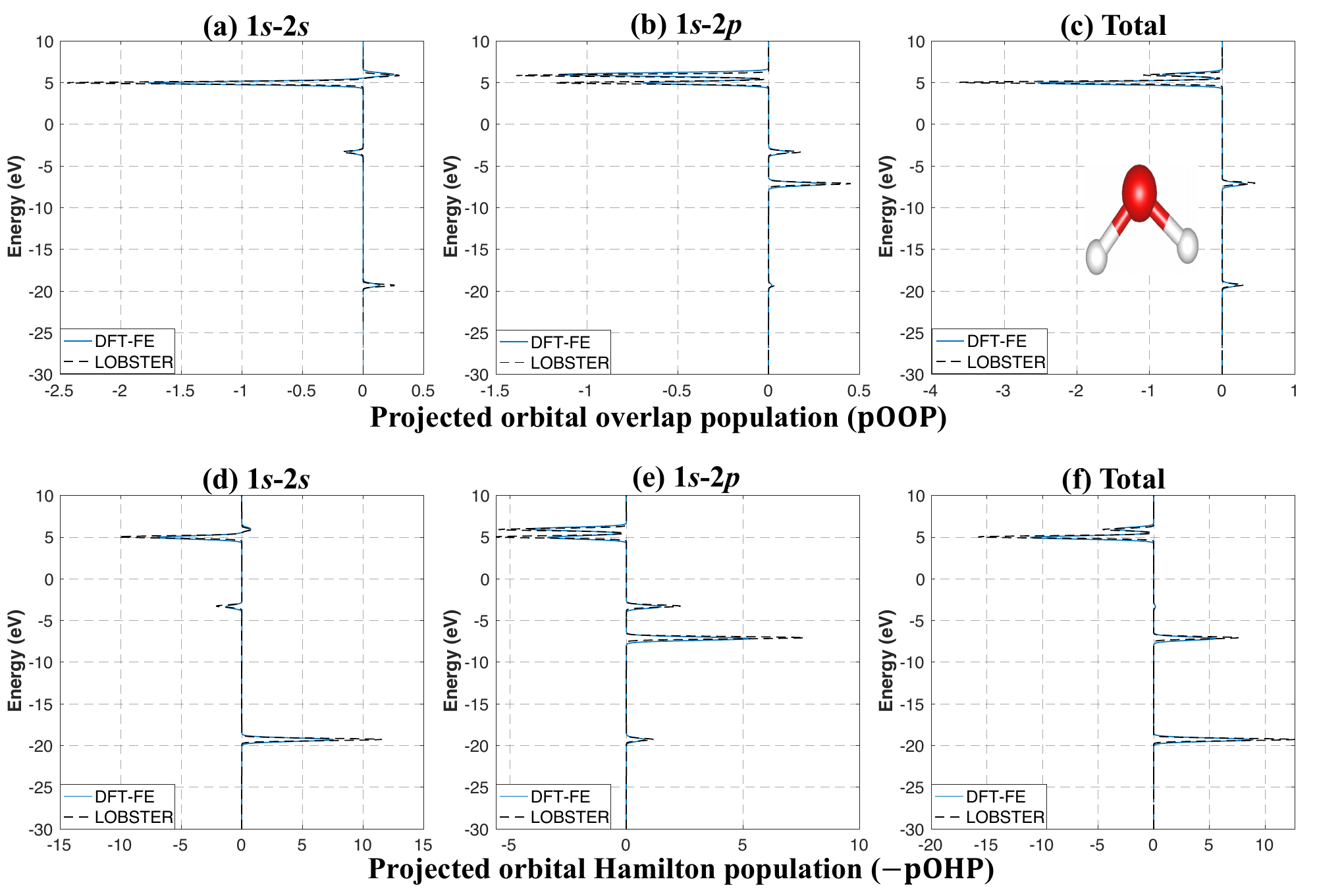}
\caption[short]{\footnotesize{Comparison between \pfop~implemented in \DFTFE~using STO basis by Bunge, and \Lob~for H-O atom pair in H\textsubscript{2}O molecule. The top row shows the projected orbital overlap population, with the  sub-figures (a) and (b) in this row showing the contributions of H$_{1s}$-O$_{2s}$ and H$_{1s}$-O$_{2p}$ to the total orbital overlap population that is plotted in sub-figure (c). The bottom row shows negative of the projected orbital Hamilton population. The sub-figures in the bottom row (d) and (e) show the contributions of H$_{1s}$-O$_{2s}$ and H$_{1s}$-O$_{2p}$ to the total orbital Hamilton population that is plotted in sub-figure (f). Energy-scale is shifted such that Fermi level ($\epsilon_F$) is zero.  \textbf{Case study:} H\textsubscript{2}O molecule}}\label{fig:H2O Bunge comparison}
\end{figure}

\begin{figure}[H]
\includegraphics[scale=0.325]{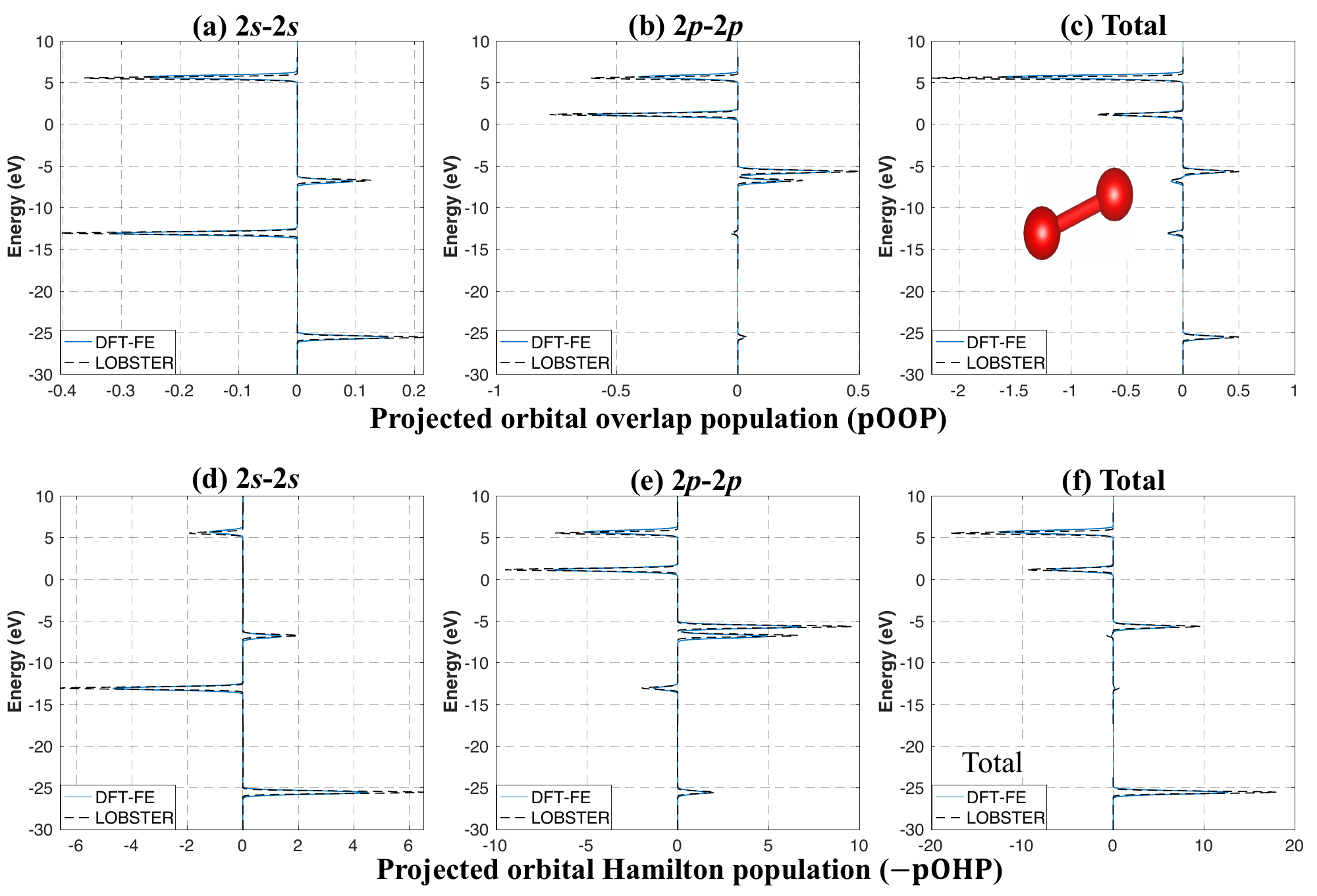}
\caption[short]{\footnotesize{Comparison between \pfop~implemented in \DFTFE~using \pa~orbitals, and \Lob~for O-O atom pair in O\textsubscript{2} molecule. The top row shows the projected orbital overlap population, with the  sub-figures (a) and (b) in this row showing the contributions of O$_{2s}$-O$_{2s}$ and O$_{2p}$-O$_{2p}$ to the total orbital overlap population that is plotted in sub-figure (c). The bottom row shows negative of the projected orbital Hamilton population. The sub-figures in the bottom row (d) and (e) show the contributions of O$_{2s}$-O$_{2s}$ and O$_{2p}$-O$_{2p}$ to the total orbital Hamilton population that is plotted in sub-figure (f). Energy-scale is shifted such that Fermi level ($\epsilon_F$) is zero. \textbf{Case study:} spin-polarized O\textsubscript{2}~$\uparrow$}}\label{fig:O2 PA up comparison}
\end{figure}
\begin{figure}[H]
\includegraphics[scale=0.325]{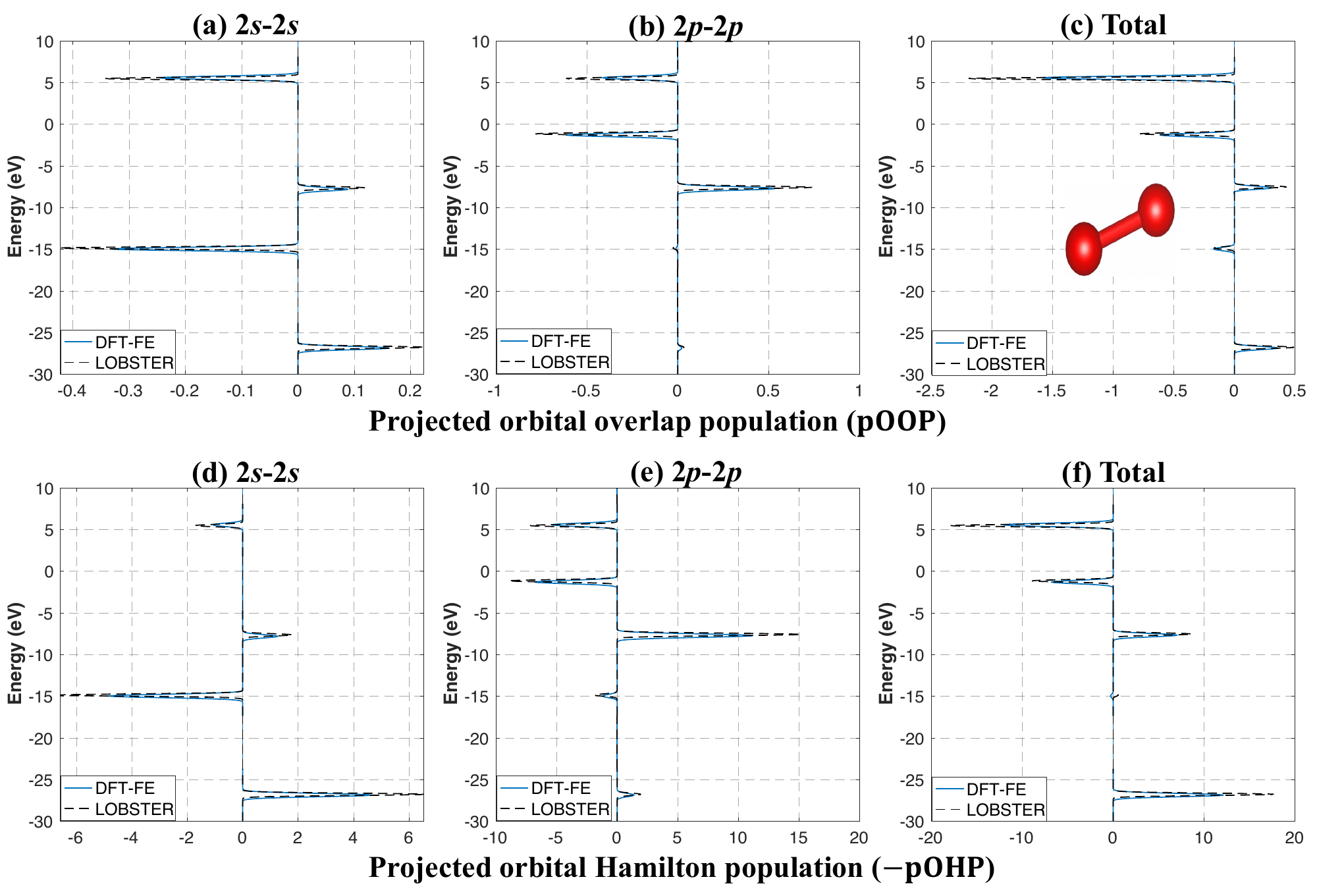}
\caption[short]{\footnotesize{Comparison between \pfop~implemented in \DFTFE~using \pa~orbitals, and \Lob~for O-O atom pair in O\textsubscript{2} molecule. The top row shows the projected orbital overlap population, with the  sub-figures (a) and (b) in this row showing the contributions of O$_{2s}$-O$_{2s}$ and O$_{2p}$-O$_{2p}$ to the total orbital overlap population that is plotted in sub-figure (c). The bottom row shows negative of the projected orbital Hamilton population. The sub-figures in the bottom row (d) and (e) show the contributions of O$_{2s}$-O$_{2s}$ and O$_{2p}$-O$_{2p}$ to the total orbital Hamilton population that is plotted in sub-figure (f). Energy-scale is shifted such that Fermi level ($\epsilon_F$) is zero. \textbf{Case study:} spin-polarized O\textsubscript{2}~$\downarrow$}}\label{fig:O2 PA down comparison}
\end{figure}
\begin{figure}[H]
\includegraphics[scale=0.325]{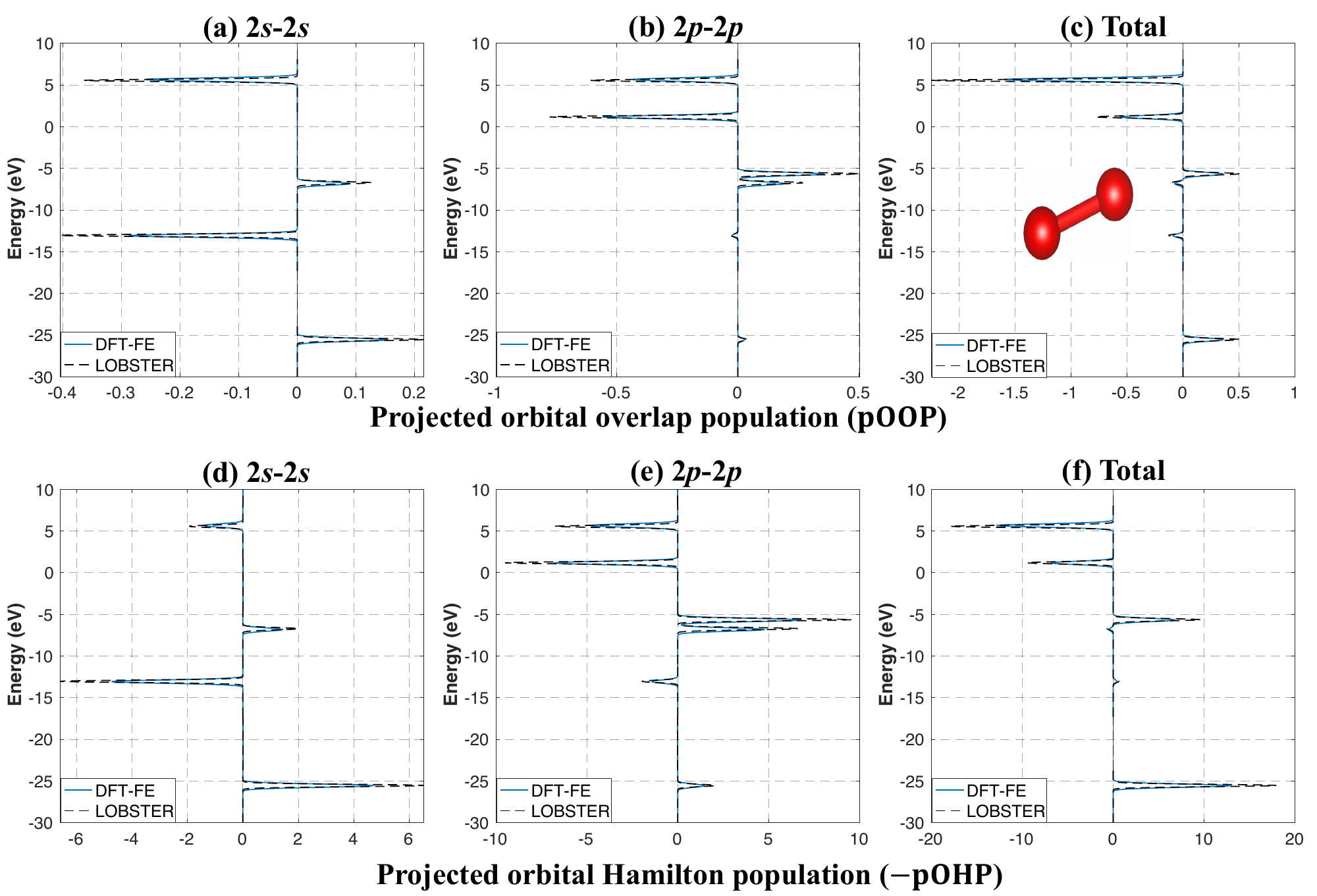}
\caption[short]{\footnotesize{Comparison between \pfop~implemented in \DFTFE~using STO basis by Bunge, and \Lob~for O-O atom pair in O\textsubscript{2} molecule. The top row shows the projected orbital overlap population, with the  sub-figures (a) and (b) in this row showing the contributions of O$_{2s}$-O$_{2s}$ and O$_{2p}$-O$_{2p}$ to the total orbital overlap population that is plotted in sub-figure (c). The bottom row shows negative of the projected orbital Hamilton population. The sub-figures in the bottom row (d) and (e) show the contributions of O$_{2s}$-O$_{2s}$ and O$_{2p}$-O$_{2p}$ to the total orbital Hamilton population that is plotted in sub-figure (f). Energy-scale is shifted such that Fermi level ($\epsilon_F$) is zero. \textbf{Case study:} spin-polarized O\textsubscript{2}~$\uparrow$}}\label{fig:O2 Bunge up comparison}
\end{figure}
\begin{figure}[H]
\includegraphics[scale=0.325]{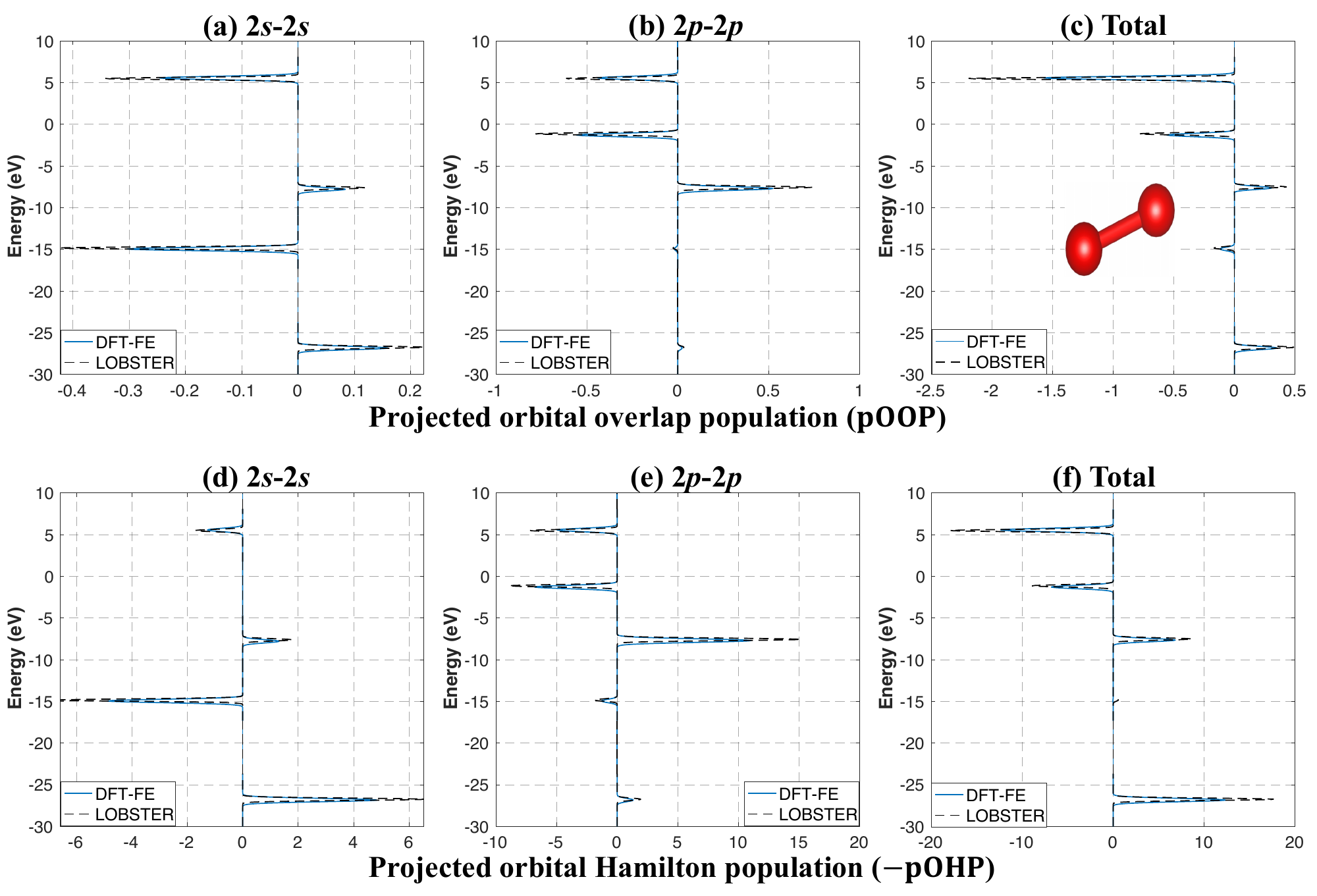}
\caption[short]{\footnotesize{Comparison between \pfop~implemented in \DFTFE~using STO basis by Bunge, and \Lob~for O-O atom pair in O\textsubscript{2} molecule. The top row shows the projected orbital overlap population, with the  sub-figures (a) and (b) in this row showing the contributions of O$_{2s}$-O$_{2s}$ and O$_{2p}$-O$_{2p}$ to the total orbital overlap population that is plotted in sub-figure (c). The bottom row shows negative of the projected orbital Hamilton population. The sub-figures in the bottom row (d) and (e) show the contributions of O$_{2s}$-O$_{2s}$ and O$_{2p}$-O$_{2p}$ to the total orbital Hamilton population that is plotted in sub-figure (f). Energy-scale is shifted such that Fermi level ($\epsilon_F$) is zero. \textbf{Case study:} spin-polarized O\textsubscript{2}~$\downarrow$}}\label{fig:O2 Bunge down comparison}
\end{figure}

\begin{figure}[H]
\includegraphics[scale=0.32]{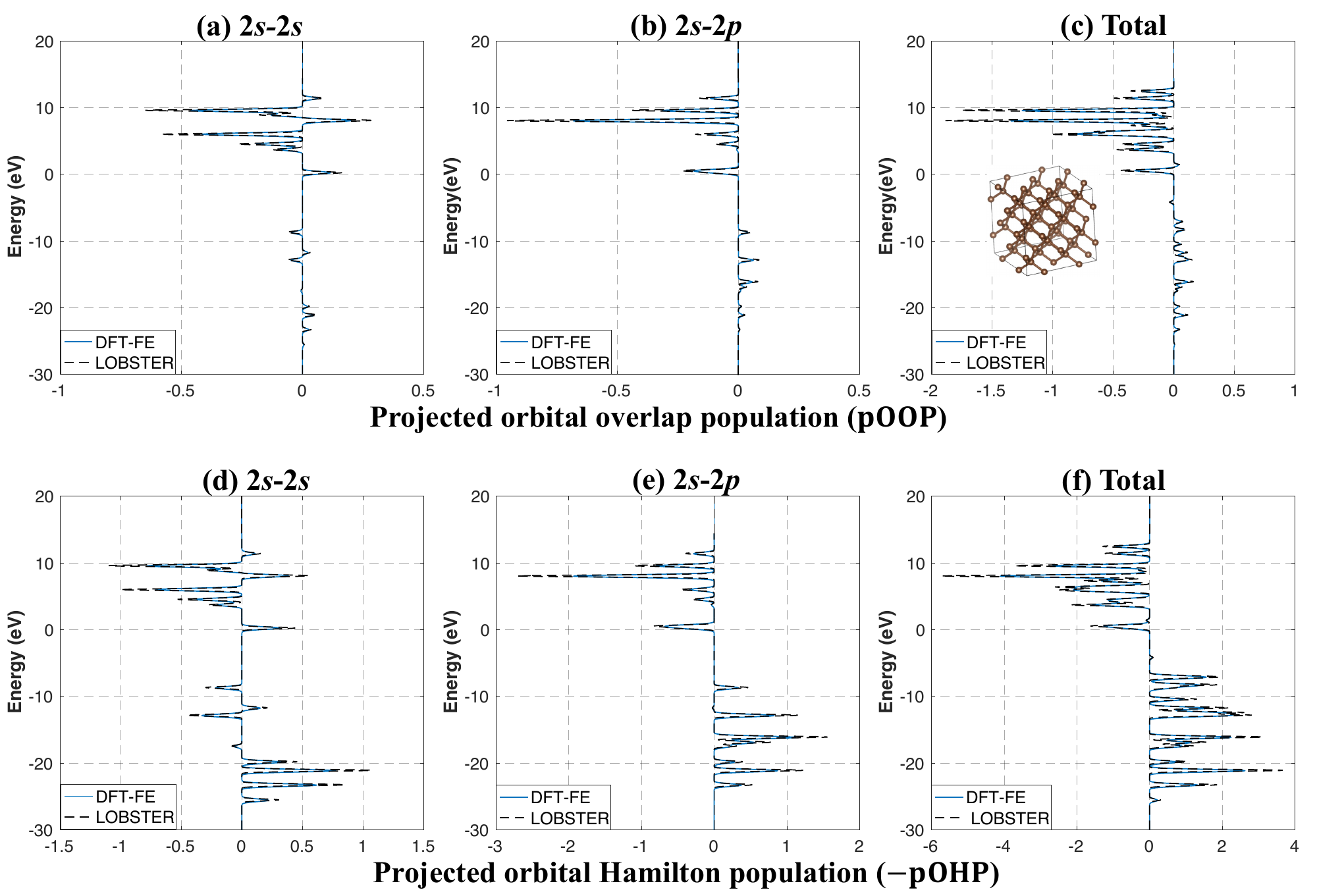}
\caption[short]{\footnotesize{Comparison between \pfop~implemented in \DFTFE~using STO basis by Bunge, and \Lob~for nearest neighbour C-C atom pair in carbon diamond supercell. The top row shows the orbital overlap population, with the sub-figures (a) and (b) in this row showing the contributions of C$_{2s}$-C$_{2s}$ and C$_{2s}$-C$_{2p}$ to the total orbital overlap population that is plotted in sub-figure (c). The bottom row shows the negative of the orbital Hamilton population. The sub-figures in the bottom row (d) and (e) show the contributions of C$_{2s}$-C$_{2s}$ and C$_{2s}$-C$_{2p}$ to the total orbital Hamilton population that is plotted in sub-figure (f). Energy-scale is shifted such that Fermi level ($\epsilon_F$) is zero. \textbf{Case study:} $2\cross2\cross2$ carbon diamond supercell with periodic boundary conditions using a $\Gamma$ point.}}\label{fig:C periodic Bunge comparison}
\end{figure}

\begin{figure}[H]
\includegraphics[scale=0.32]{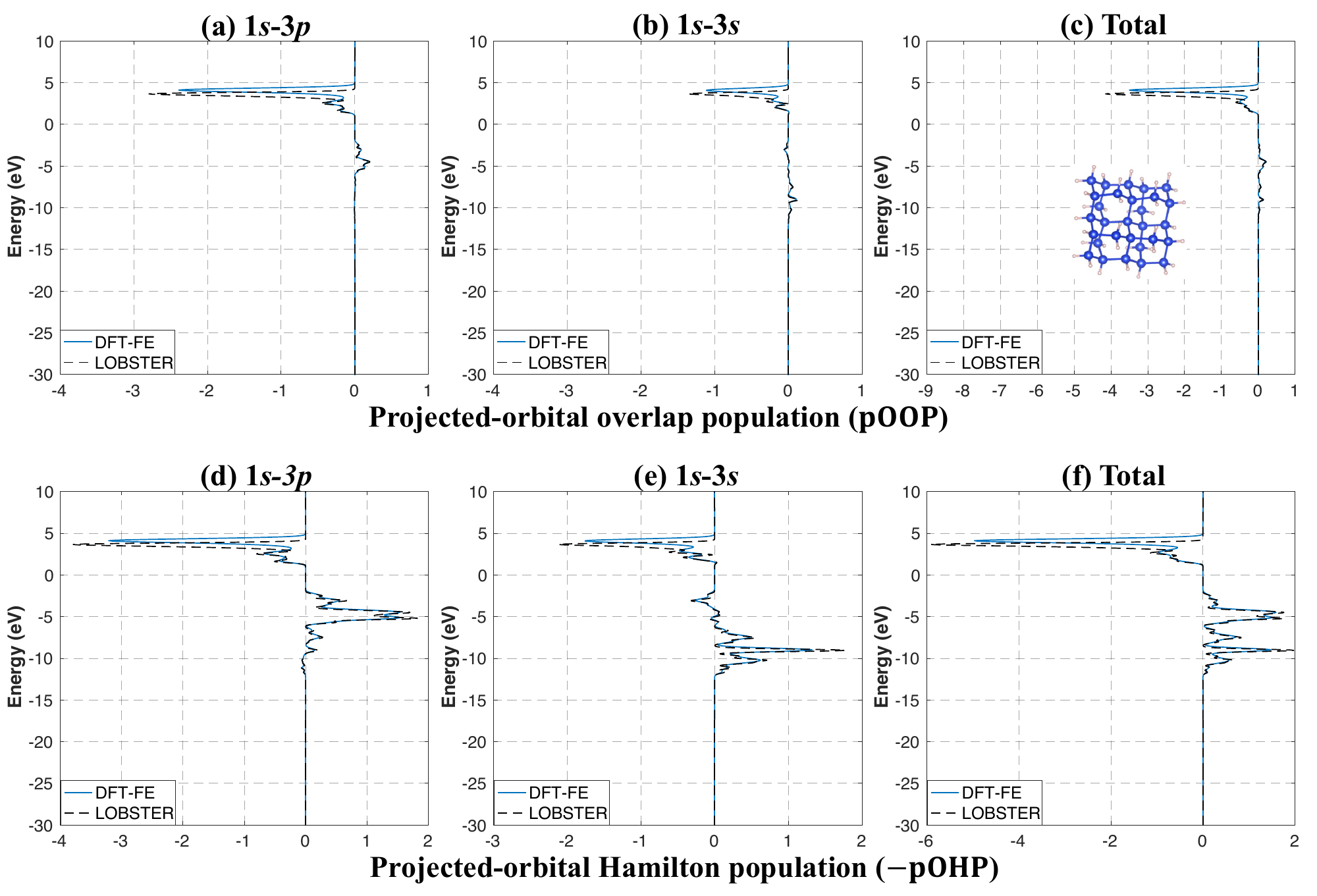}
\caption[short]{\footnotesize{Comparison between \pfop~implemented in \DFTFE~using STO basis by Bunge, and \Lob~for nearest neighbor Si-H atom pair. The top row shows the orbital overlap population, with the  sub-figures (a) and (b) in this row showing the contributions of H$_{1s}$-Si$_{3p}$ and H$_{1s}$-Si$_{3s}$ to the total orbital overlap population that is plotted in sub-figure (c). The bottom row shows negative of the orbital Hamilton population. The sub-figures in the bottom row (d) and (e) show the contributions of H$_{1s}$-Si$_{3p}$ and H$_{1s}$-Si$_{3s}$ to the total orbital Hamilton population that is plotted in sub-figure (f). Energy-scale is shifted such that Fermi level ($\epsilon_F$) is zero. \textbf{Case study:} Single-fold Si\textsubscript{29}H\textsubscript{36} nanoparticle}}\label{fig:Si29H36 Bunge comparison}
\end{figure}
 \paragraph{$\bk$-dependent population analysis in 1x1x1 orthogonal unit-cell of carbon:}\label{subsection: kdependentOPA}
 We discuss here the comparison of $\bk$-dependent population analysis with \Lob. We consider the 8 atom orthogonal unit-cell of C diamond crystal with lattice constant $3.573$\angstrom. The nearest C-C bond length is 1.547\angstrom. We first compute the self-consistent converged ground-state for the benchmark study on a $4\times4\times4$~$\bk$-point grid in both \DFTFE~and \QE. Subsequently, we perform a non self-consistent calculation separately on 3 high symmetry k-points: a) M:(0.5,0.5,0) b) R:(0.5,0.5,0.5) c) X:(0.0,0.5,0.0) in reciprocal space using  both \DFTFE~and \QE. Finally, we perform the population analysis with \Lob and \DFTFE~(using \pfop) for each of these k-points.  Figure~\ref{fig:C periodic complex comparison} shows the comparison of the total \oop($\bk$) and total \ohp($\bk$) with total $\mathtt{COOP}(\bk)$ and total $\mathtt{COHP}(\bk)$ computed using \Lob~for a pair of nearest neighbouring C atoms and we note that the results show a close agreement between \DFTFE~and \Lob.
\begin{figure}[H]
\centering
\includegraphics[scale=0.375]{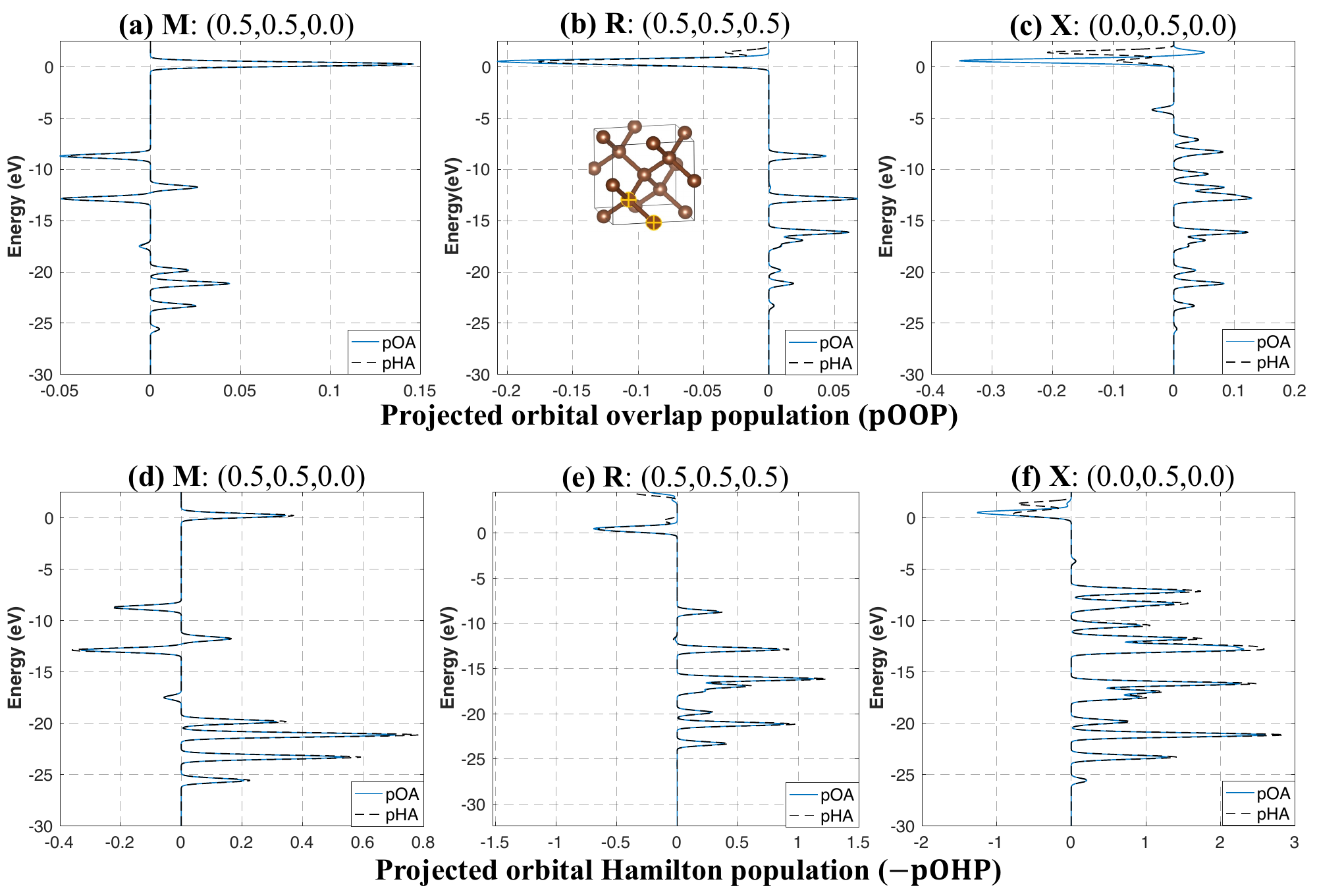}
\centering
\caption[short]{\footnotesize{Comparison between \pfop~implemented in \DFTFE~using \pa~orbitals, and \Lob~for nearest neighbour C-C atom pair in carbon diamond cubic unit cell. The top row shows the total projected orbital overlap population, with the sub-figures (a), (b) and (c) in this row showing the overlap population for $M:(0.5,0.5,0.0)$, $R:(0.5,0.5,0.5)$ and $X:(0.0,0.5,0.0)$ points respectively in the Brillouin zone.  The bottom row shows the negative of the total projected orbital Hamilton population. The sub-figures in the bottom row (d), (e) and (f) in this row show the Hamilton population for $M:(0.5,0.5,0.0)$, $R:(0.5,0.5,0.5)$ and $X:(0.0,0.5,0.0)$ points respectively in the Brillouin zone. Energy-scale is shifted such that Fermi level ($\epsilon_F$) is zero. \textbf{Case study:} Carbon diamond $1\cross1\cross1$ cubic-unit cell with periodic boundary conditions}}\label{fig:C periodic complex comparison}
\end{figure}

\subsection{Projected Hamiltonian population (\pfhp)} \label{subsection: pfhpResults}
In this section, we illustrate the comparisons between \pfop~and \pfhp~on few representative material systems not discussed in the main manuscript. To this end, we compare the population energy diagrams of CO molecule in Figure~\ref{fig:CO new comparison}, H\textsubscript{2}O molecule in Figure~\ref{fig:H2O new comparison} and spin-polarized O\textsubscript{2} molecule in Figures~\ref{fig:O2 down new comparison} and \ref{fig: O2 up new comparison}. We observe a very good agreement between the two methods. 
\begin{figure}[H]
\includegraphics[scale=0.375]{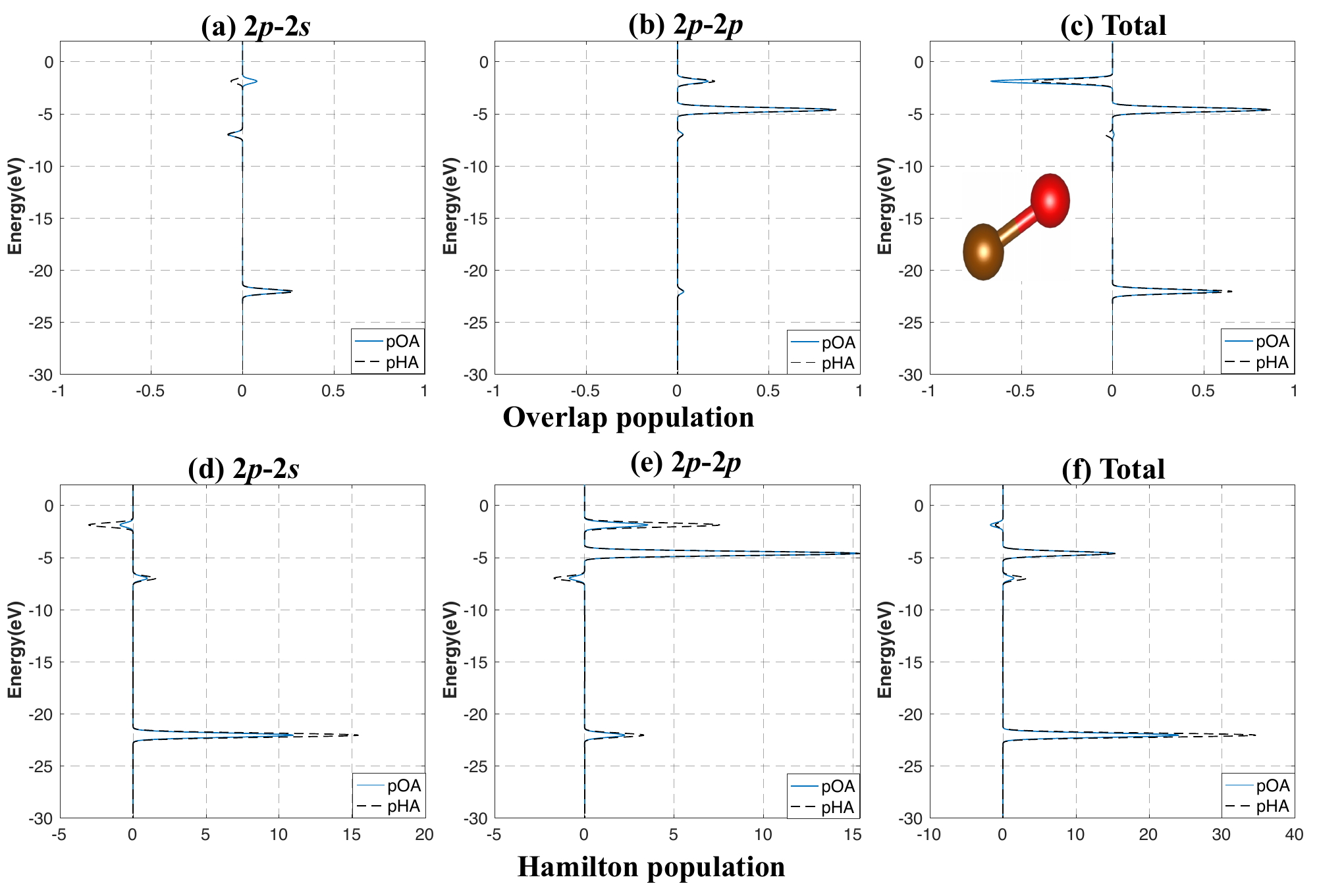}
\caption[short]{\footnotesize{Comparison of overlap and Hamilton populations between the two proposed methods of projected population analysis (\pfop~and \pfhp) for C-O atom pair in CO molecule. The top row shows the overlap population obtained using both these methods. The sub-figures (a) and (b) in this row show the contributions of C$_{2p}$-O$_{2s}$ and C$_{2p}$-O$_{2p}$ to the total overlap population that is plotted in sub-figure (c). The bottom row shows the negative of the Hamilton population for both methods. The sub-figures in this bottom row (d) and (e) show the contributions of C$_{2p}$-O$_{2s}$ and C$_{2p}$-O$_{2p}$ to the total Hamilton population that is plotted in sub-figure (f). Energy-scale is shifted such that Fermi level ($\epsilon_F$) is zero. \textbf{Case study:} CO molecule}}\label{fig:CO new comparison}
\end{figure}
\begin{figure}[H]
\includegraphics[scale=0.325]{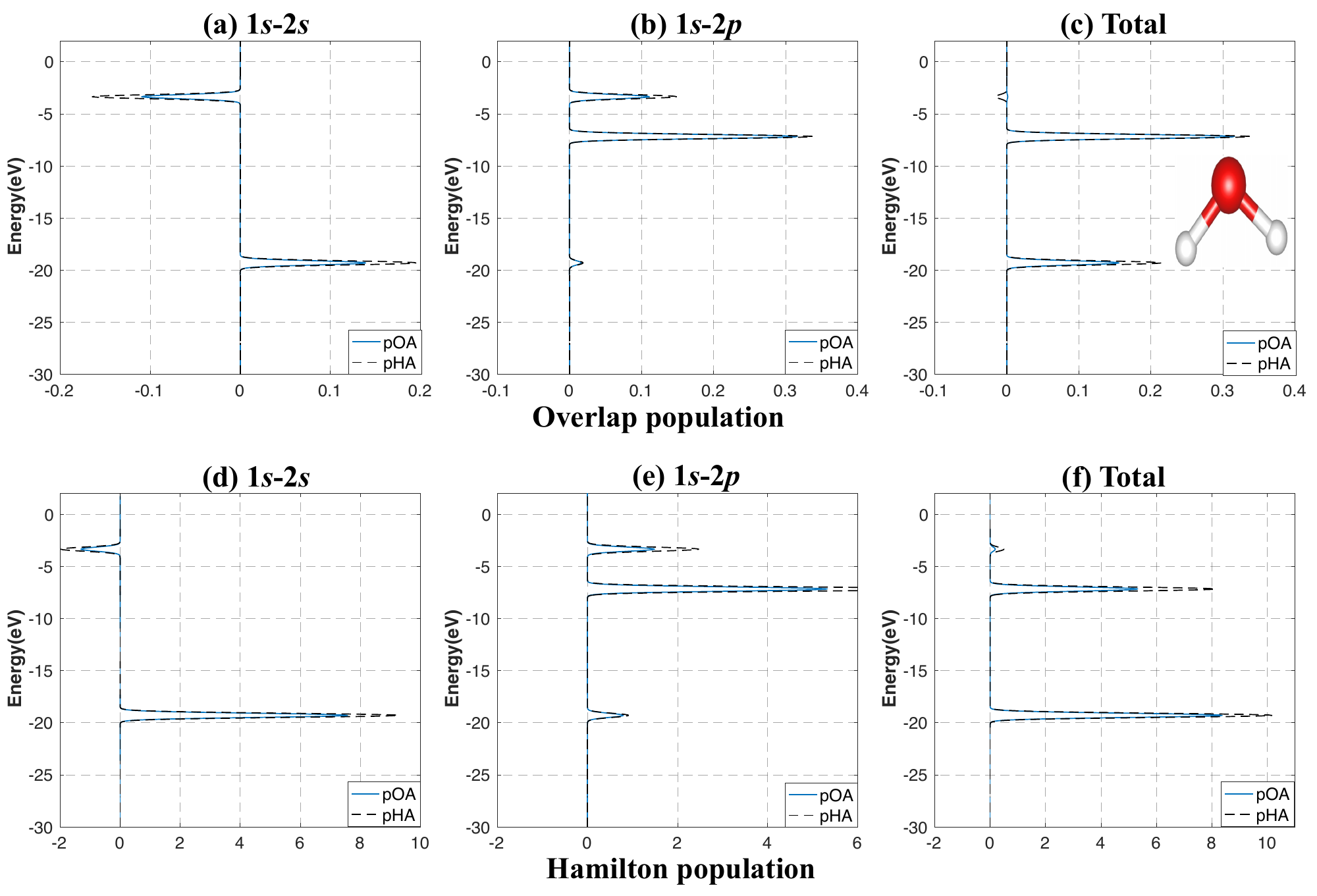}
\caption[short]{\footnotesize{Comparison of overlap and Hamilton populations between the two proposed methods of projected population analysis (\pfop~and \pfhp) for H-O atom pair in H\textsubscript{2}O molecule. The top row shows the overlap population obtained using both these methods. The sub-figures (a) and (b) in this row show the contributions of H$_{1s}$-O$_{2s}$ and H$_{1s}$-O$_{2p}$ to the total overlap population that is plotted in sub-figure (c). The bottom row shows the negative of the Hamilton population for both methods. The sub-figures in this bottom row (d) and (e) show the contributions of C$_{2s}$-C$_{2s}$ and C$_{2s}$-C$_{2p}$ to the total Hamilton population that is plotted in sub-figure (f). Energy-scale is shifted such that Fermi level ($\epsilon_F$) is zero. \textbf{Case study:} H\textsubscript{2}O molecule}}\label{fig:H2O new comparison}
\end{figure}
\begin{figure}[H]
\includegraphics[scale=0.325]{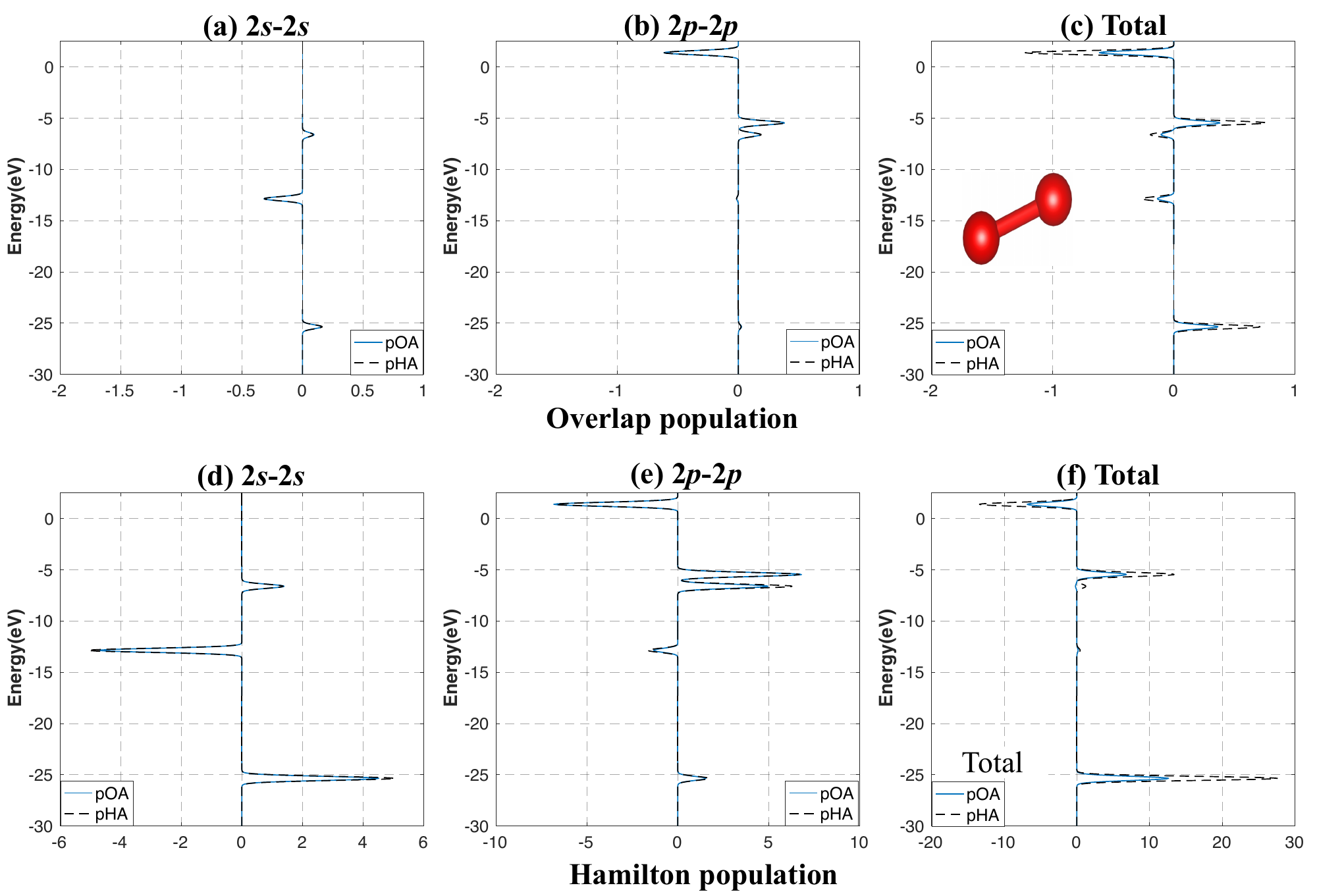}
\caption[short]{\footnotesize{Comparison of overlap and Hamilton populations between the two proposed methods (\pfop~and \pfhp) for O-O atom pair in O\textsubscript{2} molecule. The top row shows the overlap population obtained using both these methods. The sub-figures (a) and (b) in this row show the contributions of O$_{2s}$-O$_{2s}$ and O$_{2p}$-O$_{2p}$ to the total overlap population that is plotted in sub-figure (c). The bottom row shows the negative of the Hamilton population for both methods. The sub-figures in this bottom row (d) and (e) show the contributions of O$_{2s}$-O$_{2s}$ and O$_{2p}$-O$_{2p}$ to the total Hamilton population that is plotted in sub-figure (f). Energy-scale is shifted such that Fermi level ($\epsilon_F$) is zero. \textbf{Case study:} Spin-polarized O\textsubscript{2}~$\uparrow$}}\label{fig: O2 up new comparison}
\end{figure}
\begin{figure}[H]
\includegraphics[scale=0.375]{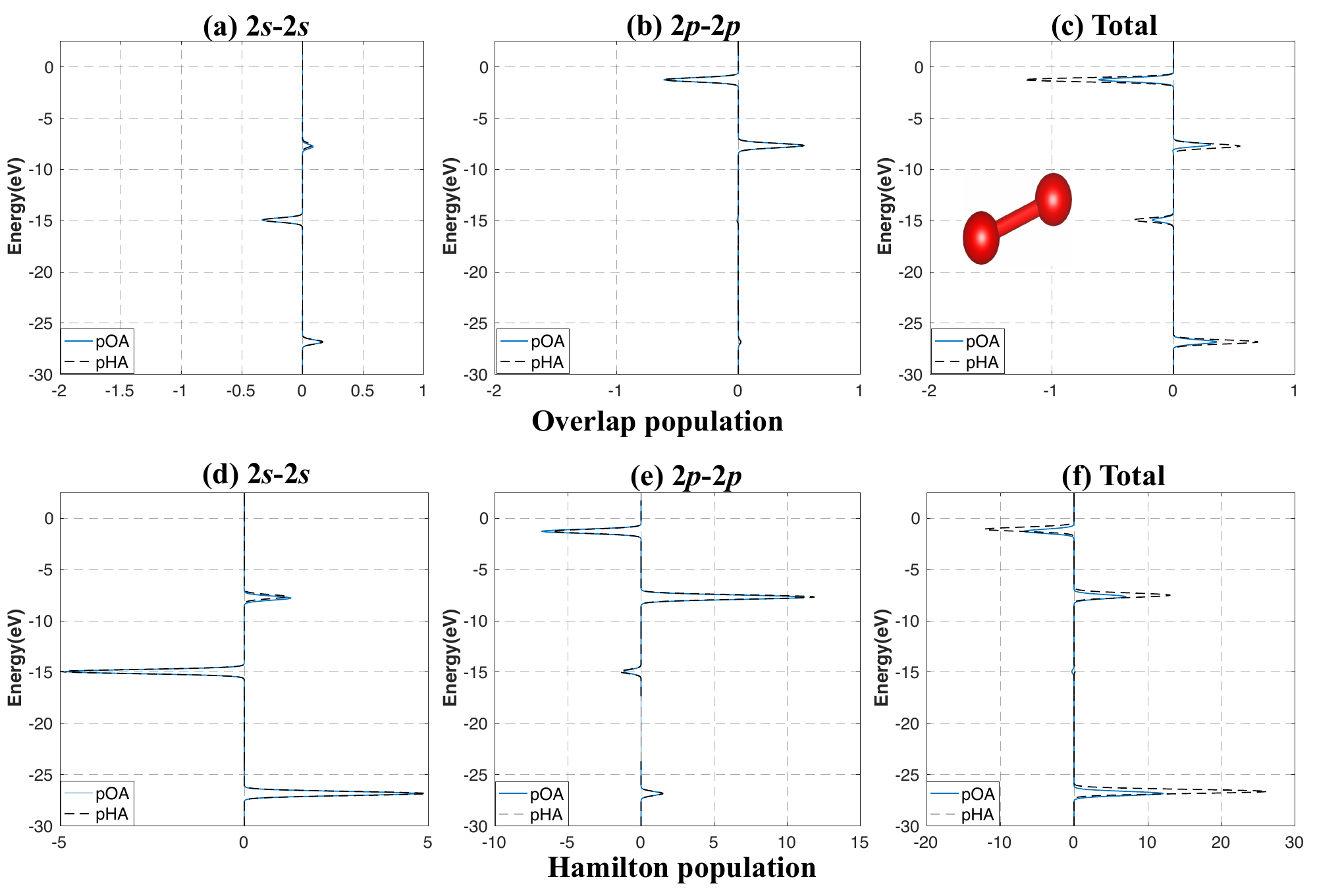}
\caption[short]{\footnotesize{Comparison of overlap and Hamilton populations between the two proposed methods of projected population analysis (\pfop~and \pfhp) for O-O atom pair in O\textsubscript{2} molecule. The top row shows the overlap population obtained using both these methods. The sub-figures (a) and (b) in this row show the contributions of O$_{2s}$-O$_{2s}$ and O$_{2p}$-O$_{2p}$ to the total overlap population that is plotted in sub-figure (c). The bottom row shows the negative of the Hamilton population for both methods. The sub-figures in this bottom row (d) and (e) show the contributions of O$_{2s}$-O$_{2s}$ and O$_{2p}$-O$_{2p}$ to the total Hamilton population that is plotted in sub-figure (f). Energy-scale is shifted such that Fermi level ($\epsilon_F$) is zero. \textbf{Case study:} Spin-polarized O\textsubscript{2}~$\downarrow$}}\label{fig:O2 down new comparison}
\end{figure}

\end{suppinfo}

\pagebreak
\bibliography{JCTC}
\end{document}